\documentclass[prx,twocolumn,english,superscriptaddress,floatfix,longbibliography]{revtex4-2}
\usepackage{amsmath,amsfonts,amssymb,fullpage,color,graphicx,xcolor}
\usepackage[colorlinks]{hyperref}
\newcommand{\CC}{\mathbb{C}}
\newcommand{\RR}{\mathbb{R}}

\newcommand{\set}[1]{\lbrace #1 \rbrace}
\newcommand{\abs}[1]{\lvert #1 \rvert}
\newcommand{\norm}[1]{\lVert #1 \rVert}

\newcommand{\bra}[1]{\langle #1 \rvert}
\newcommand{\ket}[1]{\lvert #1 \rangle}

\newcommand{\Tr}{ {\rm{Tr}}}
\DeclareMathOperator{\EE}{\mathbb{E}}
\DeclareMathOperator{\diag}{diag}
\DeclareMathOperator{\uvec}{uvec}

\newtheorem{theorem}{Theorem}

\numberwithin{equation}{section}
\numberwithin{theorem}{section}
\numberwithin{figure}{section}

\begin{document}

\title{Compressed Sensing Measurement of Long-Range Correlated Noise}
\date{\today}
\author{Alireza Seif}

\affiliation{Department of Physics and the Joint Quantum Institute (JQI), University of Maryland, College Park, MD 20742} 
\affiliation{Joint  Center  for  Quantum  Information  and  Computer  Science (QuICS), University  of  Maryland,  College  Park,  MD  20742}
\affiliation{Pritzker School of Molecular Engineering, University of Chicago, Chicago, IL 60637}

\author{Mohammad Hafezi}
\affiliation{Department of Physics and the Joint Quantum Institute (JQI), University of Maryland, College Park, MD 20742} 
\affiliation{Department of Electrical and Computer Engineering, University of Maryland, College Park, MD 20742}

\author{Yi-Kai Liu}
\affiliation{Joint  Center  for  Quantum  Information  and  Computer  Science (QuICS), University  of  Maryland,  College  Park,  MD  20742}
\affiliation{Applied and Computational Mathematics Division, National  Institute  of  Standards  and  Technology,  Gaithersburg,  MD  20899}

\begin{abstract}
Long-range correlated errors can severely impact the performance of NISQ (noisy intermediate-scale quantum) devices, and fault-tolerant quantum computation. Characterizing these errors is important for improving the performance of these devices, via calibration and error correction, and to ensure correct interpretation of the results.  We propose a compressed sensing method for detecting two-qubit correlated dephasing errors, assuming only that the correlations are \textit{sparse} (i.e., at most $s$ pairs of qubits have correlated errors, where  $s \ll n(n-1)/2$, and $n$ is the total number of qubits). In particular, our method can detect \textit{long-range} correlations between any two qubits in the system (i.e., the correlations are not restricted to be geometrically local). 

Our method is highly scalable: it requires as few as $m = O(s \log n)$ measurement settings, and efficient classical postprocessing based on convex optimization. In addition, when $m = O(s \log^4 n)$, our method is highly robust to noise, and has sample complexity $O(\max(n,s)^2 \log^4(n))$, which can be compared to conventional methods that have sample complexity $O(n^3)$. Thus, our method is advantageous when the correlations are sufficiently sparse, that is, when $s \leq O(n^{3/2} / \log^2 n)$. Our method also performs well in numerical simulations on small system sizes, and has some resistance to state-preparation-and-measurement (SPAM) errors. The key ingredient in our method is a new type of compressed sensing measurement, which works by preparing entangled Greenberger-Horne-Zeilinger states (GHZ states) on random subsets of qubits, and measuring their decay rates with high precision. 
\end{abstract}

\maketitle


\section{Introduction}
\label{sec-intro}
The development of noisy intermediate-scale quantum information processors (NISQ devices) has the potential to advance many areas of computational science \cite{preskill2018quantum}.
An important problem is the characterization of noise processes in these devices, in order to improve their performance (via calibration and error correction), and to ensure correct interpretation of the results \cite{eisert2019quantum}. 
The challenge here is to characterize all of the noise processes that are likely to occur in practice, using some experimental procedure that is efficient and can scale up to large numbers of qubits. 

Compressed sensing \cite{candes2008intro} offers one approach to solving this problem. Here one uses specially-designed measurements (and classical postprocessing) to learn an unknown signal that has some prescribed structure. For example, the unknown signal can be a sparse vector or a low-rank matrix, the measurements can consist of random projections sampled from various distributions, and the classical postprocessing can consist of solving a convex optimization problem (e.g., minimizing the $\ell_1$ or trace norm), using efficient algorithms. This approach has been used in several previous works on quantum state and process tomography, and estimation of Hamiltonians and Lindbladians \cite{gross2010quantum, shabani2011efficient, shabani2011estimation, flammia2012quantum, roth2018recovering}. 

From a theoretical perspective, one of the main challenges in this line of work is to design \textit{measurements} that have the mathematical properties needed for compressed sensing, and can be implemented efficiently on a quantum device. There has been a substantial amount of work in this area, which can be broadly classified into two approaches: ``sparsity-based'' and ``low-rank'' compressed sensing. For compressed sensing of ``low-rank'' objects (e.g., low-rank density matrices and quantum processes), there seem to be a few natural choices for measurements, including random Pauli measurements, and fidelities with random Clifford operations \cite{gross2010quantum, liu2011universal, roth2018recovering}. For compressed sensing of ``sparse'' objects (e.g., sparse Hamiltonians, Lindbladians, or Pauli channels), however, the situation is more complicated, as a larger number of different measurement schemes and classical postprocessing methods have been proposed, and the optimal type of measurement seems to depend on the situation at hand \cite{shabani2011efficient, shabani2011estimation, harper2020fast}. This complicated state of affairs can be explained in part because ``sparsity'' occurs in a wider variety of situations than ``low-rankness.'' 

\begin{figure}[h!]
	\centering
	\includegraphics[width=\columnwidth]{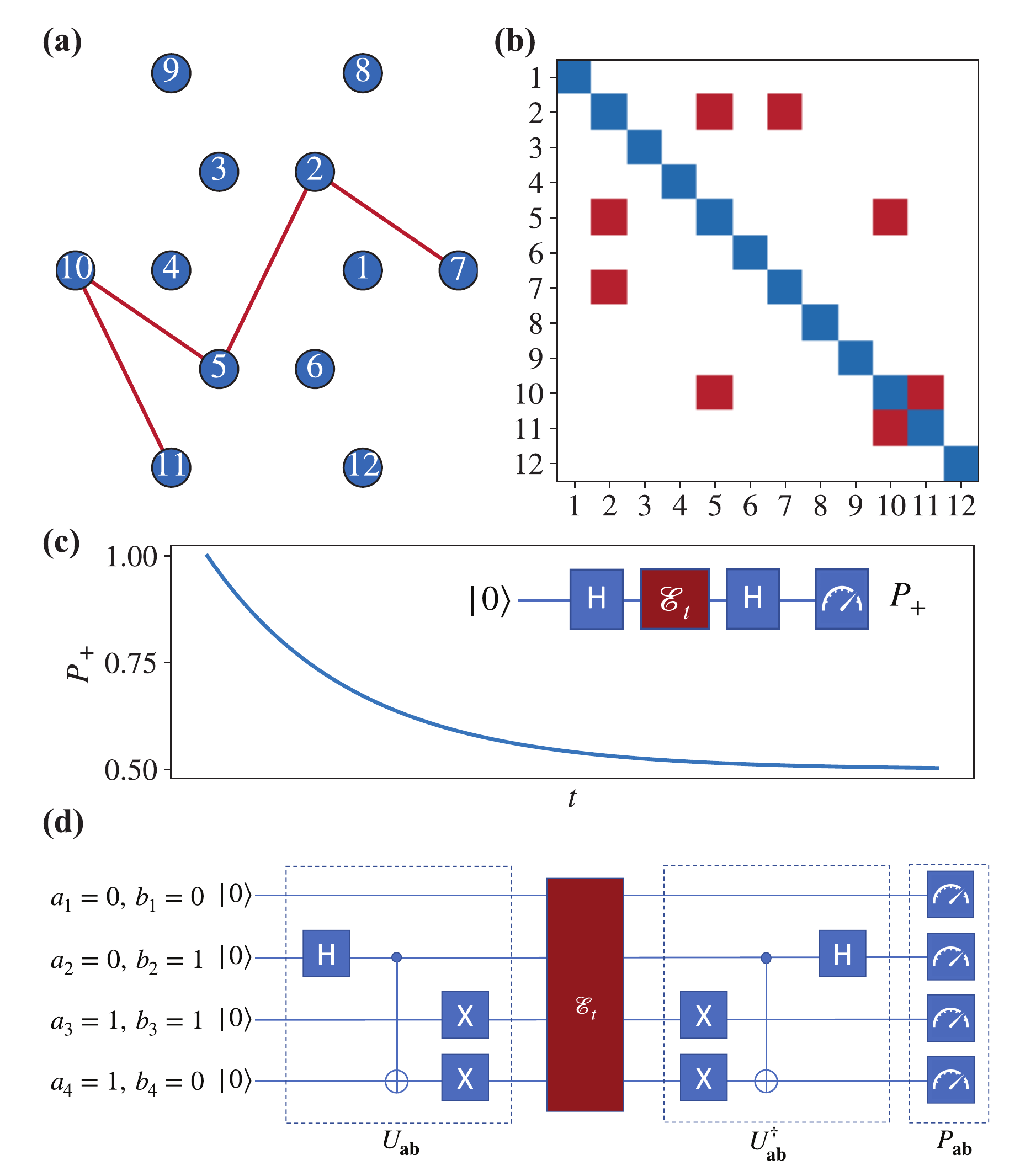}
	\caption[Illustration of the noise model and the protocol.]{Illustration of the noise model and the protocol. (a) The qubits experience correlated Markovian dephasing. The red lines correspond to non-zero $c_{ij}$, indicating correlated noise affecting the pairs of qubits connected by those lines. (b) The $C$ matrix corresponding to the correlation graph in panel (a). The diagonal elements correspond to single qubit dephasing whereas the off-diagonal elements indicate correlated dephasing noise. (c) Single qubit Ramsey spectroscopy. The plot shows the overlap $P_+$, which decays exponentially (towards 1/2) with decay rate $\Gamma$, as a function of time $t$. The inset shows the Ramsey protocol, where a superposition of qubit states is prepared with the first Hadamard gate $\mathsf{H}$, the system undergoes dephasing (represented by the noise channel $\mathcal{E}_t$) for time $t$, and the second $\mathsf{H}$ followed by a measurement in the computational basis measures the overlap $P_+$. (d) The generalized measurement protocol involves generating vectors $\mathbf{a}$ and $\mathbf{b}$ whose elements are randomly chosen from $\{0,1\}$. The operation $U_\mathbf{ab}$ prepares the state $\frac{1}{\sqrt{2}} (\ket{\mathbf{a}} + \ket{\mathbf{b}})$. The system  evolves under dephasing noise for time $t$, represented by $\mathcal{E}_t$. Finally, we apply $U^\dagger_\mathbf{ab}$ and perform a computational basis measurement. The probability of obtaining the outcome 0, $P_{\mathbf{ab}}$, decays exponentially (towards 1/2) as $t$ increases. By measuring the decay rate $\Gamma_{\mathbf{ab}}$ for various $\mathbf{a}$'s and $\mathbf{b}$'s we can recover $C$. }  
	\label{fig:protocol}
\end{figure}

In this paper, we extend the theory of ``sparsity-based'' quantum compressed sensing, and apply it to a physical problem that is relevant to the development of NISQ devices: detecting long-range correlated dephasing errors. We use a simple model of correlated dephasing, which is described by a Markovian master equation:
\begin{equation}\label{eqn-master}
\frac{d\rho}{dt} = \mathcal{L}(\rho) = \sum_{j,k=1}^n c_{jk} \bigl( Z_k\rho Z_j - \frac{1}{2} \lbrace Z_j Z_k, \rho \rbrace \bigr).
\end{equation}
Here the system consists of $n$ qubits, and $Z_j$ and $Z_k$ are Pauli $\sigma_z$ operators that act on the $j$'th and $k$'th qubits, respectively. The noise is then completely described by the correlation matrix $C = (c_{jk}) \in \RR^{n\times n}$, see Fig.~\ref{fig:protocol}(a) and (b). (We will also consider generalizations of this model with complex coefficients $c_{jk}$, and an additional environment-induced Hamiltonian.) 

Here, the diagonal elements $c_{jj}$ show the rates at which single qubits dephase, and the off-diagonal elements $c_{jk}$ show the rates at which pairs of qubits undergo correlated dephasing. Typically, the diagonal of $C$ will be nonzero, while the off-diagonal part may be dense or sparse, depending on the degree of connectivity between the qubits and their environment. 

This master equation describes a number of physically plausible scenarios, such as spin-1/2 particles coupling to a shared bosonic bath \cite{breuer2002theory} (see Appendix \ref{sec:physderiv}). It has also been studied as an example of how collective decoherence can affect physical implementations of quantum computation \cite{palma1996quantum, duan1998reducing, reina2002decoherence, qeccor2005} and quantum sensing \cite{layden2018spatial, layden2018ancilla}. 

This model of correlated dephasing is quite different from other models of crosstalk that are based on quantum circuits or Pauli channels \cite{sarovar2020detecting, harper2020efficient, govia2020bootstrapping, harper2020fast}. Roughly speaking, our model describes crosstalk that arises from the physical coupling of the qubits to their shared environment. This has a different character from crosstalk that arises from performing imperfect two-qubit gates, or correlated errors that result when the physical noise processes are symmetrized by applying quantum operations such as Pauli twirls. Nonetheless, there are intriguing parallels between our results, and some of these other works, particularly on the estimation of sparse Pauli channels \cite{harper2020fast}. We will discuss this later in this section.

In this paper, we show how our model of correlated dephasing can be learned efficiently when the off-diagonal part of the correlation matrix $C$ is \textit{sparse}. We assume that $C$ has at most $s \ll n(n-1)/2$ nonzero entries above the diagonal, but these entries may be distributed arbitrarily; in particular, long-range correlated errors are allowed. (Note that physical constraints imply that $C$ is positive semidefinite, hence $C = C^\dagger$ \cite{jeske2013derivation}.) 

This model is applicable in a number of scenarios, including experimental NISQ devices, which are often engineered to have long-range interactions, in order to perform quantum computations more efficiently \cite{linke2017experimental}; the execution of quantum circuits on distributed quantum computers, where long-range correlations are generated when qubits are moved or teleported from one location to another \cite{beals2013efficient}; and quantum sensor arrays, where a detection event at a location $(j,k)$ in the array may be registered as a pairwise correlation between a qubit that is coupled to row $j$ and a qubit that is coupled to column $k$ in the array.

Our main technical contribution is a new method for performing compressed sensing of the coupling matrix $C$. At a high level, our method works by performing $m = O(s\log n)$ or $m = O(s\log^4 n)$ random linear measurements of the coupling matrix $C$, where each measurement can be understood as a generalized Ramsey measurement\cite{ramsey1950measurement,capellaro2017} (Fig.~\ref{fig:protocol}(c) and (d)). At an abstract level, each measurement has the following form: choose two vectors $\mathbf{a}, \mathbf{b} \in \set{0,1}^n$ uniformly at random, and estimate the quantity $\mathbf{r}^T C \mathbf{r}$, where $\mathbf{r} = \mathbf{b} - \mathbf{a} \in \set{1,0,-1}^n$. 

These kinds of measurements can be realized experimentally, using techniques from noise spectroscopy and quantum sensing \cite{capellaro2017,layden2018spatial, layden2018ancilla}: prepare an $n$-qubit state $\ket{\psi_\mathbf{ab}} = \frac{1}{\sqrt{2}} (\ket{\mathbf{a}}+\ket{\mathbf{b}}) \in (\CC^2)^{\otimes n}$ (assuming $\mathbf{a} \neq \mathbf{b}$), and allow it to evolve according to equation (\ref{eqn-master}) for some time $t$ to get a state $\rho(t)$. A straightforward calculation shows that $\rho(t)$ has the form 
\begin{equation}
\label{eqn-coherences-decay}
\rho(t) = \frac{1}{2} \Bigl( \ket{\mathbf{a}}\bra{\mathbf{a}} + e^{-\Gamma_\mathbf{ab} t}  \ket{\mathbf{a}}\bra{\mathbf{b}} +   e^{-\Gamma_\mathbf{ab} t}  \ket{\mathbf{b}}\bra{\mathbf{a}} + \ket{\mathbf{b}}\bra{\mathbf{b}} \Bigr),
\end{equation}
where the off-diagonal elements decay exponentially at a rate $\Gamma_\mathbf{ab} = 2 \mathbf{r}^T C \mathbf{r}$. One can estimate $\Gamma_\mathbf{ab}$ from experiments, which then gives us the desired quantity $\mathbf{r}^T C \mathbf{r}$. 

This data can be viewed as an estimate of the vector $\Phi(C)$ returned by the ``measurement operator''
\begin{equation}
\Phi:\: C \mapsto [ (\mathbf{r}^{(1)})^T C\mathbf{r}^{(1)},\, \ldots,\, (\mathbf{r}^{(m)})^T C\mathbf{r}^{(m)} ],
\end{equation}
where $\mathbf{r}^{(1)}, \ldots, \mathbf{r}^{(m)}$ are random vectors sampled independently from the same distribution as $\mathbf{r}$. 
Given a (noisy) estimate of $\Phi(C)$, the matrix $C$ can then be reconstructed (up to some small error) by using techniques from compressed sensing, e.g., by solving a convex optimization problem such as constrained $\ell_1$-minimization or $\ell_1$-regularized least-squares regression \cite{chen2001atomic}. We describe the complete method in Sections \ref{sec-ramsey} and \ref{sec-learning-sparse}. 

\begin{figure*}[t]
	\centering
	\includegraphics[width=1\linewidth]{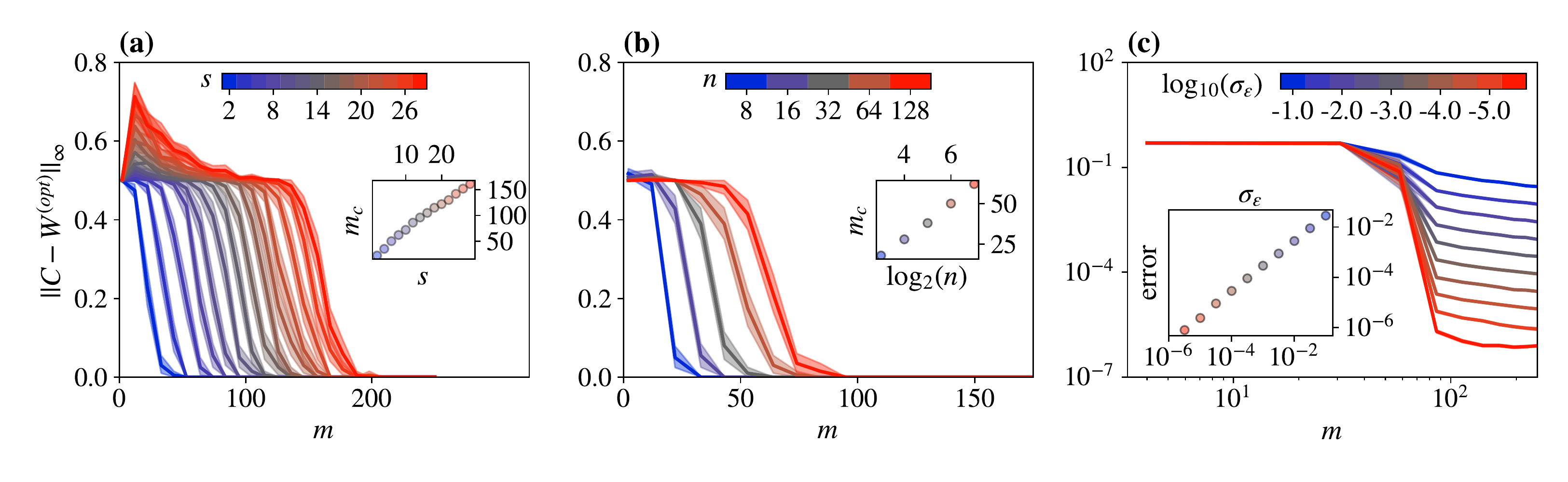}
	\caption[Scaling of the reconstruction error $\norm{W^{(opt)} - C}_\infty$ under various circumstances.]{Scaling of the reconstruction error $\norm{W^{(opt)} - C}_\infty$ under various circumstances. Here $W^{(opt)}$ denotes the estimate of $C$ obtained via compressed sensing. The solid lines are the average errors over 100 random instances of the problem, and the shaded region is their 95\% confidence interval obtained by bootstrapping. (a) The reconstruction error as a function of the number of measurement settings $m$ (assuming noiseless measurements) for various values of sparsity $s$ (and $n=64$ qubits). The errors go through a phase transition whose location $m_c$ scales linearly with $s$. (b) The reconstruction error as a function of the number of measurement settings $m$ (assuming noiseless measurements) for various numbers of qubits $n$ (and sparsity $s=12$). The phase transition point $m_c$ scales logarithmically with $n$.  (c)  The reconstruction error as a function of the number of measurement settings $m$, for different values of added noise strength $\sigma_\epsilon$, with fixed parameters $(n,s)=(64,12)$. The inset shows that the recovery errors (when $m>m_c$, i.e., after the transition point) scale linearly with $\sigma_\epsilon$, as expected. See Section \ref{sec-num-examples} for details.}
	\label{fig:scaling}
\end{figure*}	

Numerical simulations show that our method performs well, and readily scales up to 128 qubits or more (see Fig.~\ref{fig:scaling} and Section \ref{sec-num-examples}). But the reasons for this success are not at all obvious, because our method is substantially different from previous work on quantum compressed sensing. 
In particular, the linear measurement $C \mapsto \mathbf{r}^T C \mathbf{r}$ has an unusual form: it is an inner product between $C$ and a random \textit{rank-1} matrix $\mathbf{r}\mathbf{r}^T$. In the context of compressed sensing, this means that our method does not fit into the framework of Gaussian (or sub-Gaussian) random measurements \cite{vershynin2010introduction}, because of this rank-1 structure (or more concretely, because the matrix $\mathbf{r}\mathbf{r}^T$ only involves $n$ rather than $n^2$ independent random variables). 

In the context of sparse Hamiltonian estimation, this means that the theoretical analysis of our method requires different, considerably stronger techniques than those shown in \cite{shabani2011efficient, shabani2011estimation}. (In particular, our method requires probabilistic proof techniques that account for correlations between related random variables, such as ``generic chaining'' \cite{candes2006near, rudelson2008sparse, foucart2013mathematical}, in contrast to simpler techniques that neglect such correlations, such as the union bound used in \cite{shabani2011efficient, shabani2011estimation}.) 

Instead, our method turns out to have surprising connections to random Fourier measurements (and random measurements in incoherent bases) in compressed sensing \cite{candes2006near, candes2006stable, rudelson2008sparse, ripless} (see also \cite{vershynin2010introduction, davenport2012, foucart2013mathematical}). The key observation is that the off-diagonal elements of the matrix $\mathbf{r}\mathbf{r}^T$ (subject to a suitable normalization condition) are random variables that are \textit{centered} around 0 and \textit{bounded} independent of the dimension $n$. These properties do not hold for the diagonal elements of $\mathbf{r}\mathbf{r}^T$, but those diagonal elements are irrelevant, because we are only trying to detect the off-diagonal part of $C$, which encodes correlations between \textit{different} qubits. These observations imply that our method fits into the theoretical framework of compressed sensing using ``bounded orthonormal systems'' \cite{candes2006near, candes2006stable, rudelson2008sparse, ripless, foucart2013mathematical}. 

Using this theoretical framework, we prove several results about the accuracy of our compressed sensing method, and the physical resources needed to implement it in an experiment. In Section~\ref{sec:recguarantee}, we show two different recovery guarantees for our method. First, we prove a ``RIPless'' recovery guarantee (Section~\ref{sec-ripless}), which shows that our method can reconstruct $C$ accurately when the number of measurement settings is $m = O(s\log n)$, just slightly above the information-theoretic lower bound. (Here, the ``RIP'' refers to the restricted isometry property, a standard proof technique in the theory of compressed sensing.) 

Second, we prove a ``RIP-based'' recovery guarantee (Section~\ref{sec-rip-based}), which shows that the recovery of $C$ using our method is highly robust to noise in the measurements, provided that $m$ is slightly larger, say $m = O(s\log^4 n)$. 
In addition, the ``RIP-based'' result shows \textit{universal} recovery, meaning that a single fixed measurement operator $\Phi$ 
is capable of recovering all possible sparse matrices $C$ (up to some unavoidable error due to the noise in the measurements). 

\begin{figure*}
\begin{tabular}{|l|c|c|}
\hline
Reconstruction & \# meas.      & Total sample \\
method           & settings $m$ & complexity \\
\hline
\hline
Naive method & $O(n^2)$ & $O(n^3/\delta^2)$ \\
\hline
CS (RIP-based) & $O(s \log^4 n)$ & $O(\max(n,s)^2 \log^4(n) / \delta^2)$ \\
\hline
CS (RIPless*) & $O(s \log n)$ & $O(s^3 \max(n,s) \log^6(n) / \delta^2)$ \\
\hline
\end{tabular}
\caption{Sample complexity of different methods for reconstructing a correlation matrix $C$, of size $n \times n$, with $2s$ nonzero elements off the diagonal. The naive method is to measure each element of $C$ separately. ``CS'' refers to the compressed sensing method, and ``RIP-based'' and ``RIPless'' refer to different analytical bounds on the accuracy of the reconstruction of $C$. The asterisk (*) indicates that the results using the RIPless bound hold under a technical assumption that the diagonal of $C$ does not have any unusually large elements, see equation (\ref{eqn-diagC-assumption}). The different methods are parameterized in such a way that they reconstruct the diagonal of $C$ up to an additive error of size $\delta \norm{\diag(C)}_2$, and they reconstruct the off-diagonal part of $C$ up to an additive error of size $\delta \norm{C}_F$. Each method makes use of single-qubit spectroscopy (with $n$ experimental configurations or ``measurement settings''), as well as multi-qubit spectroscopy (with $m$ measurement settings, where $m$ varies between $O(n^2)$ and $O(s\log n)$). The total sample complexity, shown in the table, includes both single-qubit and multi-qubit spectroscopy; for a more detailed accounting, see Figure \ref{fig-sample-complexity-full}. For the CS method, the number of measurement settings can be as low as $m = O(s \log n)$, using the RIPless bound, but the best sample complexity is achieved when $m = O(s \log^4 n)$, using the RIP-based bound. This sample complexity, $O(\max(n,s)^2 \log^4(n))$, compares favorably with the sample complexity of the naive method, which is $O(n^3)$; we see that compressed sensing has an advantage over the naive method whenever $s \leq O(n^{3/2}/\log^2 n)$. See Section \ref{sec-performance} for details.}
\label{fig-sample-complexity}
\end{figure*}

In Section \ref{sec-performance}, we study the performance of our compressed sensing method, and compare it to the naive method where one measures each element of $C$ independently. We can make a rigorous comparison, since we have error bounds for each of these methods. These can be summarized as follows:
\begin{widetext}
\begin{align}
\text{Naive: } \qquad \norm{\widehat{C} - C}_F 
&\leq O(\sqrt{n} (\delta_1+\delta_2) \norm{\diag(C)}_2 + \delta_2 \norm{C'}_F), \\
\text{RIP-based: } \norm{W^{(opt)}-C}_F 
&\leq O(\sqrt{n} \delta_1 \norm{\diag(C)}_2 + \delta_2 (\sqrt{n} \norm{\diag(C)}_2 + \sqrt{2s} \norm{C'}_F)), \\
\text{RIPless: } \norm{W^{(opt)}-C}_F 
&\leq O(s \log^{5/2}(n) [\delta_1 \norm{\diag(C)}_\infty \sqrt{sn\log(n)} 
 + \delta_2 \sqrt{n} \norm{\diag(C)}_2 + \delta_2 \sqrt{2s} \norm{C'}_F]).
\end{align}
\end{widetext}
Here $\widehat{C}$ is the estimate of $C$ using the naive method, and $W^{(opt)}$ is the estimate of $C$ using compressed sensing. (This is a simplification of the notation used in Section \ref{sec-performance}, which defines separate estimators for the diagonal and off-diagonal parts of $C$.) We write $\diag(C)$ to denote the diagonal of $C$, and $C'$ to denote the off-diagonal part of $C$. We set $m = O(s\log^4 n)$ for the RIP-based bound, and $m = O(s\log n)$ for the RIPless bound. We then use (\ref{eqn-direct-error-bound-L2-diag-whp}) to bound the error in estimating $\diag(C)$, and (\ref{eqn-direct-error-bound-L2-whp}), (\ref{eqn-error-bound-cs-slog4n}) and (\ref{eqn-error-bound-cs-slogn}) to bound the error in estimating $C'$. Finally, $\delta_1$ and $\delta_2$ are parameters that control the accuracy of the single-qubit and multi-qubit spectroscopy procedures. Using this theory, we can prove rigorous bounds on the sample complexity that is required for each method to reconstruct $C$ with comparable accuracy. The results are summarized in Figure \ref{fig-sample-complexity}. 

For the compressed sensing method, the number of measurement settings can be as low as $m = O(s \log n)$, using the RIPless bound, but the best sample complexity is achieved when $m = O(s \log^4 n)$, using the RIP-based bound. This sample complexity, $O(\max(n,s)^2 \log^4(n))$, compares favorably with the sample complexity of the naive method, which is $O(n^3)$. Thus we see that compressed sensing has an advantage over the naive method whenever $s \leq O(n^{3/2}/\log^2 n)$.

Our method has some similarities with more recent work on phase retrieval, particularly the PhaseLift algorithm, although there are significant differences \cite{candes2013phaselift, ohlsson2012cprl, li2013sparse, jaganathan2013sparse, kueng2017low} (see also \cite{roth2018recovering, suess2020rapid}). We discuss this in detail in Section \ref{sec-reconstructing-C}. We are not aware of any previous work on phase retrieval that directly addresses the situation studied in this paper; however, it is an interesting question whether techniques based on phase retrieval can be used to re-derive or improve on our results.

Our method has a second novel feature, which concerns the estimation of the decay rate $\Gamma_\mathbf{ab} = 2 \mathbf{r}^T C \mathbf{r}$. In Sections~\ref{sec-estimating-decay-rates} and \ref{sec-evolution-time}, we show a procedure for estimating $\Gamma_\mathbf{ab}$ with high precision (i.e., with small multiplicative error), by allowing the system to evolve for a time $t \sim 1/\Gamma_\mathbf{ab}$. Here the time $t$ is chosen by an adaptive procedure that starts with an initial guess $\tau_0$ and converges (provably) after $\sim \abs{\log(\Gamma_\mathbf{ab}\tau_0)}$ steps (see Fig.~\ref{fig:chooset}). To achieve high precision, we treat the time evolution operator $e^{t\mathcal{L}}$ exactly, in contrast to previous work on sparse Hamiltonian estimation \cite{shabani2011efficient, shabani2011estimation}, which used a linear approximation that is only valid for short times $t$. 

This feature of our method is helpful in experimental setups where measurements are time-consuming and the values of $\Gamma_\mathbf{ab}$ span several orders of magnitude, thus making it difficult to determine the appropriate evolution time and accurately measure the decay rates using conventional methods. This feature can also be used as a standalone technique to estimate the relaxation  ($T_1$) and decoherence ($T_2$) times of any quantum system~\cite{capellaro2017}.

In Section~\ref{sec-spam-errors}, we show that since our method relies on estimating exponential decay rates, it can be made at least partially robust to state-preparation-and-measurement (SPAM) errors (see Fig.~\ref{fig:gateerrors}). This follows by some of the same approaches used in randomized benchmarking and gate set tomography \cite{emerson2005scalable, knill2008randomized, greenbaum2015introduction}.

Finally in Section~\ref{sec-complex-c} we sketch a generalization of our method that includes unitary evolution according to some Hamiltonian $H_S$, and allows the matrix $C$ to have complex entries $c_{jk}$. We show how a similar approach can be used to estimate the Hamiltonian $H_S$, as well as the imaginary part of $c_{jk}$. 

It is interesting to compare our method with the recent work of \cite{harper2020fast} on estimation of sparse Pauli channels. At a technical level, these two works are very different: they are measuring different types of noise (correlated dephasing versus Pauli errors), using different types of measurements (generalized Ramsey spectroscopy versus quantum Clifford circuits), and different reconstruction algorithms (convex optimization over a continuous domain, versus a combinatorial ``peeling decoder''). 

But from a broader perspective, these two methods do share certain general features. Both methods assume that the noise is sparse (albeit with very different mathematical representations), meaning that the noise is supported on a subset $S$ of the domain, where $S$ is small, but unknown. Both methods avoid making additional assumptions about the structure of $S$ (e.g., in our work, $S$ is allowed to contain long-range correlations, and in \cite{harper2020fast}, $S$ is allowed to contain high-weight Pauli errors). Finally, both methods utilize measurements of ``decay rates'' (albeit with very different types of experiments), in order to obtain results that are robust in the presence of state preparation and measurement (SPAM) errors. 

This suggests that similar techniques, for SPAM-robust estimation of sparse noise models, can be used to characterize other kinds of correlated noise processes in many-body quantum systems.

\subsection{Notation}

In this paper we use the following notation: Vectors are written in boldface, and matrices are denoted by capital letters. For a vector $\mathbf{v}$, $\norm{\mathbf{v}}_p$ denotes the $\ell_p$ norm (we will be mainly interested in the cases $p = 1, 2, \infty$). 

For a matrix $M$, $\norm{M}_F = (\sum_{jk} \abs{M_{jk}}^2)^{1/2}$ denotes the Frobenius norm (i.e., the Schatten 2-norm, or the $\ell_2$ norm of the vector containing the entries of $M$), $\norm{M}$ denotes the operator norm (i.e., the Schatten $\infty$-norm), $\norm{M}_\text{tr}$ denotes the trace norm (i.e., the Schatten $1$-norm, or the nuclear norm), and $\norm{M}_{\ell_1} = \sum_{jk} \abs{M_{jk}}$ denotes the $\ell_1$ norm of the vector containing the entries of $M$. 

Given an $n\times n$ matrix $M$, let $\uvec(M) = (M_{ij})_{i<j}$ denote the vector (of dimension $n(n-1)/2$) containing the entries in the upper-triangular part of $M$, excluding the diagonal.

Asymptotic bounds are written using big-O notation, such as $O(s\log n)$. Polylogarithmic factors are written in a compact way as follows: $\log^c n = (\log n)^c$. A ``universal constant'' is a quantity whose value is fixed once and for all, and does not depend on any other variable. 

Statistical estimators are written with a hat superscript, e.g., $\hat{\Gamma}$ is a random variable that represents an estimator for some unknown quantity $\Gamma$. For a random variable $\hat{\Gamma}$, $\norm{\hat{\Gamma}}_{\psi_2}$ denotes the sub-Gaussian norm, and $\norm{\hat{\Gamma}}_{\psi_1}$ denotes the subexponential norm, in the sense of \cite{vershynin2010introduction}.


\section{Generalized Ramsey Spectroscopy}
\label{sec-ramsey}

We begin by describing a general form of Ramsey spectroscopy using entangled states on multiple qubits. We will also describe a simple method for directly measuring the correlation matrix $C$, by performing spectroscopy on every pair of qubits. This simple method serves as a baseline for measuring the performance of our compressed sensing method, which we will introduce in Section \ref{sec-learning-sparse}. 

Here we assume that the entries in the matrix $C$ are real (i.e., with imaginary part equal to zero). This holds true in a number of important cases, for instance, when the qubits are coupled to a bath at high temperature (see Appendix \ref{sec:physderiv}). When $C$ is complex, it can be handled using a generalization of our method, described in Section \ref{sec-complex-c}.

Note that physical constraints imply that $C$ is positive semidefinite \cite{jeske2013derivation}; hence we have $C=C^T$ in the real case, and $C = C^\dagger$ in the complex case.

	\subsection{Dephasing of Entangled States \label{sec:prelims}}
	\label{subsec-dephasing-ghz-like}
	We begin by describing a procedure that allows us to measure certain linear functions of the correlation matrix $C$. This procedure is very general, and includes single- and two-qubit Ramsey spectroscopy as special cases. Consider an $n$-qubit state of the form 
	\begin{equation}
		\label{eqn-psi}
		\ket{\psi_\mathbf{ab}} = \frac{1}{\sqrt{2}} (\ket{\mathbf{a}}+\ket{\mathbf{b}}) \in (\CC^2)^{\otimes n}, 
	\end{equation}
	where $\mathbf{a}, \mathbf{b} \in \set{0,1}^n$, $\mathbf{a} \neq \mathbf{b}$, $\ket{\mathbf{a}} = \ket{a_1, a_2, \ldots, a_n}$ and $\ket{\mathbf{b}} = \ket{b_1, b_2, \ldots, b_n}$. By choosing $\mathbf{a}$ and $\mathbf{b}$ appropriately, one can make $\ket{\psi_\mathbf{ab}}$ be a single-qubit $\ket{+}$ state, a two-qubit Bell state, or a many-qubit 
Greenberger-Horne-Zeilinger state (GHZ state) (while the other qubits are in a tensor product of $\ket{0}$ and $\ket{1}$ states).
	
	Say we prepare the state $\ket{\psi_\mathbf{ab}}$, then allow it to evolve for time $t$ under the Lindbladian \eqref{eqn-master}. Let $\rho(t)$ be the resulting density matrix. As can be seen in Eq.~\eqref{eqn-coherences-decay},  the coherences (that is, the off-diagonal elements $\ket{\mathbf{a}}\bra{\mathbf{b}}$) of $\rho(t)$ decay as $\exp(-\Gamma_\mathbf{ab}t)$, where the decay rate $\Gamma_\mathbf{ab} \in \RR$ is defined so that $\mathcal{L}(\ket{\mathbf{a}}\bra{\mathbf{b}}) = -\Gamma_\mathbf{ab} \ket{\mathbf{a}}\bra{\mathbf{b}}$. This decay rate can be estimated by allowing the system to evolve for a suitable amount of time $t$, and then measuring in the basis $\tfrac{1}{\sqrt{2}}(\ket{\mathbf{a}} \pm \ket{\mathbf{b}})$ (see Section \ref{sec-estimating-decay-rates} for details). 
	
	The decay rate $\Gamma_\mathbf{ab}$ tells us a certain linear function of the correlation matrix $C$, which can be written explicitly as follows. Let $\alpha_i = (-1)^{a_i}$ and $\beta_i=(-1)^{b_i}$ denote the expectation values of $Z_i$ corresponding to the states $\ket{\mathbf{a}}$ and $\ket{\mathbf{b}}$, respectively. In addition, define the vectors $\boldsymbol{\alpha} = (\alpha_1,\ldots,\alpha_n)$ and $\boldsymbol{\beta} = (\beta_1,\ldots,\beta_n)$. We can then see that 
	\begin{align}
		\Gamma_\mathbf{ab} &= -\sum_{ij} c_{ij} [\alpha_i \beta_j - \tfrac{1}{2} \alpha_i \alpha_j-\tfrac{1}{2} \beta_i \beta_j ] \\
		&= 2 \mathbf{r}^T C \mathbf{r} \label{eqn-rgamma},
	\end{align}
	where we recall the definition of $\mathbf{r}$, 
	\begin{equation}\label{eqn-r}
		\mathbf{r} = \mathbf{b} - \mathbf{a},
	\end{equation}
	we note that $\mathbf{r} =  \frac{\boldsymbol{\alpha}-\boldsymbol{\beta}}{2}$, and we use the fact that $C=C^T$ to symmetrize the equation. 
	
	Single-qubit Ramsey spectroscopy (Fig.~\ref{fig:protocol}(c)) is a special case of this procedure, where we set $\mathbf{a} = (0,0,\ldots,0)$ and $\mathbf{b} = (0,\ldots,0,1,0,\ldots,0)$ (where the 1 appears in the $j$'th position). Then $\ket{\psi_\mathbf{ab}}$ is a $\ket{+}$ state on the $j$'th qubit, and $\Gamma_\mathbf{ab} = 2c_{jj}$ tells us the rate of dephasing on the $j$'th qubit. 
	
	Two-qubit generalized spectroscopy is another special case, where we set $\mathbf{a} = (0,0,\ldots,0)$ and $\mathbf{b} = (0,\ldots,0,1,0,\ldots,0,1,0,\ldots,0)$ (where the 1's appear in the $j$'th and $k$'th positions). Then $\ket{\psi_\mathbf{ab}}$ is a maximally-entangled state on qubits $j$ and $k$, and $\Gamma_\mathbf{ab} = 2(c_{jj}+c_{jk}+c_{kj}+c_{kk})$ gives us information about the rate of correlated dephasing on qubits $j$ and $k$. 


\subsection{Estimating Decay Rates}
\label{sec-estimating-decay-rates}

There are many possible ways to estimate the decay rate $\Gamma_\mathbf{ab}$. For concreteness, we describe one simple and rigorous method here:
	\begin{enumerate}
		\item Choose some evolution time $t \geq 0$ such that $\tfrac{1}{2} \leq \Gamma_\mathbf{ab} t \leq 2$.  

This can be done in various ways, for instance, by starting with an initial guess $t = \tau_0$ and performing binary search, using $\sim \abs{\log(\Gamma_\mathbf{ab} \tau_0)}$ trials of the experiment (see Section \ref{sec-evolution-time} for details).
		\item Repeat the following experiment $N_\text{trials}$ times: (we can set $N_\text{trials}$ using equation (\ref{eqn-0-Ntrials}) below)
		\begin{enumerate}
			\item Prepare the state $\ket{\psi_\mathbf{ab}} = \frac{1}{\sqrt{2}} (\ket{\mathbf{a}} + \ket{\mathbf{b}})$, allow the state to evolve for time $t$, then measure in the basis $\frac{1}{\sqrt{2}} (\ket{\mathbf{a}} \pm \ket{\mathbf{b}})$. \\
		\end{enumerate}
		Let $N_+$ and $N_-$ be the number of $\frac{1}{\sqrt{2}} (\ket{\mathbf{a}} + \ket{\mathbf{b}})$ and $\frac{1}{\sqrt{2}} (\ket{\mathbf{a}} - \ket{\mathbf{b}})$ outcomes, respectively. 
		Note that the probabilities of these outcomes are given by $P_+ = \tfrac{1}{2}(1+e^{-\Gamma_\mathbf{ab} t})$ and $P_- = \tfrac{1}{2}(1-e^{-\Gamma_\mathbf{ab} t})$. 
		\item Define: 
		\begin{equation}
			\Delta = \frac{N_+ - N_-}{N_\text{trials}}. 
		\end{equation}
		Note that $\Delta$ is an unbiased estimator for $P_+-P_-$, that is, $\EE(\Delta) = P_+-P_- = e^{-\Gamma_\mathbf{ab} t}$. 
		This motivates our definition of an estimator for $\Gamma_\mathbf{ab}$:
		\begin{equation}
			\hat{\Gamma}_\mathbf{ab} = -\frac{1}{t} \ln(\Delta). 
		\end{equation}
	\end{enumerate}
	
	We now state some bounds on the accuracy of the estimator $\hat{\Gamma}_\mathbf{ab}$. To do this, we introduce the notion of a sub-Gaussian random variable (roughly speaking, a random variable whose moments and tail probabilities behave like those of a Gaussian distribution) \cite{vershynin2010introduction}. Formally, we say that a real-valued random variable $X$ is \textit{sub-Gaussian} if there exists a real number $K_2$ such that, 
	\begin{equation}
	\label{eqn-sub-gaussian-k-2}
		\text{for all $p \geq 1$, $(\EE(\abs{X}^p))^{1/p} \leq K_2 \sqrt{p}$.} 
	\end{equation}
	The \textit{sub-Gaussian norm} of $X$, denoted $\norm{X}_{\psi_2}$, is defined to be the smallest choice of $K_2$ in (\ref{eqn-sub-gaussian-k-2}), i.e., 
	\begin{equation}
		\norm{X}_{\psi_2} = \sup_{p \geq 1} p^{-1/2} (\EE(\abs{X}^p))^{1/p}.
	\end{equation}
	In addition, it is known that $X$ is sub-Gaussian if and only if there exists a real number $K_1$ such that, 
	\begin{equation}
	\label{eqn-sub-gaussian-k-1}
		\text{for all $t \geq 0$, $\Pr[\abs{X} > t] \leq \exp(1 - t^2/K_1^2)$.}
	\end{equation}
	The smallest choice of $K_1$ in (\ref{eqn-sub-gaussian-k-1}) is equivalent to the sub-Gaussian norm $\norm{X}_{\psi_2}$, in the following sense: there is a universal constant $c$ such that, for all sub-Gaussian random variables $X$, the smallest choice of $K_1$ in (\ref{eqn-sub-gaussian-k-1}) differs from $\norm{X}_{\psi_2}$ by at most a factor of $c$.
	
	We now show that $\hat{\Gamma}_\mathbf{ab} - \Gamma_\mathbf{ab}$ is a sub-Gaussian random variable, whose sub-Gaussian norm is bounded by 
	\begin{equation}\label{eqn-psi2-norm-gamma-hat}
		\norm{\hat{\Gamma}_\mathbf{ab} - \Gamma_\mathbf{ab}}_{\psi_2} \leq C_0 \frac{\Gamma_\mathbf{ab}}{\sqrt{N_\text{trials}}}, 
	\end{equation}
	where $C_0$ is a universal constant. (This $1/\sqrt{N_\text{trials}}$ scaling is familiar from classical statistics, and is not novel. The novelty of this paper will appear later, when we analyze the sample complexity of our compressed sensing estimators, in Section \ref{sec-performance}.)
	
	In particular, the accuracy of $\hat{\Gamma}_\mathbf{ab}$ can be controlled by setting $N_\text{trials}$ appropriately: for any $\delta > 0$ and $\epsilon > 0$, if we set 
	\begin{equation}\label{eqn-0-Ntrials}
		N_\text{trials} \geq \tfrac{2}{\delta^2} \ln(\tfrac{2}{\epsilon}), 
	\end{equation}
	then $\hat{\Gamma}_\mathbf{ab}$ satisfies the following error bound: with probability at least $1-\epsilon$, 
	\begin{equation}\label{eqn-error-bound-gamma-hat}
		\abs{\hat{\Gamma}_\mathbf{ab} - \Gamma_\mathbf{ab}} \leq 2\delta e^2 \Gamma_\mathbf{ab}.
	\end{equation}
	This shows that the error in $\hat{\Gamma}_\mathbf{ab}$ is at most a small fraction of the true value of $\Gamma_\mathbf{ab}$, independent of the magnitude of $\Gamma_\mathbf{ab}$. 
	
	We can also use this to derive an error bound that involves $\hat{\Gamma}_\mathbf{ab}$ rather than $\Gamma_\mathbf{ab}$, and hence can be computed from the observed value of $\hat{\Gamma}_\mathbf{ab}$. To show such an error bound, use the triangle inequality to write $\abs{\hat{\Gamma}_\mathbf{ab} - \Gamma_\mathbf{ab}} \leq 2\delta e^2 (\abs{\hat{\Gamma}_\mathbf{ab}} + \abs{\hat{\Gamma}_\mathbf{ab} - \Gamma_\mathbf{ab}})$, and divide by $(1-2\delta e^2)$ to get: 
	\begin{equation}
		\abs{\hat{\Gamma}_\mathbf{ab} - \Gamma_\mathbf{ab}} \leq \frac{2\delta e^2}{1-2\delta e^2} \abs{\hat{\Gamma}_\mathbf{ab}}.
	\end{equation}
	
	It remains to prove (\ref{eqn-psi2-norm-gamma-hat}) and (\ref{eqn-error-bound-gamma-hat}). First use Hoeffding's inequality to show that $\Delta$ is close to its expectation value:
	\begin{equation}
		\label{eqn-gammaerror}
		\Pr[|\Delta - (P_+-P_-)| \geq \delta] \leq 2\exp(-N_\text{trials} \delta^2 / 2). 
	\end{equation}
	When $|\Delta - (P_+-P_-)| \leq \delta$, we can bound the error in $\hat{\Gamma}_\mathbf{ab}$ as follows: (using the fact that $P_+-P_- = e^{-\Gamma_\mathbf{ab} t} \geq e^{-2}$ and $t \geq \frac{1}{2\Gamma_\mathbf{ab}}$)
	\begin{equation}
		\begin{split}
			-\hat{\Gamma}_\mathbf{ab} &\leq \frac{1}{t} \ln(P_+-P_-+\delta) \\
			&\leq \frac{1}{t} [\ln(P_+-P_-) + \delta e^2] \\
			&\leq -\Gamma_\mathbf{ab} + 2\Gamma_\mathbf{ab} \delta e^2,
		\end{split}
	\end{equation}
	\begin{equation}
		\begin{split}
			-\hat{\Gamma}_\mathbf{ab} &\geq \frac{1}{t} \ln(P_+-P_--\delta) \\
			&\geq \frac{1}{t} [\ln(P_+-P_-) - \delta e^2] \\
			&\geq -\Gamma_\mathbf{ab} - 2\Gamma_\mathbf{ab} \delta e^2.
		\end{split}
	\end{equation}
	Hence we have:
	\begin{equation}
		\Pr[\abs{\hat{\Gamma}_\mathbf{ab} - \Gamma_\mathbf{ab}} \geq 2\delta e^2 \Gamma_\mathbf{ab}] \leq 2\exp(-N_\text{trials} \delta^2 / 2). 
	\end{equation}
	This implies (\ref{eqn-psi2-norm-gamma-hat}) and (\ref{eqn-error-bound-gamma-hat}). 


\subsection{Direct Estimation of the Correlation Matrix}
\label{sec-direct-estimation}

There is a simple way to estimate the correlation matrix $C$ directly, by performing single-qubit spectroscopy to measure the diagonal elements $c_{jj}$, and performing two-qubit spectroscopy to measure the off-diagonal elements $c_{jk}$. We describe this method here. We will use this method as a baseline, to measure the performance of the compressed sensing method that we will introduce in Section \ref{sec-learning-sparse}.

For simplicity, we consider the case where $C$ is real. Since $C$ is positive definite, this implies that $c_{jj} \geq 0$ and $c_{jk} = c_{kj}$.

First, we estimate the diagonal elements $c_{jj}$, for $j = 1,\ldots,n$, as follows:
	\begin{enumerate}
		\item Let $\mathbf{a} = (0,0,\ldots,0)$ and $\mathbf{b} = (0,\ldots,0,1,0,\ldots,0)$ (where the 1 appears in the $j$'th position). 
		\item Construct an estimate $\hat{\Gamma}_\mathbf{ab}$ of the decay rate $\Gamma_\mathbf{ab} = 2c_{jj}$ (for instance, by using the procedure in Section \ref{sec-estimating-decay-rates}). Define $g_j = \hat{\Gamma}_\mathbf{ab}/2$.
	\end{enumerate}

To write this in a compact form, we define 
\begin{equation}
\diag(C) = (c_{11},\ldots,c_{nn}), 
\end{equation}
that is, the diagonal of $C$. We then define 
\begin{equation}\label{eqn-g}
\mathbf{g} = (g_1,\ldots,g_n), 
\end{equation}
and we view this as an estimator for $\diag(C)$. 

Next, we estimate the off-diagonal elements $c_{jk}$, for $1\leq j<k\leq n$, as follows:
	\begin{enumerate}
		\item Let $\mathbf{a} = (0,0,\ldots,0)$ and $\mathbf{b} = (0,\ldots,0,1,0,\ldots,0,1,0,\ldots,0)$ (where the 1's appears in the $j$'th and $k$'th positions). 
		\item Construct an estimate $\hat{\Gamma}_\mathbf{ab}$ of the decay rate $\Gamma_\mathbf{ab} = 2(c_{jj}+2c_{jk}+c_{kk})$ (for instance, by using the procedure in Section \ref{sec-estimating-decay-rates}). Define $h_{jk} = \tfrac{1}{4}\hat{\Gamma}_\mathbf{ab} - \tfrac{1}{2}g_j - \tfrac{1}{2}g_k$.
	\end{enumerate}

To write this in a compact form, we define $C'$ to be the matrix $C$ with the diagonal entries replaced by zeroes, 
\begin{equation}
C'_{jk} = \begin{cases}
c_{jk} &\text{ if $j\neq k$,}\\
0 &\text{ if $j=k$.}
\end{cases}
\end{equation}
We call this the ``off-diagonal part'' of $C$. We then construct an estimator $\widehat{C}'$ for $C'$, as follows:
\begin{equation}
\widehat{C}'_{jk} = \begin{cases}
h_{jk} &\text{ if $j<k$,}\\
h_{kj} &\text{ if $j>k$,}\\
0 &\text{ if $j=k$.}
\end{cases}
\end{equation}

We can analyze the accuracy of these estimators as follows. Choose two parameters $\delta_1$ and $\delta_2$. Suppose that, during the measurement of the diagonal elements $c_{jj}$, the decay rates $\Gamma_\mathbf{ab}$ are estimated with accuracy
\begin{equation}\label{eqn-delta-1-gamma-ab}
		\norm{\hat{\Gamma}_\mathbf{ab} - \Gamma_\mathbf{ab}}_{\psi_2} \leq \delta_1 \Gamma_\mathbf{ab}, 
\end{equation}
and during the measurement of the off-diagonal elements $c_{jk}$, the decay rates $\Gamma_\mathbf{ab}$ are estimated with accuracy
\begin{equation}\label{eqn-delta-2-gamma-ab}
		\norm{\hat{\Gamma}_\mathbf{ab} - \Gamma_\mathbf{ab}}_{\psi_2} \leq \delta_2 \Gamma_\mathbf{ab}.
\end{equation}

Bounds of the form (\ref{eqn-delta-1-gamma-ab}) and (\ref{eqn-delta-2-gamma-ab}) can be obtained from equation (\ref{eqn-psi2-norm-gamma-hat}), by setting $N_\text{trials} \sim 1/\delta_1^2$ and $N_\text{trials} \sim 1/\delta_2^2$, respectively. Here, we are neglecting to count those trials of the experiment that are used to choose the evolution time $t$, because the number of those trials grows only logarithmically with $\Gamma_\mathbf{ab}$. We allow $\delta_1$ and $\delta_2$ to be different, because in many experimental scenarios, measurements of $c_{jj}$ take less time than measurements of $c_{jk}$.

Using (\ref{eqn-delta-1-gamma-ab}) and (\ref{eqn-delta-2-gamma-ab}), we can easily show bounds on the accuracy of $g_j$ and $h_{jk}$:
\begin{equation}\label{eqn-direct-delta-1}
\norm{g_j-c_{jj}}_{\psi_2} \leq \delta_1 c_{jj},
\end{equation}
\begin{equation}\label{eqn-direct-delta-2}
\norm{h_{jk}-c_{jk}}_{\psi_2} \leq \delta_2 c_{jk} + \tfrac{1}{2} (\delta_1+\delta_2) (c_{jj}+c_{kk}).
\end{equation}
These bounds on the sub-Gaussian norm imply bounds on the moments, such as $\EE\abs{X} \leq \norm{X}_{\psi_2}$ and $\EE(X^2) \leq 2\norm{X}_{\psi_2}^2$ (see \cite{vershynin2010introduction} for details). 

We now use these results to bound the accuracy of the estimators $\mathbf{g}$ and $\widehat{C}'$. For $\mathbf{g}$, we have the following bounds, using the $\ell_1$ and $\ell_2$ norms:
\begin{equation}\label{eqn-direct-error-bound-L1-diag}
\EE(\norm{\mathbf{g} - \diag(C)}_1) \leq \delta_1 \norm{\diag(C)}_1,
\end{equation}
\begin{equation}\label{eqn-direct-error-bound-L2-diag}
\EE(\norm{\mathbf{g} - \diag(C)}_2^2) \leq 2\delta_1^2 \norm{\diag(C)}_2^2.
\end{equation}

For $\widehat{C}'$, we have the following bounds, using the $\ell_1$ vector norm and the Frobenius matrix norm:
\begin{equation}\label{eqn-direct-error-bound-L1}
\begin{split}
\EE(\norm{&\widehat{C}' - C'}_{\ell_1}) \\
&\leq 2\sum_{j<k} (\delta_2 c_{jk} + \tfrac{1}{2} (\delta_1+\delta_2) (c_{jj}+c_{kk})) \\
&= \delta_2 \sum_{j\neq k} c_{jk} + \tfrac{1}{2} (\delta_1+\delta_2) \sum_{j\neq k} (c_{jj}+c_{kk}) \\
&\leq (n-1) (\delta_1+\delta_2) \norm{\diag(C)}_{\ell_1} + \delta_2 \norm{C'}_{\ell_1},
\end{split}
\end{equation}
\begin{equation}\label{eqn-direct-error-bound-L2}
\begin{split}
\EE(&\norm{\widehat{C}' - C'}_F^2) \\
&\leq 2\sum_{j<k} 2[ \delta_2 c_{jk} + \tfrac{1}{2} (\delta_1+\delta_2) (c_{jj}+c_{kk}) ]^2 \\
&\leq 2\sum_{j<k} 6[ \delta_2^2 c_{jk}^2 + \tfrac{1}{4} (\delta_1+\delta_2)^2 c_{jj}^2 + \tfrac{1}{4} (\delta_1+\delta_2)^2 c_{kk}^2 ] \\
&\leq 6 \delta_2^2 \norm{C'}_F^2 + \tfrac{3}{2} (\delta_1+\delta_2)^2 \sum_{j\neq k} (c_{jj}^2 + c_{kk}^2) \\
&\leq 3(n-1) (\delta_1+\delta_2)^2 \norm{\diag(C)}_2^2 + 6\delta_2^2 \norm{C'}_F^2.
\end{split}
\end{equation}
Note that, in the second step of (\ref{eqn-direct-error-bound-L2}), we used the fact that for any real numbers $a_1,a_2,a_3$, $(a_1+a_2+a_3)^2 \leq 3(a_1^2+a_2^2+a_3^2)$.

These are bounds on the expected error of $\mathbf{g}$ and $\widetilde{C}'$. One can then use Markov's inequality to prove bounds that hold with high probability. For instance, using (\ref{eqn-direct-error-bound-L2-diag}), we get that with probability at least $1-\eta$, 
\begin{equation}\label{eqn-direct-error-bound-L2-diag-whp}
\norm{\mathbf{g} - \diag(C)}_2 \leq \tfrac{1}{\sqrt{\eta}} \sqrt{2} \delta_1 \norm{\diag(C)}_2,
\end{equation}
and using (\ref{eqn-direct-error-bound-L2}), we get that with probability at least $1-\eta$, 
\begin{equation}\label{eqn-direct-error-bound-L2-whp}
\begin{split}
\norm{&\widehat{C}' - C'}_F \\
&\leq \tfrac{1}{\sqrt{\eta}} [3(n-1) (\delta_1+\delta_2)^2 \norm{\diag(C)}_2^2 + 6\delta_2^2 \norm{C'}_F^2]^{1/2} \\
&\leq \tfrac{1}{\sqrt{\eta}} [\sqrt{3} \sqrt{n-1} (\delta_1+\delta_2) \norm{\diag(C)}_2 + \sqrt{6}\delta_2 \norm{C'}_F].
\end{split}
\end{equation}
In fact, one can prove tighter bounds on the failure probability, by replacing Markov's inequality with a sharp concentration bound for sub-Gaussian random variables \cite{vershynin2010introduction}. However, these tighter bounds are not needed for the purposes of this paper.

The bounds (\ref{eqn-direct-error-bound-L2-diag-whp}) and (\ref{eqn-direct-error-bound-L2-whp}) give a rough sense of how well this estimator performs. In particular, these bounds show that the error in estimating $C'$ depends on the magnitude of $\diag(C)$, as well as the magnitude of $C'$. This is due to the fact that our procedure for estimating the off-diagonal matrix element $c_{jk}$ also involves the diagonal matrix elements $c_{jj}$ and $c_{jk}$. Later in the paper, we will use these bounds as a baseline to understand the performance of our compressed sensing estimators (see Section \ref{sec-performance}).


	
	\section{Learning Sparse Correlations via Compressed Sensing\label{sec-learning-sparse}}
	
	Our main contribution in this paper is an efficient method for learning the off-diagonal part of the correlation matrix $C$, under the assumption that it is \textit{sparse}, i.e., the part of $C$ that lies above the diagonal has at most $s$ nonzero elements, where $s \ll n(n-1)/2$. (Since $C$ is Hermitian, it is sufficient to learn the part that lies above the diagonal; this then determines the part that lies below the diagonal.)
	
	For simplicity, we first consider the special case where the entries in the matrix $C$ are real (i.e., with zero imaginary part), which occurs in a number of physical situations (for example, when the system is coupled to a bath at high temperature, see Appendix \ref{sec:physderiv}). Later in Section~\ref{sec-complex-c} we will show how our method can be extended to handle complex matrices $C$.
	
	
	Our method consists of two steps: first, we perform single-qubit Ramsey spectroscopy in order to learn the diagonal elements of $C$; second, we apply techniques from compressed sensing (e.g., random measurements, and $\ell_1$-minimization) in order to recover the off-diagonal elements of $C$. 
	
	\subsection{Random Measurements of the Correlation Matrix \label{sec:randommeas}}
	We now describe our method in more detail. First, we estimate each of the diagonal elements $c_{jj}$, for $j = 1,\ldots,n$, using single-qubit Ramsey spectroscopy, as described in Fig.~\ref{fig:protocol}(c).
	Let $\mathbf{g} = (g_1,\ldots,g_n) \in \RR^n$ be the output of this procedure (this is the same notation used in Section \ref{sec-direct-estimation}). We can view $\mathbf{g}$ as an estimate of a ``sensing operator'' that returns the diagonal elements of the matrix $C$, 
	\begin{equation}
		\mathbf{g} \approx \diag(C) = (c_{11}, c_{22}, \ldots, c_{nn}).
	\end{equation}
	(Note that $c_{jj} \geq 0$, since $C$ is positive semidefinite.) 
	
	In order to estimate the off-diagonal part of $C$, we will use a compressed sensing technique, which involves a certain type of generalized Ramsey measurement with random GHZ-type states, see Fig.~\ref{fig:protocol}(d). First, we choose a parameter $m$, which can be roughly $m \sim s \log n$ or $m \sim s \log^4 n$, which controls the number of different measurements. (The particular choice of $m$ is motivated by the theoretical recovery guarantees in Section \ref{sec:recguarantee}.) Now, for $j = 1,\ldots,m$, we perform the following procedure:
	\begin{enumerate}
		\item Choose vectors $\mathbf{a}, \mathbf{b} \in \set{0,1}^n$ uniformly at random. As in equation (\ref{eqn-r}), define 
		\begin{equation}\label{eqn-r-distrib}
		\mathbf{r} = \mathbf{b} - \mathbf{a}.
		\end{equation}
		\item Prepare the state $\ket{\psi_\mathbf{ab}} = \frac{1}{\sqrt{2}} (\ket{\mathbf{a}} + \ket{\mathbf{b}})$. This is a GHZ state on a subset of the qubits, with some bit flips. It can be created by preparing those qubits $i$ where $a_i = b_i$ in the state $\ket{a_i}$, preparing those qubits $i$ where $a_i \neq b_i$ in a GHZ state, and applying a Pauli $X$ operator on those qubits $i$ where $a_i > b_i$. (This requires a quantum circuit of depth $\lceil\log_2(n)\rceil+2$.)
		\item Construct an estimate $\hat{\Gamma}_\mathbf{ab}$ of the decay rate $\Gamma_\mathbf{ab} = 2\mathbf{r}^T C\mathbf{r}$ (for instance, using the procedure in Section \ref{sec-estimating-decay-rates}). Define $h_j = \hat{\Gamma}_\mathbf{ab}$.
	\end{enumerate}
	Let $\mathbf{h} = (h_1,\ldots,h_m) \in \RR^m$ be the output of the above procedure. Again, we can view $\mathbf{h}$ as an estimate of a ``sensing operator'' $\Phi:\: \RR^{n\times n} \rightarrow \RR^m$ that is applied to the matrix $C$, 
	\begin{align}
		\mathbf{h} \approx \Phi(C) &= \bigl[ \Phi_j(C) \bigr]_{j=1,\ldots,m} \\
		\Phi_j(C) &=  2(\mathbf{r}^{(j)})^T C\mathbf{r}^{(j)} 		\label{eqn-Phi}
	\end{align}
	where $\mathbf{r}^{(1)}, \mathbf{r}^{(2)}, \ldots, \mathbf{r}^{(m)} \in \set{1,0,-1}^n$ are independent random vectors chosen from the same distribution as $\mathbf{r}$ (described above). Note that $\Phi_j(C) \geq 0$, since $C$ is positive semidefinite. The factor of 2 is chosen to ensure that $\Phi$ has a certain isotropy property, which will be discussed in Section \ref{sec-isotropy-incoherence}.

	\subsection{Reconstructing the Correlation Matrix}
	\label{sec-reconstructing-C}
	
	We now show how to reconstruct the correlation matrix $C \in \RR^{n\times n}$. We are promised that $C$ is positive semidefinite, due to physical constraints \cite{jeske2013derivation}, and its off-diagonal part is sparse, with at most $s \ll n(n-1)/2$ nonzero elements above the diagonal. In general, this sparsity constraint leads to an optimization problem that is computationally intractable. However, in this particular case, this problem can be solved using a strategy from compressed sensing: given $\mathbf{g} \approx \diag(C) \in \RR^n$ and $\mathbf{h} \approx \Phi(C) \in \RR^m$, we will recover $C$ by solving a convex optimization problem, where we minimize the $\ell_1$ (vector) norm of the off-diagonal part of the matrix. We will show that this strategy succeeds when $m \geq c_0 s\log n$, and is highly robust to noise when $m \geq c_0 s\log^4 n$, where $c_0$ is some universal constant. 
	
	We now describe this approach in more detail. First, we consider the case where the measurements are noiseless, i.e., $\mathbf{g} = \diag(C)$ and $\mathbf{h} = \Phi(C)$. We solve the following convex optimization problem:
	\begin{equation}\label{eqn-min-W}
	\begin{split}
		&\text{Find $W \in \RR^{n\times n}$ that minimizes $\sum_{i\neq j} |W_{ij}|$,} \\
		&\text{such that:}
	\end{split}
	\end{equation}
	\begin{equation}\label{eqn-diag-W}
		\diag(W) = \mathbf{g}, 
	\end{equation}
	\begin{equation}\label{eqn-Phi-W}
		\Phi(W) = \mathbf{h}, 
	\end{equation}
	\begin{equation}\label{eqn-W-psd}
		W \succeq 0.
	\end{equation}
	Here, $W \succeq 0$ means that $W$ is positive semidefinite, which implies that $W = W^T$. As a sanity check, note that $W = C$ is a feasible solution to this problem. (Recall that $C$ is positive semidefinite.)
	
	We remark that this scheme bears some resemblance to the PhaseLift algorithm for phase retrieval \cite{candes2013phaselift, ohlsson2012cprl, li2013sparse, jaganathan2013sparse, kueng2017low}. In phase retrieval, one wishes to estimate an unknown vector $\mathbf{x}$ from measurements of the form $\abs{\mathbf{r}^T\mathbf{x}}^2$. The PhaseLift algorithm works by ``lifting'' the unknown vector $\mathbf{x}$ to a matrix $X = \mathbf{x}\mathbf{x}^T$, so that the problem becomes one of learning a rank-1 matrix $X$ from measurements of the form $\mathbf{r}^TX\mathbf{r}$; then one solves a convex relaxation of this problem. In cases where the unknown vector $\mathbf{x}$ is sparse (``compressive phase retrieval''), variants of the PhaseLift algorithm (as well as other approaches) can also be used \cite{ohlsson2012cprl, li2013sparse, jaganathan2013sparse}. 

The main difference between our method and PhaseLift is that, in our method, the unknown matrix $C$ is almost always full-rank (because every qubit has a nonzero dephasing rate), whereas in PhaseLift, the unknown matrix $X$ has rank 1. In our situation, physical constraints imply that $C$ is positive semidefinite, so it can be factored as $C = BB^T$, which is superficially similar to $X = \mathbf{x}\mathbf{x}^T$; however, an important difference is that $B$ is a square matrix, whereas $\mathbf{x}$ is a vector. Methods like PhaseLift have been extended to handle low-rank matrices $X$, albeit without taking advantage of sparsity \cite{kueng2017low}, and it is an interesting question whether one can use this approach to re-derive or improve on our method, where sparsity plays a crucial role.

	
	\subsection{Reconstruction from Noisy Measurements}
	\label{sec-reconstructing-C-noisy}
	
	In the case where the measurements of $\mathbf{g}$ and $\mathbf{h}$ are noisy, we need to modify the above convex optimization problem, by relaxing the constraints (\ref{eqn-diag-W}) and (\ref{eqn-Phi-W}). This leads to some technical complications, due to the fact that we are reconstructing two variables that have different characteristics: the diagonal part of $C$ (which is not sparse), and the off-diagonal part of $C$ (which is sparse). 
	
	To deal with these issues, we propose two different ways of performing this reconstruction, when the measurements are noisy: (1) simultaneous reconstruction of both parts of $C$, and (2) sequential reconstruction of the diagonal part of $C$, followed by the off-diagonal part of $C$. The former approach is arguably more natural, but the latter approach allows for more rigorous analysis of the accuracy of the reconstruction (see Section \ref{sec:recguarantee}). 
	
	Suppose we have bounds on the $\ell_2$ norms of the noise terms (which we denote $\mathbf{u}$ and $\mathbf{v}$), that is, 
	\begin{equation}\label{eqn-eps-1}
		\mathbf{g} = \diag(C) + \mathbf{u}, \quad \norm{\mathbf{u}}_2 \leq \epsilon_1, 
	\end{equation}
	\begin{equation}\label{eqn-eps-2}
		\mathbf{h} = \Phi(C) + \mathbf{v}, \quad \norm{\mathbf{v}}_2 \leq \epsilon_2.
	\end{equation}
	(We do not assume anything about the distribution of $\mathbf{u}$ and $\mathbf{v}$. We will describe how to set $\epsilon_1$ and $\epsilon_2$ below, for some typical measurement procedures.) 
	
	\textit{Simultaneous reconstruction of the diagonal and off-diagonal parts of $C$:}
	Here we relax the constraints (\ref{eqn-diag-W}) and (\ref{eqn-Phi-W}) in the simplest possible way, by replacing them with:
	\begin{equation}\label{eqn-diag-W-2}
		\norm{\diag(W) - \mathbf{g}}_2 \leq \epsilon_1,
	\end{equation}
	\begin{equation}\label{eqn-Phi-W-2}
		\norm{\Phi(W) - \mathbf{h}}_2 \leq \epsilon_2.
	\end{equation}
	This leads to a convex optimization problem that attempts to reconstruct both the diagonal part of $C$, which is not necessarily sparse, and the off-diagonal part of $C$, which is assumed to be sparse. (Note that $W = C$ is a feasible solution to this problem.) This method often works quite well in practice. 
	
	Unfortunately, the behavior of this reconstruction algorithm can be complicated, because it involves two different estimators (an $\ell_1$-regularized estimator for the off-diagonal part of $C$, and a least-squares estimator for the diagonal part of $C$). These two estimators are coupled together (through the constraint on $\Phi(W)$, and the positivity constraint $W \succeq 0$). 
	
	Therefore, this method can behave quite differently, depending on whether the dominant source of error is $\mathbf{g}$ or $\mathbf{h}$. When $\mathbf{g}$ is the dominant source of error, this method will behave like a least-squares estimator, whose accuracy scales according to the dimension $n$; when $\mathbf{h}$ is the dominant source of error, this method will behave like an $\ell_1$-regularized estimator, whose accuracy scales according to the sparsity $s$ (neglecting log factors). From a theoretical point of view, this makes it more difficult to prove recovery guarantees for this method.
	
	\textit{Sequential reconstruction of the diagonal part of $C$, followed by the off-diagonal part of $C$:}
	In practice, one is often interested in the regime where $\mathbf{g}$ is known with high precision, and $\mathbf{h}$ is the dominant source of error. This is because measurements of $\mathbf{g}$ are relatively easy to perform, because they only require single-qubit state preparations and measurements; whereas measurements of $\mathbf{h}$ are more costly, because they require the preparation and measurement of entangled states on many qubits. So measurements of $\mathbf{g}$ can often be performed more quickly, and measurements on different qubits can be performed simultaneously in parallel; hence one can repeat the measurements more times, to obtain more accurate estimates of $\mathbf{g}$. 
	
	In this regime, it is natural to try to recover the diagonal part of $C$ \textit{directly} from $\mathbf{g}$, and then use $\ell_1$-minimization to recover \textit{only} the off-diagonal part of $C$. This leads to a convex optimization problem which is arguably less natural, but it makes it easier to prove rigorous guarantees on the accuracy of the reconstruction of $C$ (see Section \ref{sec:recguarantee}). 
	
	We now describe this approach in detail. We take the convex optimization problem ((\ref{eqn-min-W})-(\ref{eqn-W-psd})) for the noiseless case, and we relax the last two constraints to get: 
	\begin{equation}\label{eqn-min-W-3}
	\begin{split}
		&\text{Find $W \in \RR^{n\times n}$ that minimizes $\sum_{i\neq j} |W_{ij}|$,} \\
		&\text{such that:} 
	\end{split}
	\end{equation}
	\begin{equation}\label{eqn-diag-W-3}
		\diag(W) = \mathbf{g}, 
	\end{equation}
	\begin{equation}\label{eqn-Phi-W-3}
		\norm{\Phi(W) - \mathbf{h}}_2 \leq \epsilon_2 + \epsilon_1 \sqrt{mn}, 
	\end{equation}
	\begin{equation}\label{eqn-W-psd-3}
		d(W, K_+) \leq \epsilon_1, \text{ and } W = W^T.
	\end{equation}
	Here $K_+ = \set{W' \in \RR^{n\times n} \;|\; W' \succeq 0}$ denotes the (real) positive semidefinite cone, and we define 
	\begin{equation}
		d(W, K_+) = \min_{W' \in K_+} \norm{W-W'}_F
	\end{equation}
	to be the minimum distance from $W$ to a point $W'$ in $K_+$, measured in Frobenius norm; note that this is a convex function. While this convex optimization problem looks complicated, it follows from a simple underlying idea: since the diagonal elements of $W$ are fixed by the constraint (\ref{eqn-diag-W-3}), this is simply an $\ell_1$-regularized estimator for the sparse, off-diagonal part of $W$. 
	
	The attentive reader will notice two potential concerns with this approach. First, in general, $C$ will not be a feasible solution to this convex optimization problem, since the diagonal elements of $C$ will not satisfy (\ref{eqn-diag-W-3}). However, we claim that $C$ lies close to a feasible solution. To see this, let $\widetilde{C}$ be the matrix whose off-diagonal elements agree with $C$, and whose diagonal elements agree with $\mathbf{g}$. Then $C$ is within distance $\epsilon_1$ of $\widetilde{C}$ (in Frobenius norm), and we claim that $\widetilde{C}$ is a feasible solution. 
	To see this, we can check that $\widetilde{C}$ satisfies the constraints (\ref{eqn-diag-W-3}), (\ref{eqn-Phi-W-3}) and (\ref{eqn-W-psd-3}), since we have: 
	\begin{equation}\label{eqn-snail}
		\begin{split}
			\norm{\Phi(\widetilde{C}) - \Phi(C)}_2 &= \norm{\Phi(\diag(\mathbf{g} - \diag(C)))}_2 \\
			&\leq ( m \norm{\mathbf{g} - \diag(C)}_1^2 )^{1/2} \\
			&\leq \epsilon_1 \sqrt{mn}.
		\end{split}
	\end{equation}
	(Here, we wrote $\widetilde{C}$ in a compact form, $\widetilde{C} = C + \diag(\mathbf{g} - \diag(C))$, where the $\diag(\cdot)$ notation has the following meaning: for a matrix $M$, $\diag(M)$ is the vector containing the entries $M_{jj}$ that lie along the diagonal of $M$; and for a vector $\mathbf{v} = (v_1,\ldots,v_n)$, $\diag(\mathbf{v})$ is the diagonal matrix with $v_1,\ldots,v_n$ along the diagonal.)
	
	Second, the reader will notice that the optimal solution $W$ may violate the positivity constraint (\ref{eqn-W-psd}), making it un-physical. (Similar issues can arise when performing quantum state and process tomography.) However, $W$ can be easily corrected to get a physically-admissible solution. This follows because equation (\ref{eqn-W-psd-3}) shows that $W$ lies within distance $\epsilon_1$ of a physically-admissible solution $W' \succeq 0$, and this solution $W'$ can be obtained by truncating the negative eigenvalues of $W$.
	
	Finally, we remark that there are different ways of relaxing the positivity constraint (\ref{eqn-W-psd}), and (\ref{eqn-W-psd-3}) is not the strongest possible choice. For instance, we could have used a stronger constraint than (\ref{eqn-W-psd-3}), such as: $d'(W, K_+) \leq \epsilon_1$, where we define $d'(W, K_+) = \min_{W' \in K_+} \norm{\diag(W-W')}_2$. However, the constraint (\ref{eqn-W-psd-3}) may be simpler to implement using numerical linear algebra software.
	
	Since these are convex optimization problems, they can be solved efficiently (both in theory and in practice), for instance by using interior point algorithms. Nonetheless, some care is needed to ensure that these algorithms can scale up to solve very large instances of these problems. In particular, enforcing the positivity constraint (\ref{eqn-W-psd}), and its relaxed version (\ref{eqn-W-psd-3}), can be computationally expensive. 
	
	\subsection{Omitting the Positivity Constraint}
	
	The theoretical analysis in Section \ref{sec:recguarantee} shows that $C$ can be reconstructed by solving the convex optimization problem (\ref{eqn-min-W-3})-(\ref{eqn-W-psd-3}). We remark that this analysis holds even \textit{without} the positivity constraint (\ref{eqn-W-psd-3}). It is easy to check that the positivity constraint is not used in Section \ref{sec:recguarantee}, and indeed, most of the theory of compressed sensing applies to all sparse signals, not just positive ones, although positivity can be helpful in certain situations \cite{kalev2015quantum}.
	
	This observation has a practical consequence: by omitting the positivity constraint (\ref{eqn-W-psd-3}), one can make the convex optimization problem simpler, and thus easier to solve in practice (e.g., by using second-order cone programming, rather than semidefinite programming) \cite{kueng2021personal}. One can then take the resulting solution, and project it onto the positive semidefinite cone, as is sometimes done in quantum state tomography \cite{sugiyama2013precision, guta2020fast}, without increasing the error (in Frobenius norm). This technique may be useful for scaling up our method to extremely large numbers of qubits.
	
	\subsection{Setting the Error Parameters $\epsilon_1$ and $\epsilon_2$}
	\label{sec-epsilon}
	
	Next, we describe how to set the parameters $\epsilon_1$ and $\epsilon_2$ in equations (\ref{eqn-eps-1}) and (\ref{eqn-eps-2}). We will use an approach that is similar to the one in Sections \ref{sec-estimating-decay-rates} and \ref{sec-direct-estimation}.
	
	First, we consider $\epsilon_1$, which bounds $\mathbf{u}$, the error in $\mathbf{g}$. Note that, when $g_j$ is estimated using the procedure in Section \ref{sec-estimating-decay-rates}, we also obtain large-deviation bounds on $u_j$. In particular, for some $\delta_1 > 0$, we have that:
	\begin{equation}\label{eqn-norm-uj}
		\norm{u_j}_{\psi_2} \leq \delta_1 c_{jj},
	\end{equation}
where $\norm{\cdot}_{\psi_2}$ is the sub-Gaussian norm (in the sense of \cite{vershynin2010introduction}). (This bound can be obtained from equation (\ref{eqn-psi2-norm-gamma-hat}), by setting $N_\text{trials} \sim 1/\delta_1^2$. Here, we are neglecting to count those trials of the experiment that are used to choose the evolution time $t$, because the number of those trials grows only logarithmically with $\Gamma_\mathbf{ab}$.) 

	This implies that $\norm{\mathbf{u}}_2^2$ is a subexponential random variable, whose subexponential norm (in the sense of \cite{vershynin2010introduction}) is at most $2\delta_1^2 \norm{\diag(C)}_2^2$. 
This implies that $\norm{\mathbf{u}}_2$ is bounded with high probability: for any $\tau \geq 1$, 
	\begin{equation}
		\Pr[\norm{\mathbf{u}}_2 \geq \tau\delta_1 \norm{\diag(C)}_2] \leq e\cdot \exp(-\tau^2/2c),
	\end{equation}
where $c > 0$ is some universal constant. We then choose $\tau$ to be a sufficiently large constant, so that the failure probability is small. (Note that in some cases, one can prove stronger bounds, by taking advantage of the fact that the coordinates of $\mathbf{u}$ are independent, and using a Bernstein-type inequality \cite{vershynin2010introduction}. This bound is stronger when the diagonal elements of $C$ satisfy $\norm{\diag(C)}_\infty \ll \norm{\diag(C)}_2$, i.e., when $\mathbf{u}$ has many independent coordinates with similar magnitudes.)

	The above bound does not immediately tell us how to set $\epsilon_1$, because the bound depends on $\diag(C)$, which is not known exactly. Instead, we now derive a bound that depends on $\mathbf{g}$, which is known explicitly, and can be used to set $\epsilon_1$. To do this, we assume that $\delta_1$ is sufficiently small so that $\tau\delta_1 < 1/4$. With high probability, we have 
	\begin{equation}\label{eqn-how-to-set-eps-1}
	\begin{split}
		\norm{\mathbf{u}}_2 &\leq \tau\delta_1 \norm{\diag(C)}_2 \\
		&\leq \tau\delta_1 (\norm{\mathbf{g}}_2 + \norm{\mathbf{u}}_2) \\
		&\leq \frac{\tau\delta_1}{1-\tau\delta_1} \norm{\mathbf{g}}_2 =: \epsilon_1,
	\end{split}
	\end{equation}
where we used the triangle inequality, and some algebra. This tells us how to set $\epsilon_1$ so that equation (\ref{eqn-eps-1}) holds. 
	
	We remark that $\epsilon_1$ can also be bounded in terms of $\diag(C)$, as follows:
	\begin{equation}\label{eqn-eps-1-simple-bound}
	\begin{split}
		\epsilon_1 &\leq  \frac{\tau\delta_1}{1-\tau\delta_1} (\norm{\diag(C)}_2 + \norm{\mathbf{u}}_2) \\
		&\leq  \frac{\tau\delta_1}{1-\tau\delta_1} (1+\tau\delta_1) \norm{\diag(C)}_2 \\
		&\leq 2\tau\delta_1 \norm{\diag(C)}_2.
	\end{split}
	\end{equation}
We will use this bound in Section \ref{sec:recguarantee}, when we analyze the accuracy of our estimate of $C$.
		
	Next, we consider $\epsilon_2$, which bounds $\mathbf{v}$, the error in $\mathbf{h}$. We use the same approach as above. When $h_j$ is estimated using the procedure in Section \ref{sec-estimating-decay-rates}, we obtain a bound on the sub-Gaussian norm of $v_j$: for some $\delta_2 > 0$, 
	\begin{equation}\label{eqn-norm-vj}
		\norm{v_j}_{\psi_2} \leq \delta_2 \Phi_j(C).
	\end{equation}
(This bound can be obtained from equation (\ref{eqn-psi2-norm-gamma-hat}), by setting $N_\text{trials} \sim 1/\delta_2^2$. We allow $\delta_2$ to be different from $\delta_1$, because the measurements used to estimate $h_j$ are more costly than the measurements used to estimate $g_j$, hence one may prefer to use different values for $N_\text{trials}$ in each case.)

	This implies that $\norm{\mathbf{v}}_2^2$ is a subexponential random variable, hence $\norm{\mathbf{v}}_2$ is bounded with high probability: for any $\tau \geq 1$, 
	\begin{equation}
		\Pr[\norm{\mathbf{v}}_2 \geq \tau\delta_2 \norm{\Phi(C)}_2] \leq e\cdot \exp(-\tau^2/2c),
	\end{equation}
where $c > 0$ is some universal constant. We then choose $\tau$ to be a sufficiently large constant, so that the failure probability is small. (Note that, for typical choices of $\Phi(\cdot)$, we expect that $\norm{\Phi(C)}_\infty \ll \norm{\Phi(C)}_2$. This implies that $\mathbf{v}$ has many independent coordinates with similar magnitudes. When this occurs, one can prove a stronger bound using a Bernstein-type inequality \cite{vershynin2010introduction}. For simplicity, we do not use this more elaborate bound here.)

	The above bound does not immediately tell us how to set $\epsilon_2$, because the bound depends on $\Phi(C)$, which is not known exactly. Instead, we now derive a bound that depends on $\mathbf{h}$, which is known explicitly, and can be used to set $\epsilon_2$. To do this, we assume that $\delta_2$ is sufficiently small so that $\tau\delta_2 < 1/4$. With high probability, we have 
	\begin{equation}\label{eqn-how-to-set-eps-2}
	\begin{split}
		\norm{\mathbf{v}}_2 &\leq \tau\delta_2 \norm{\Phi(C)}_2 \\
		&\leq \tau\delta_2 (\norm{\mathbf{h}}_2 + \norm{\mathbf{v}}_2) \\
		&\leq \frac{\tau\delta_2}{1-\tau\delta_2} \norm{\mathbf{h}}_2 =: \epsilon_2.
	\end{split}
	\end{equation}
This tells us how to set $\epsilon_2$ so that equation (\ref{eqn-eps-2}) holds. 
	
	Finally, we remark that $\epsilon_2$ can also be bounded in terms of $\norm{C}_{\ell_1}$, as follows:
	\begin{equation}\label{eqn-eps-2-simple-bound}
	\begin{split}
		\epsilon_2 &\leq  \frac{\tau\delta_2}{1-\tau\delta_2} (\norm{\Phi(C)}_2 + \norm{\mathbf{v}}_2) \\
		&\leq  \frac{\tau\delta_2}{1-\tau\delta_2} (1+\tau\delta_2) \norm{\Phi(C)}_2 \\
		&\leq 2\tau\delta_2 \norm{\Phi(C)}_2 \\
		&\leq 2\tau\delta_2 \sqrt{m} \norm{\Phi(C)}_\infty \\
		&\leq 4\tau\delta_2 \sqrt{m} \norm{C}_{\ell_1}.
	\end{split}
	\end{equation}
We will use this bound in Section \ref{sec:recguarantee}, when we analyze the accuracy of our estimate of $C$.


\section{Numerical Examples \label{sec-num-examples}}

We use numerical simulations to test how well our method performs on realistic system sizes, with different levels of sparsity, and when the data contain statistical fluctuations due to finite sample sizes. We find that our method performs well overall (see Figure \ref{fig:scaling}). 

	We numerically simulate the protocol for randomly chosen $C$ matrices (see Appendix~\ref{app:numdetails} for details). In these examples we assume that the diagonal elements of $C$ are known, that is, $\epsilon_1 = 0$ in Eq.~\eqref{eqn-diag-W-2}. We then solve the convex optimization problem given by \eqref{eqn-min-W}-\eqref{eqn-Phi-W} using CVXPY, a convex optimization package for Python \cite{cvxpy,cvxpy_rewriting}. 
	
	We first investigate the case of noiseless measurements, corresponding to $\epsilon_2=0$ in Eq.~\eqref{eqn-Phi-W-2}. In Fig.~\ref{fig:scaling}(a) we show the recovery error as a function of the number of measurements, $m$, for a fixed number of qubits, $n$, and various choices of the off-diagonal sparsity, $s$. The sharp transition in the recovery error as a function of $m$ is evident. Moreover, as shown in the inset of Fig.~\ref{fig:scaling}(a), the transition point $m_c$, which we define as the point where $\norm{C-W^{(opt)}}_\infty$ drops below 0.25, 

	scales linearly with $s$, consistent with our analytical results.  In Fig.~\ref{fig:scaling}(b) we fix  $s$, vary $n$, and study the recovery error as a function of $m$. Again, we observe a phase transition as $m$ increases. In this case, $m_c$ scales polynomially with $\log(n)$ as suggested in the inset of Fig.~\ref{fig:scaling}(b). 
	
	We then investigate the effect of noisy measurements on the recovery error. We generate random $C$ matrices, with a fixed number of qubits $n$ and sparsity $s$. We simulate noise by adding a random vector $\mathbf{e}$, whose entries are independent Gaussian random variables with mean 0 and standard deviation $\sigma_\epsilon$, to measurement vector $\mathbf{h}$. We now replace \eqref{eqn-Phi-W} in the previous convex program with \eqref{eqn-Phi-W-2} and choose $\epsilon_2=\sqrt{m}\sigma_\epsilon$. The scaling of the reconstruction error $\norm{W^{(opt)} - C}_\infty$ as a function of $\sigma_\epsilon$ is shown in Fig.~\ref{fig:scaling} (c). The recovery error after the phase transition point scales linearly with $\sigma_\epsilon$, consistent with our analytical bounds.

	
	\section{Recovery Guarantees \label{sec:recguarantee}}
	
	In this section we will study the convex optimization problem (\ref{eqn-min-W-3})-(\ref{eqn-W-psd-3}), and prove rigorous recovery guarantees that show that the optimal solution $W^{(opt)}$ is close to the true correlation matrix $C$, provided that $m \geq c_0 s\log n$ (and with better robustness to noise, when $m \geq c_0 s\log^4 n$). Here, $m$ is the dimension of the measurement vector $\mathbf{h}$, $s$ is the sparsity (the number of nonzero elements) in the off-diagonal part of the matrix $C$, and $c_0$ is some universal constant. 
	
	Actually, we will prove two different results: a non-universal recovery guarantee, using the ``RIPless'' framework of \cite{ripless}, as well as a universal recovery guarantee, using RIP-based techniques \cite{candes2006near, candes2006stable, rudelson2008sparse, vershynin2010introduction, davenport2012, foucart2013mathematical}. Here, RIP refers to the ``restricted isometry property,'' a fundamental proof technique in compressed sensing. There are different advantages to the RIPless and RIP-based bounds: the RIPless bounds require slightly fewer measurements, while the RIP-based bounds are more robust when the measurements are noisy.

	Along the way, we will introduce two variants of the problem (\ref{eqn-min-W-3})-(\ref{eqn-W-psd-3}): constrained $\ell_1$-minimization and the LASSO (``least absolute shrinkage and selection operator''). Generally speaking, recovery guarantees that hold for one of these problems can be adapted to the other one, with minor modifications. Here, we follow \cite{ripless} and prove a RIPless bound for the LASSO, and we follow \cite{davenport2012, foucart2013mathematical} and prove a RIP-based bound for constrained $\ell_1$-minimization.
	
	\subsection{Simplifying the Problem}
	
	We start with the convex optimization problem (\ref{eqn-min-W-3})-(\ref{eqn-W-psd-3}). We first remove the positivity constraint $d(W, K_+) \leq \epsilon_1$; this change should only hurt the accuracy of the solution $W^{(opt)}$. We also change the objective function to sum over all $i<j$ rather than all $i\neq j$; since $W$ is symmetric, this merely changes the objective function by a factor of 2. Finally, we shift the variable $W$ by subtracting away $\diag(\mathbf{g})$, so that its diagonal elements are all zero. In similar way, we shift the measurement vector $\mathbf{h}$ to get
	\begin{equation}\label{eqn-h'}
		\mathbf{h'} = \mathbf{h} - \Phi(\diag(\mathbf{g})). 
	\end{equation}
	This gives us an equivalent problem:
	\begin{equation}\label{eqn-min-W-4}
	\begin{split}
		&\text{Find $W \in \RR^{n\times n}$ that minimizes $\sum_{i<j} |W_{ij}|$,} \\
		&\text{such that:}
	\end{split}
	\end{equation}
	\begin{equation}\label{eqn-diag-W-4}
		\diag(W) = \mathbf{0}, 
	\end{equation}
	\begin{equation}\label{eqn-Phi-W-4}
		\norm{\Phi(W) - \mathbf{h'}}_2 \leq \epsilon_2 + \epsilon_1 \sqrt{mn}, 
	\end{equation}
	\begin{equation}\label{eqn-W-symmetric}
		W = W^T.
	\end{equation}
	
	We will use the following notation. We define an operation $\diag(\cdot)$ that has two meanings: given an $n\times n$ matrix $M$, $\diag(M)$ returns an $n$-dimensional vector containing the diagonal elements of $M$; and given an $n$-dimensional vector $\mathbf{v}$, $\diag(\mathbf{v})$ returns an $n\times n$ matrix that contains $\mathbf{v}$ along the diagonal, and zeroes off the diagonal. 
	
	Let us define $C'$ to be the off-diagonal part of the correlation matrix $C$, that is, $C'$ is the matrix whose off-diagonal elements match those of $C$, and whose diagonal elements are zero. We can write this concisely as:
	\begin{equation}
		\label{eqn-C'}
		C' = C - \diag(\diag(C)).
	\end{equation}
	We can view $\mathbf{h'}$ as a measurement of $C'$, with additive error $\mathbf{z}$, 
	\begin{equation}\label{eqn-z}
		\mathbf{h'} = \Phi(C') + \mathbf{z}.
	\end{equation}
	We want to show that the solution $W^{(opt)}$ is an accurate estimate of $C'$. Note that we can write the error term $\mathbf{z}$ in the form 
	\begin{equation}\label{eqn-zuv}
		\begin{split}
			\mathbf{z} &= \mathbf{h'} - \Phi(C') \\
			&= \mathbf{h} - \Phi(C) - \Phi(\diag(\mathbf{g} - \diag(C))) \\
			&= \mathbf{v} - \Phi(\diag(\mathbf{u})),
		\end{split}
	\end{equation}
	where $\mathbf{u}$ and $\mathbf{v}$ are the noise terms in (\ref{eqn-eps-1}) and (\ref{eqn-eps-2}). Then we can bound $\mathbf{z}$ using (\ref{eqn-eps-2}) and (\ref{eqn-snail}), 
	\begin{equation}
		\norm{\mathbf{z}}_2 \leq \epsilon_2 + \epsilon_1 \sqrt{mn}.
	\end{equation}
	
	It will be convenient to write 
	\begin{equation}
		\label{eqn-D}
		D = \set{W \in \RR^{n\times n} \;|\; W^T = W,\, \diag(W) = \mathbf{0}}
	\end{equation}
	to denote the subspace of symmetric matrices whose diagonal elements are all 0. Let $\uvec:\: D \rightarrow \RR^{n(n-1)/2}$ denote the linear operator that returns the upper-triangular part of $W$, 
	\begin{equation}
	\label{eqn-uvec}
		\uvec:\: W \mapsto (W_{ij})_{i<j}.
	\end{equation}
	Let us write $\Phi_D:\: D \rightarrow \RR^m$ to denote the measurement operator $\Phi$ restricted to act on the subspace $D$ (this definition will be useful later, when we work with the adjoint operator $\Phi_D^\dagger$). Then we can rewrite our problem (\ref{eqn-min-W-4})-(\ref{eqn-W-symmetric}) in a more concise form:
	\begin{equation}\label{eqn-min-W-5}
	\begin{split}
		&\text{Find $W \in D$ that minimizes $\norm{\uvec(W)}_1$,} \\
		&\text{such that:}
	\end{split}
	\end{equation}
	\begin{equation}\label{eqn-Phi-W-5}
		\norm{\Phi_D(W) - \mathbf{h'}}_2 \leq \epsilon_2 + \epsilon_1 \sqrt{mn}.
	\end{equation}
	
	\subsection{LASSO Formulation}
	
	In the following discussion, we will also consider a variant of our problem, the LASSO~\cite{hastie2013elements}:
	\begin{align}\label{eqn-lasso-simple}
		\text{Find $W \in D$ that minimizes: } \nonumber \\ 
		 \tfrac{1}{2} \norm{\Phi_D(W) - \mathbf{h'}}_2^2 + \lambda \norm{\uvec(W)}_1.
	\end{align}
	This can be viewed as a Lagrangian relaxation of the previous problem ((\ref{eqn-min-W-5})-(\ref{eqn-Phi-W-5})), or as an $\ell_1$-regularized least-squares problem. In addition, the convex optimization problems that were described earlier in Section \ref{sec-reconstructing-C-noisy} can also be relaxed into a LASSO-like form, in a similar way. The choice of the regularization parameter $\lambda$ requires some care. We will discuss this next.
	
	\subsection{Setting the Regularization Parameter $\lambda$}
	\label{sec-setting-lambda}
	
	In general, the regularization parameter $\lambda$ controls the relative strength of the two parts of the objective function in (\ref{eqn-lasso-simple}). When the noise in the measurement of $\mathbf{h'}$ is strong, then $\lambda$ must be set large enough to ensure that the $\ell_1$ regularization term still has the desired effect. However, if $\lambda$ is too large, it strongly biases the solution $W^{(opt)}$, making it less accurate. 
	
	Here, we sketch one approach to setting $\lambda$, following the analysis in \cite{ripless}. Our goal is to ensure that the solution $W^{(opt)}$ converges to (a sparse approximation of) the true correlation matrix $C$. To do this, we must set $\lambda$ large enough to satisfy two constraints, which involve the noise in the measurement of $\mathbf{h'}$ (see equation (IV.1) and the equation below (IV.2) in \cite{ripless}). When these constraints are satisfied, the error in the solution $W^{(opt)}$ is bounded by equation (IV.2) in \cite{ripless}. (Note that this error bound grows with $\lambda$, hence one should choose the smallest value of $\lambda$ that satisfies the above constraints.) 
	
	We now show in detail how to carry out the above calculation, in order to set $\lambda$. First, we give precise statements of the two constraints on $\lambda$: 
	\begin{equation}\label{eqn-lasso-noise-1}
		\norm{\uvec(\Phi_D^\dagger \mathbf{z})}_\infty \leq \lambda,
	\end{equation}
	\begin{equation}\label{eqn-lasso-noise-2}
		\norm{\uvec(\Phi_{D,T^c}^\dagger (I-P) \mathbf{z})}_\infty \leq \lambda.
	\end{equation}
	Here, $\mathbf{z}$ is the noise term in the measurement of $\mathbf{h'}$ in equation (\ref{eqn-z}); $\Phi_D^\dagger:\: \RR^m \rightarrow D$ is the adjoint of the measurement operator $\Phi_D$; $T \subset \set{(j,j') \;|\; 1\leq j<j'\leq n}$ is the support of (a sparse approximation of) the true correlation matrix $C$; $\Phi_{D,T}$ is the sub-matrix of $\Phi_D$ that contains those columns of $\Phi_D$ whose indices belong to the set $T$; $P$ is the projection onto the range of $\Phi_{D,T}$; $T^c$ is the complement of the set $T$; and $\Phi_{D,T^c}$ is the sub-matrix of $\Phi_D$ that contains those columns of $\Phi_D$ whose indices belong to the set $T^c$.
	
	In order to set $\lambda$, we need to compute the quantities in equations (\ref{eqn-lasso-noise-1}) and (\ref{eqn-lasso-noise-2}), and to do this, we need to have some bounds on the noise $\mathbf{z}$. We now demonstrate two ways of obtaining such bounds. 

One straightforward way is as follows. 
We can use equation (\ref{eqn-zuv}) to write
	\begin{equation}\label{eqn-zuv-2}
		\mathbf{z} = \mathbf{v} - \Phi(\diag(\mathbf{u})),
	\end{equation}
	where $\mathbf{u}$ and $\mathbf{v}$ are the noise terms in the measurements of $\mathbf{g}$ and $\mathbf{h}$, respectively. Also, recall that we previously showed bounds on $\mathbf{u}$ and $\mathbf{v}$ in Section \ref{sec-epsilon}, 
see equations (\ref{eqn-how-to-set-eps-1}) and (\ref{eqn-how-to-set-eps-2}). These imply bounds on $\mathbf{z}$, via an elementary calculation.

However, one can get better bounds on $\mathbf{z}$ by using a more sophisticated approach, starting with bounds on the sub-Gaussian norms of $u_j$ and $v_j$, such as equations (\ref{eqn-norm-uj}) and (\ref{eqn-norm-vj}). We describe this latter approach in detail.
	
	We assume that $\mathbf{g}$ and $\mathbf{h}$ are measured using the procedures described in Sections \ref{sec-estimating-decay-rates} and \ref{sec-epsilon}. Then equations (\ref{eqn-norm-uj}) and (\ref{eqn-norm-vj}) give us bounds on the sub-Gaussian norms of $u_j$ and $v_j$:
	\begin{equation}
	\label{eqn-uj-psi2}
		\norm{u_j}_{\psi_2} \leq \delta_1 c_{jj},
	\end{equation}
	\begin{equation}
	\label{eqn-vj-psi2}
		\norm{v_j}_{\psi_2} \leq \delta_2 \Phi_j(C).
	\end{equation}
	
	Using these bounds, we can then set $\lambda$ so that it satisfies (\ref{eqn-lasso-noise-1}) and (\ref{eqn-lasso-noise-2}) with high probability. More precisely, let us set
	\begin{equation}\label{eqn-lambda}
		\lambda := \bigl(\epsilon'''_1\cdot 4\sqrt{mn} + \epsilon'''_2\bigr) 4\sqrt{m} (1+\sqrt{2}) \sqrt{\ln(n)/c'},
	\end{equation}
	where 
	\begin{equation}
		\epsilon'''_1 := \frac{\delta_1}{1-\epsilon''_1} \norm{\mathbf{g}}_\infty, \quad \epsilon''_1 := 2\sqrt{\ln(n)/c_0} \; \delta_1,
	\end{equation}
	\begin{equation}
		\epsilon'''_2 := \frac{\delta_2}{1-\epsilon''_2} \norm{\mathbf{h}}_\infty, \quad \epsilon''_2 := 2\sqrt{\ln(m)/c_0} \; \delta_2.
	\end{equation}
	Here we are assuming that $\delta_1$ and $\delta_2$ are sufficiently small (for instance, $\delta_1 \lesssim 1/\sqrt{\ln(n)}$ and $\delta_2 \lesssim 1/\sqrt{\ln(n)}$) to ensure that $\epsilon''_1 < 1/2$ and $\epsilon''_2 < 1/2$. Also, here $c'$ and $c_0$ are universal constants (which are defined in the proof below). 
	
	Now let the correlation matrix $C$ and measurement operator $\Phi_D$ be fixed, and note that the noise term $\mathbf{z}$ is stochastic. Then we claim that, with high probability (over the random realization of $\mathbf{z}$), equations (\ref{eqn-lasso-noise-1}) and (\ref{eqn-lasso-noise-2}) will be satisfied; here the failure probability is at most $(e/n^3) + (e/m^3) + (2e/n^{1+2\sqrt{2}})$. 
	We prove this claim in Appendix~ \ref{sec-proof-lasso-lambda}.
	
	Finally, we have the following simple upper bounds on $\epsilon'''_1$, $\epsilon'''_2$ and $\lambda$ (these follow from the definitions of $\epsilon'''_1$ and $\epsilon'''_2$, the definitions of $\mathbf{g}$ and $\mathbf{h}$, equations (\ref{eqn-u-ell-infty}) and (\ref{eqn-v-ell-infty}), and the definition of $\Phi(\cdot)$):
	\begin{equation}
		\epsilon'''_1 \leq 3\delta_1 \norm{\diag(C)}_\infty,
	\end{equation}
	\begin{equation}
		\epsilon'''_2 \leq 6\delta_2 \norm{C}_{\ell_1},
	\end{equation}
	\begin{equation}\label{eqn-lambda-simple-bound}
		\begin{split}
			\lambda &\leq O\bigl( \sqrt{m\ln(n)} \bigl(\delta_1 \norm{\diag(C)}_\infty \sqrt{mn} + \delta_2 \norm{C}_{\ell_1}\bigr) \bigr).
		\end{split}
	\end{equation}
	We will make use of these bounds later, when we analyze the accuracy of our estimate of $C$.

	\subsection{Isotropy and Incoherence of the Measurement Operator}
	\label{sec-isotropy-incoherence}
	
	We will show that the rows of the measurement operator $\Phi_D$ have two properties, \textit{isotropy} and \textit{incoherence}, which play a fundamental role in compressed sensing (see, e.g., \cite{vershynin2010introduction, ripless}). Let $Q$ be the matrix (of size $m$ by $n(n-1)/2$) that represents the action of $\Phi_D$ (using the fact that the subspace $D$ is isomorphic to $\RR^{n(n-1)/2}$); that is, $Q$ and $\Phi_D$ are related by the equation: 
	\begin{equation}\label{eqn-phi-and-Q}
		\Phi_D(C) = Q\cdot\uvec(C), \quad \forall C \in D.
	\end{equation}
	The rows of $Q$ are chosen independently at random, and each row has the form 

	\begin{equation}\label{eqn-def-q}
		\mathbf{q} = 4\uvec(\mathbf{r}\mathbf{r}^T) \in \RR^{n(n-1)/2},
	\end{equation}
	where $\mathbf{r}$ is sampled from the distribution described in (\ref{eqn-r-distrib}). We say that $\mathbf{q}$ is \textit{centered} if it has mean $\EE(\mathbf{q}) = \mathbf{0}$, and we say that $\mathbf{q}$ is \textit{isotropic} if its covariance matrix is the identity:
	\begin{equation}
		\EE(\mathbf{q}\mathbf{q}^T) = I.
	\end{equation}
	It is straightforward to check that $\mathbf{q}$ is centered and isotropic (up to a normalization factor of 2), since:
	\begin{align}
		\label{eqn:isotropicQ}
		\begin{cases}
			\EE[r_i r_j] = 0 & i<j\\
			\EE[r_i r_j r_k r_l] = 0 & \parbox[t]{0.6 \columnwidth}{ $i<j$ and $k<l$ 
				and $\{i,j\}\cap\{k,l\}=\emptyset$}\\
			\EE[r_i r_j r_k r_l] = 0 & \parbox[t]{0.6\columnwidth}{  $i<j$ and $k<l$ 
				and $|\{i,j\}\cap\{k,l\}|=1$}\\
			\EE[r_i r_j r_k r_l] = \tfrac{1}{4} &\parbox[t]{0.6\columnwidth}{ $i<j$ and $k<l$ 
				and $i=k$ and $j=l$}
		\end{cases}.
	\end{align}
	(Note that in the last line of Eq.~\ref{eqn:isotropicQ}, we cannot have a case with $i=l$ and $j=k$, as the requirements of $i<j$ and $k<l$ lead to a contradiction.) 
	
	In addition, we say that $\mathbf{q}$ is \textit{incoherent} with parameter $\mu > 0$ if, with probability 1, all of its coordinates are small:
	\begin{equation} \label{eqn-incoherence}
		\norm{\mathbf{q}}_\infty^2 \leq \mu.
	\end{equation}
	In order for this to be useful for compressed sensing, one needs $\mu$ to be small, say, at most polylogarithmic in the dimension of $\mathbf{q}$. In our case, it is easy to see that $\mathbf{q}$ is incoherent with parameter $\mu = 16$.

	\subsection{Non-universal (RIPless) Recovery Guarantee\label{sec-ripless}}
	
	We begin by proving a non-universal recovery guarantee, using the ``RIPless'' framework of \cite{ripless}, which in turn relies on the isotropy and incoherence properties shown in the preceding section. 
	
	Let $C \in \RR^{n\times n}$ be a correlation matrix, and let $C' \in D$ be its off-diagonal part (see equation (\ref{eqn-C'})). We will assume that $C'$ is approximately $s$-sparse, i.e., there exists a matrix $C^{(s)} \in D$ that has at most $s$ nonzero entries above the diagonal, and that approximates $C'$ in the (vector) $\ell_1$ norm, up to an error of size $\eta_s$. This can be written compactly as:
	\begin{equation}
	\label{eqn-ripless-eta-s}
		\norm{\uvec(C' - C^{(s)})}_1 \leq \eta_s,
	\end{equation}
	where $\uvec(\cdot)$ was defined in equation (\ref{eqn-uvec}). (Recall that both $C'$ and $C^{(s)}$ are symmetric, with all zeroes on the diagonal. Hence it suffices to consider those matrix entries that lie above the diagonal.)
	
	We now choose the measurement operator $\Phi$ at random (see equation (\ref{eqn-Phi})). We assume that $m$ (the dimension of $\mathbf{h}$) satisfies the bound:
	\begin{equation}
	\label{eqn-ripless-m}
		m \geq \tilde{c}_0 (1+\beta)\cdot 4s \log \tfrac{n(n-1)}{2}.
	\end{equation}
	Here, $\tilde{c}_0$ is a universal constant, and $\beta > 0$ is a parameter that can be chosen freely by the experimenter. Note that $m$ scales linearly with the sparsity $s$, but only logarithmically with the dimension $n$ of the matrix $C$. This scaling is close to optimal, in an information-theoretic sense. This is one way of quantifying the advantage of compressed sensing, when compared with measurements that do not exploit the sparsity of $C$. 
	
	We measure $\mathbf{g} \approx \diag(C)$ and $\mathbf{h} \approx \Phi(C)$, and we calculate $\mathbf{h'}$ (see equation (\ref{eqn-h'})). We let $\delta_1$ and $\delta_2$ quantify the noise in the measurements of $\mathbf{g}$ and $\mathbf{h}$, as described in equations (\ref{eqn-uj-psi2}) and (\ref{eqn-vj-psi2}). We then solve the LASSO problem in (\ref{eqn-lasso-simple}), setting the regularization parameter $\lambda$ according to (\ref{eqn-lambda}). Let $W^{(opt)}$ be the solution of this optimization problem. 
	
	We now have the following recovery guarantee, which shows that $W^{(opt)}$ gives a good approximation to $C'$, in both the Frobenius ($\ell_2$) norm, and the vector $\ell_1$ norm. This follows directly from Theorem 1.2 in \cite{ripless} (and the extension of that theorem to more general classes of noisy measurements in Section IV in \cite{ripless}).
	\begin{theorem}\label{thm-ripless}
		For any correlation matrix $C$ satisfying the sparsity condition (\ref{eqn-ripless-eta-s}), with probability at least $1 - \tfrac{12}{n(n-1)} - 6e^{-\beta}$ (over the random choice of the measurement operator $\Phi$, with $m$ set according to (\ref{eqn-ripless-m})), the solution $W^{(opt)}$ is close to $C'$, with an error that is bounded by:
		\begin{align}
			\norm{W^{(opt)} - C'}_F &\leq \sqrt{2}c (1+\alpha) \biggl[ \frac{\eta_s}{\sqrt{s}} + \lambda \sqrt{s} \biggr], \\
			\norm{W^{(opt)} - C'}_{\ell_1} &\leq 2c (1+\alpha) \bigl[ \eta_s + \lambda s \bigr],
		\end{align}
		where $c$ is a universal constant, and $\alpha \leq \log^{3/2} \bigl( \tfrac{n(n-1)}{2} \bigr)$.
	\end{theorem}
	In these bounds, the first term upper-bounds the error that results from approximating $C'$ by a sparse matrix, and the second term upper-bounds the error due to noise in the measurements of $\mathbf{g}$ and $\mathbf{h}$. 
	
	To make the second term more transparent, we can combine it with the bound on $\lambda$ from equation (\ref{eqn-lambda-simple-bound}):
	\begin{equation}
	\lambda \leq O\bigl( \sqrt{m\ln(n)} \bigl(\delta_1 \norm{\diag(C)}_\infty \sqrt{mn} + \delta_2 \norm{C}_{\ell_1}\bigr) \bigr),
	\end{equation}
	where $\delta_1$ and $\delta_2$ quantify the noise in the measurements of $\mathbf{g}$ and $\mathbf{h}$, as described earlier.
	
	Also, it is useful to consider the special case where $C'$ is \textit{exactly} $s$-sparse, so $\eta_s = 0$, and where we use as few measurement settings as possible, by setting $m = O(s\log n)$. In this case, we have:
\begin{widetext}
\begin{align}
\norm{&W^{(opt)}-C'}_F \leq O(\log^{3/2}(n) \lambda \sqrt{s}) \\
&\leq O(\sqrt{s} \log^{3/2}(n) \sqrt{m\log(n)} 
 \cdot [\delta_1 \norm{\diag(C)}_\infty \sqrt{mn} + \delta_2 \norm{C}_{\ell_1}]) \\
&\leq O(s \log^{5/2}(n) [\delta_1 \norm{\diag(C)}_\infty \sqrt{sn\log(n)} 
 + \delta_2 \norm{C}_{\ell_1}]) \\
&\leq O(s \log^{5/2}(n) [\delta_1 \norm{\diag(C)}_\infty \sqrt{sn\log(n)} 
 + \delta_2 \sqrt{n} \norm{\diag(C)}_2 + \delta_2 \sqrt{2s} \norm{C'}_F]).
\label{eqn-error-bound-cs-slogn}
\end{align}
\end{widetext}

	This can be compared with the error bound (\ref{eqn-direct-error-bound-L2-whp}) for the naive method, and the RIP-based error bound (\ref{eqn-error-bound-cs-slog4n}) for compressed sensing. Generally speaking, compressed sensing has an advantage over the naive method when $s$ is small, and the RIPless bound is useful in the regime between $m = O(s\log n)$ and $m = O(s\log^4 n)$, where the RIP-based bound does not apply. (When $m$ is $O(s\log^4 n)$ or larger, the RIP-based bound applies, and gives better results than the RIPless bound.) We will carry out a more detailed comparison between the naive method and compressed sensing in Section \ref{sec-performance}.

	\subsection{Universal (RIP-based) Recovery Guarantee \label{sec-rip-based}}
	
	Next, we prove a universal recovery guarantee, using an older approach based on the restricted isometry property (RIP) \cite{candes2006near, candes2006stable, rudelson2008sparse, vershynin2010introduction, davenport2012, foucart2013mathematical}. This also relies on the isotropy and incoherence properties shown above. As these techniques are fairly standard in compressed sensing, we will simply sketch the proof. 
	
	First, we set the number of measurement settings to be 
	\begin{equation}
	\label{eqn-rip-m}
	m \geq c_0 s\log^4 n, 
	\end{equation}
	where $s$ is the sparsity parameter, and $c_0$ is some universal constant. (Note that $m$ is slightly larger, by a poly($\log n$) factor, compared to the RIPless case.) Also, recall that $D$ is the subspace of symmetric matrices whose diagonal elements are all 0 (see equation (\ref{eqn-D})), and $\Phi_D$ is the measurement operator restricted to act on this subspace.
	
	We claim that, with high probability (over the random choice of $\Phi_D$), the normalized measurement operator $\Phi_D/\sqrt{m}$ satisfies the RIP (for sparsity level $2s$). To see this, we recall the isotropy and incoherence properties shown above. These properties imply that the measurement operator $\Phi_D$ is sampling at random from a ``bounded orthonormal system.'' Such operators are known to satisfy the RIP, via a highly nontrivial proof \cite{candes2006near, rudelson2008sparse}; a recent exposition can be found in Chapter 12 in \cite{foucart2013mathematical}. 
	
	
	From this point onwards, we let the measurement operator $\Phi_D$ be fixed. We will show that $\Phi_D$ is capable of reconstructing the off-diagonal parts of \textit{all} sparse matrices $C$, i.e., $\Phi_D$ can perform ``universal'' recovery. 
	
	As in the previous section, let $C \in \RR^{n\times n}$ be a correlation matrix, and let $C' \in D$ be its off-diagonal part (see equation (\ref{eqn-C'})). We will assume that $C'$ is approximately $s$-sparse, i.e., there exists a matrix $C^{(s)} \in D$ that has at most $s$ nonzero entries above the diagonal, and that approximates $C'$ in the (vector) $\ell_1$ norm, up to an error of size $\eta_s$. This can be written compactly as:
	\begin{equation}
	\label{eqn-rip-eta-s}
		\norm{\uvec(C' - C^{(s)})}_1 \leq \eta_s,
	\end{equation}
	where $\uvec(\cdot)$ was defined in equation (\ref{eqn-uvec}). (Recall that both $C'$ and $C^{(s)}$ are symmetric, with all zeroes on the diagonal. Hence it suffices to consider those matrix entries that lie above the diagonal.)
	
	We measure $\mathbf{g} \approx \diag(C)$ and $\mathbf{h} \approx \Phi(C)$, and we calculate $\mathbf{h'}$ (see equation (\ref{eqn-h'})). We assume that the noise in the measurements of $\mathbf{g}$ and $\mathbf{h}$ is bounded by $\delta_1$ and $\delta_2$, as described in (\ref{eqn-norm-uj}) and (\ref{eqn-norm-vj}). We then solve the $\ell_1$-minimization problem in (\ref{eqn-min-W-5}) and (\ref{eqn-Phi-W-5}), setting the parameters $\epsilon_1$ and $\epsilon_2$ according to (\ref{eqn-how-to-set-eps-1}) and (\ref{eqn-how-to-set-eps-2}). Let $W^{(opt)}$ be the solution of this problem. 
	
	We now have the following recovery guarantee, which shows that $W^{(opt)}$ gives a good approximation to $C'$, in the Frobenius ($\ell_2$) norm, and in the $\ell_1$ vector norm. This follows directly from Theorem 1.9 in \cite{davenport2012}, and Theorem 6.12 in \cite{foucart2013mathematical}. (There is one subtle point: the convex optimization problem in (\ref{eqn-min-W-5}) and (\ref{eqn-Phi-W-5}) uses the unnormalized measurement operator $\Phi_D$, while the error bounds in \cite{davenport2012, foucart2013mathematical} apply to the normalized measurement operator $\Phi_D/\sqrt{m}$. Hence, the noise is smaller by a factor of $\sqrt{m}$ in these error bounds.)
	\begin{theorem} \label{thm:compsens}
		With high probability (over the choice of the measurement operator $\Phi$, with $m$ set according to (\ref{eqn-rip-m})), for all correlation matrices $C$ (satisfying the sparsity condition (\ref{eqn-rip-eta-s})), the solution $W^{(opt)}$ satisfies the following error bounds:
		\begin{equation}\label{eqn-recoveryerror}
			\norm{W^{(opt)} - C'}_F \leq c_1 \frac{\eta_s}{\sqrt{s}} + c_2 \Bigl(\frac{\epsilon_2}{\sqrt{m}} + \epsilon_1 \sqrt{n}\Bigr),
		\end{equation}
		\begin{equation}\label{eqn-recoveryerror-L1}
			\norm{W^{(opt)} - C'}_{\ell_1} \leq c_1 \eta_s + c_2 \sqrt{s} \Bigl(\frac{\epsilon_2}{\sqrt{m}} + \epsilon_1 \sqrt{n}\Bigr),
		\end{equation}
		where $c_1$ and $c_2$ are universal constants.
	\end{theorem}
	In these bounds, the first term upper-bounds the error that results from approximating $C'$ by a sparse matrix, and the second term upper-bounds the error due to noise in the measurements of $\mathbf{g}$ and $\mathbf{h}$. 
	
	In order to apply these bounds, one needs to know the values of $\epsilon_1$ and $\epsilon_2$. These can be obtained from Section \ref{sec-epsilon}, equations (\ref{eqn-eps-1-simple-bound}) and (\ref{eqn-eps-2-simple-bound}): 
	\begin{align}
	\epsilon_1 &\leq O(\delta_1 \norm{\diag(C)}_2), \\
	\epsilon_2 &\leq O(\delta_2 \sqrt{m} \norm{C}_{\ell_1}).
	\end{align}
	
	Also, it is useful to consider the special case where $C'$ is \textit{exactly} $s$-sparse, so $\eta_s = 0$, and where we use as few measurement settings as possible, by setting $m = O(s\log^4 n)$. In this case, we have:
\begin{widetext}
\begin{align}
\norm{&W^{(opt)}-C'}_F \\
&\leq O(\sqrt{n} \delta_1 \norm{\diag(C)}_2 + \delta_2 \norm{C}_{\ell_1}) \\
&\leq O(\sqrt{n} \delta_1 \norm{\diag(C)}_2 + \delta_2 (\sqrt{n} \norm{\diag(C)}_2 + \sqrt{2s} \norm{C'}_F)).
\label{eqn-error-bound-cs-slog4n}
\end{align}
\end{widetext}

	This can be compared with the error bound (\ref{eqn-direct-error-bound-L2-whp}) for the naive method, and the RIPless error bound (\ref{eqn-error-bound-cs-slogn}) for compressed sensing. Generally speaking, compressed sensing has an advantage over the naive method when $s$ is small, and the RIP-based bound has better scaling (as a function of $n$, $s$, $\diag(C)$ and $C'$) than the RIPless bound, although it requires $m$ to be slightly larger. We will carry out a more detailed comparison between the naive method and compressed sensing in Section \ref{sec-performance}.
	


\section{Performance Evaluation}
\label{sec-performance}

In this section we study the performance of our compressed sensing method, for a typical measurement scenario. We consider both the accuracy of the method, and the experimental resources required to implement it. We investigate the asymptotic scaling of our method, and compare it to the naive method, direct estimation of the correlation matrix, introduced in Section \ref{sec-direct-estimation}. 

Overall, we find that our compressed sensing method has asymptotically better sample complexity, whenever the off-diagonal part of the correlation matrix $C$ is sufficiently sparse. In particular, for a system of $n$ qubits, our method is advantageous whenever the number of correlated pairs of qubits, $s$, is at most $O(n^{3/2})$ (ignoring log factors). These results are summarized in Figure \ref{fig-sample-complexity-full}.

\begin{figure*}
\begin{tabular}{|l|c|c|c|}
\hline
\textbf{Reconstruction method:} & Naive & CS (RIP-based) & CS (RIPless*) \\
\hline
\hline
\textbf{Single-qubit}    & & & \\
\textbf{spectroscopy:} & & & \\
\hline
\# of meas. settings & $n$ & $n$ & $n$ \\
\hline
\# of samples per setting & $O(n/\delta^2)$ & $O(n/\delta^2)$ & $O(s^3 \log^6(n) / \delta^2)$ \\
\hline
Total \# of samples & $O(n^2/\delta^2)$ & $O(n^2/\delta^2)$ & $O(n s^3 \log^6(n) / \delta^2)$ \\
\hline
\hline
\textbf{Multi-qubit}     & & & \\
\textbf{spectroscopy:} & & & \\
\hline
\# of meas. settings $m$ & $O(n^2)$ & $O(s \log^4 n)$ & $O(s \log n$) \\
\hline
\# of samples per setting & $O(n/\delta^2)$ & $O(\max(n,s) / \delta^2)$ & $O(s^2 \max(n,s) \log^5(n) / \delta^2)$ \\
\hline
Total \# of samples & $O(n^3/\delta^2)$ & $O(s \max(n,s) \log^4(n) / \delta^2)$ & $O(s^3 \max(n,s) \log^6(n) / \delta^2)$ \\
\hline
\hline
\textbf{Total sample} & & & \\
\textbf{complexity:}   & $O(n^3/\delta^2)$ & $O(\max(n,s)^2 \log^4(n) / \delta^2)$ & $O(s^3 \max(n,s) \log^6(n) / \delta^2)$ \\
\hline
\end{tabular}
\caption{Sample complexity of different methods for reconstructing a correlation matrix $C$, of size $n \times n$, with $2s$ nonzero elements off the diagonal. The naive method is to measure each element of $C$ separately. ``CS'' refers to the compressed sensing method, and ``RIP-based'' and ``RIPless'' refer to different analytical bounds on the accuracy of the reconstruction of $C$. The asterisk (*) indicates that the results using the RIPless bound hold under a technical assumption that the diagonal of $C$ does not have any unusually large elements, see equation (\ref{eqn-diagC-assumption}). The different methods are parameterized in such a way that they reconstruct the diagonal of $C$ up to an additive error of size $\delta \norm{\diag(C)}_2$, and they reconstruct the off-diagonal part of $C$ up to an additive error of size $\delta \norm{C}_F$. Each method makes use of single-qubit spectroscopy (with $n$ experimental configurations or ``measurement settings''), as well as multi-qubit spectroscopy (with $m$ measurement settings, where $m$ varies between $O(n^2)$ and $O(s\log n)$). The CS method has lower sample complexity than the naive method, when $s \ll n^2$. In particular, using the RIP-based bound with $m = O(s\log^4 n)$, the CS method is advantageous whenever $s \leq O(n^{3/2} / \log^2 n)$. Using the RIPless bound with $m = O(s\log n)$, the CS method is advantageous whenever $s \leq O(n^{2/3} / \log^2 n)$.}
\label{fig-sample-complexity-full}
\end{figure*}

We now explain these results in detail. We let $C'$ be the off-diagonal part of the correlation matrix $C$, that is, $C'$ is the matrix whose off-diagonal elements match those of $C$, and whose diagonal elements are zero. We are promised that $C'$ has at most $s$ nonzero elements above the diagonal (and, by symmetry, at most $s$ nonzero elements below the diagonal). Our goal is to estimate both $C'$ and $\diag(C)$.

Compressed sensing allows the possibility of adjusting the number of measurement settings, $m$, over a range from $\sim s\log n$ to $n^2$. (Note that $m \sim s\log n$ is just slightly above the information-theoretic lower bound, while $m = n^2$ is the number of measurement settings used by the naive method.) 
Compressed sensing works across this whole range, but the error bounds vary depending on $m$. There are two cases: (1) For $m \gtrsim s\log^4 n$, both the RIP-based and RIPless error bounds are available, and the RIP-based error bounds are asymptotically stronger. (2) For $m$ between $\sim s\log n$ and $\sim s\log^4 n$, only the RIPless error bound is available. 

To make a fair comparison between compressed sensing and the naive method, we need to quantify the accuracy of these methods in a consistent way. This is a nontrivial task, because the error bounds for the different methods have different dependences on the parameters $n$, $s$, $\diag(C)$ and $C'$; this can be seen by comparing equation (\ref{eqn-direct-error-bound-L2-whp}), and Theorems \ref{thm-ripless} and \ref{thm:compsens}. 

We choose a simple way of quantifying the accuracy of all of these methods: given some $\delta > 0$, we require that each method return an estimate $\widehat{C}'$ that satisfies 
\begin{equation}\label{eqn-delta-normCF}
\norm{\widehat{C}' - C'}_F \leq \delta \norm{C}_F.
\end{equation}
Here, we use the Frobenius matrix norm, which is equivalent to the vector $\ell_2$ norm. We write $C$ (rather than $C'$) on the right hand side of the inequality, in order to allow the recovery error to depend on both the diagonal and the off-diagonal elements of $C$. 

In addition, we require that each method return an estimate $\mathbf{g}$ of $\diag(C)$ that satisfies
\begin{equation}\label{eqn-delta-normdiagC2}
\norm{\mathbf{g} - \diag(C)}_2 \leq \delta \norm{\diag(C)}_2.
\end{equation}
For both compressed sensing as well as the naive method, $\mathbf{g}$ is obtained in the same way, by performing single-qubit spectroscopy as in (\ref{eqn-g}), and the error in $\mathbf{g}$ satisfies the same bound (\ref{eqn-direct-error-bound-L2-diag-whp}).

We also need to account for the cost of implementing each method using real experiments. This cost depends on a number of factors. One factor is the total number of experiments that have to be performed, often called the \textit{sample complexity}. This is the number of measurement settings, times the number of repetitions of the experiment with each measurement setting. Another factor is the difficulty of performing a single run of the experiment. This involves both the difficulty of preparing entangled states (random $n$-qubit GHZ states for the compressed sensing method, and 2-qubit Bell states for the naive method), and the length of time that one has to wait in order to observe dephasing. 

Here, we study a scenario where we expect our compressed sensing method to perform well. We consider an advanced quantum information processor, where $n$-qubit GHZ states are fairly easy to prepare (using $O(\log n)$-depth quantum circuits), and dephasing occurs at low rates, so that the main cost of running each experiment is the amount of time needed to observe dephasing. In this scenario, it is reasonable to use the sample complexity as a rough measure of the total cost of implementing the compressed sensing method, as well as the naive method.

We now calculate the sample complexity for three methods of interest: (1) the naive method with $m = n^2$, (2) compressed sensing with $m \sim s\log^4 n$, and (3) compressed sensing with $m \sim s\log n$. We find that method (2) outperforms the naive method whenever $s \leq O(n^{3/2} / \log^2 n)$, and method (3) outperforms the naive method whenever $s \leq O(n^{2/3} / \log^2 n)$.

In addition, for each of these methods, we show the number of samples where single-qubit spectroscopy is performed, and the number of samples where multi-qubit spectroscopy is performed. (Recall that all of these methods use single-qubit spectroscopy to estimate the diagonal of $C$, and then use multi-qubit spectroscopy to estimate the off-diagonal part of $C$.) Both of these numbers can be important: multi-qubit spectroscopy is more expensive to implement on essentially all experimental platforms, and requires more samples when $s$ is large; but it is possible for single-qubit spectroscopy to dominate the overall sample complexity, when $s$ is small. 


\subsection{Naive method with $m = n^2$} 

As in Section \ref{sec-direct-estimation}, we use two parameters, $\delta_1$ and $\delta_2$, to quantify the accuracy of the measurements, as in equations (\ref{eqn-direct-delta-1}) and (\ref{eqn-direct-delta-2}). Then we get an estimate $\widehat{C}'$ of $C'$, whose error is bounded by equation (\ref{eqn-direct-error-bound-L2-whp}): with probability at least $1-\eta$, 
\begin{equation}
\begin{split}
\norm{&\widehat{C}' - C'}_F^2 \\
&\leq \tfrac{1}{\eta} [3(n-1) (\delta_1+\delta_2)^2 \norm{\diag(C)}_2^2 + 6\delta_2^2 \norm{C'}_F^2].
\end{split}
\end{equation}

For simplicity, we set $\eta$ to be some universal constant, say $\eta = 0.001$. Now, given any $\delta > 0$, we can ensure that 
\begin{equation}
\begin{split}
\norm{\widehat{C}' - C'}_F^2 &\leq \delta^2 \norm{\diag(C)}_2^2 + (\delta^2/n) \norm{C'}_F^2 \\
&\leq \delta^2 \norm{C}_F^2, 
\end{split}
\end{equation}
by setting $\delta_1 = \delta_2 = O(\delta/\sqrt{n})$. This satisfies the requirement (\ref{eqn-delta-normCF}).

In addition, one can easily check that the estimate $\mathbf{g}$ for $\diag(C)$ satisfies the requirement (\ref{eqn-delta-normdiagC2}).

Then the sample complexity is as follows (see the discussion preceding (\ref{eqn-direct-delta-1}) and (\ref{eqn-direct-delta-2})): the method performs single-qubit spectroscopy on $O(n/\delta_1^2) = O(n^2/\delta^2)$ samples, and multi-qubit spectroscopy on $O(n^2 / \delta_2^2) = O(n^3/\delta^2)$ samples. Hence the total sample complexity is $O(n^3/\delta^2)$.


\subsection{Compressed sensing with $m \sim s\log^4 n$} 

Here we consider the $\ell_1$-minimization problem in (\ref{eqn-min-W-5}) and (\ref{eqn-Phi-W-5}), whose solution $W^{(opt)}$ satisfies the RIP-based bound in Theorem \ref{thm:compsens}. We use two parameters, $\delta_1$ and $\delta_2$, to quantify the accuracy of the measurements, as in equations (\ref{eqn-norm-uj}) and (\ref{eqn-norm-vj}). Section \ref{sec-epsilon} then explains how to set the parameters $\epsilon_1$ and $\epsilon_2$ that appear in (\ref{eqn-min-W-5}) and (\ref{eqn-Phi-W-5}).

The estimator $W^{(opt)}$ satisfies the following error bound (see Theorem \ref{thm:compsens}, and equation (\ref{eqn-error-bound-cs-slog4n})):
\begin{widetext}
\begin{align}
\norm{&W^{(opt)}-C'}_F 
\leq O(\sqrt{n} \delta_1 \norm{\diag(C)}_2 + \delta_2 (\sqrt{n} \norm{\diag(C)}_2 + \sqrt{2s} \norm{C'}_F)).
\end{align}
\end{widetext}

Now, given any $\delta > 0$, we can ensure that 
\begin{equation}
\begin{split}
\norm{W^{(opt)}-C'}_F &\leq \tfrac{1}{\sqrt{2}} \delta \norm{\diag(C)}_2 + \tfrac{1}{\sqrt{2}} \delta \norm{C'}_F \\
&\leq \delta \norm{C}_F, 
\end{split}
\end{equation}
by setting $\delta_1 = O(\delta/\sqrt{n})$ and 
\begin{equation}
\delta_2 = O(\delta/\sqrt{\max(n,s)}). 
\end{equation}
This satisfies the requirement (\ref{eqn-delta-normCF}).

In addition, one can easily check that the estimate $\mathbf{g}$ for $\diag(C)$ satisfies the requirement (\ref{eqn-delta-normdiagC2}).

Then the sample complexity is as follows (see the discussion following (\ref{eqn-norm-uj}) and (\ref{eqn-norm-vj})): the method performs single-qubit spectroscopy on $O(n / \delta_1^2) = O(n^2 / \delta^2)$ samples, and multi-qubit spectroscopy on 
\begin{equation}
O(m / \delta_2^2) \leq O(s \max(n,s) \log^4(n) / \delta^2)
\end{equation}
samples. Hence the total sample complexity is at most 
\begin{equation}
O(\max(n,s)^2 \log^4(n) / \delta^2).
\end{equation}
This is less than the sample complexity of the naive method, provided the off-diagonal part of the correlation matrix is sufficiently sparse, i.e., when $s \leq O(n^{3/2} / \log^2 n)$. 


\subsection{Compressed sensing with $m \sim s\log n$} 

Here we consider the LASSO optimization problem in (\ref{eqn-lasso-simple}), whose solution $W^{(opt)}$ satisfies the RIPless bound in Theorem \ref{thm-ripless}. We use two parameters, $\delta_1$ and $\delta_2$, to quantify the accuracy of the measurements, as in equations (\ref{eqn-norm-uj}) and (\ref{eqn-norm-vj}). Section \ref{sec-setting-lambda} then explains how to set the LASSO regularization parameter $\lambda$.

In the following, we assume that the diagonal elements of $C$ satisfy a bound of the form 
\begin{equation}\label{eqn-diagC-assumption}
\norm{\diag(C)}_\infty \leq O\bigl( \tfrac{1}{\sqrt{n}} \norm{\diag(C)}_2 \bigr).
\end{equation}
We will first discuss the situations when this assumption holds; then we will use this assumption to get a stronger error bound for $W^{(opt)}$. 

Roughly speaking, the assumption (\ref{eqn-diagC-assumption}) says that none of the diagonal elements $c_{jj}$ is too much larger than the others. This is plausible for a quantum system that consists of many qubits that are constructed in a similar way. 

In order to make this intuition more precise, we can write (\ref{eqn-diagC-assumption}) in an equivalent form:
\begin{equation}\label{eqn-diagC-assumption-2}
\max_{1\leq j\leq n}(c_{jj}^2) \leq O\bigl( \tfrac{1}{n} \sum_{j=1}^n c_{jj}^2 \bigr), 
\end{equation}
which says that the largest $c_{jj}^2$ is at most a constant factor larger than the average of all of the $c_{jj}^2$. Also, it is informative to consider how (\ref{eqn-diagC-assumption}) and (\ref{eqn-diagC-assumption-2}) compare to the (arguably more natural) assumption that 
\begin{equation}\label{eqn-diagC-assumption-3}
\max_{1\leq j\leq n} \abs{c_{jj}} \leq O\bigl( \tfrac{1}{n} \sum_{j=1}^n \abs{c_{jj}} \bigr).
\end{equation}
In fact, (\ref{eqn-diagC-assumption-3}) is actually a stronger assumption, in the sense that it implies (\ref{eqn-diagC-assumption}) and (\ref{eqn-diagC-assumption-2}), via the Cauchy-Schwartz inequality. 

The estimator $W^{(opt)}$ satisfies an error bound given by Theorem \ref{thm-ripless}, and equation (\ref{eqn-error-bound-cs-slogn}). Combining this with our assumption (\ref{eqn-diagC-assumption}), we get the following:
\begin{widetext}
\begin{align}
\norm{&W^{(opt)}-C'}_F 
\leq O(s \log^{5/2}(n) [\delta_1 \sqrt{s\log(n)} \norm{\diag(C)}_2 
 + \delta_2 \sqrt{n} \norm{\diag(C)}_2 + \delta_2 \sqrt{2s} \norm{C'}_F]).
\end{align}
\end{widetext}

Now, given any $\delta > 0$, we can ensure that 
\begin{equation}
\begin{split}
\norm{W^{(opt)}-C'}_F &\leq \tfrac{1}{\sqrt{2}} \delta \norm{\diag(C)}_2 + \tfrac{1}{\sqrt{2}} \delta \norm{C'}_F \\
&\leq \delta \norm{C}_F, 
\end{split}
\end{equation}
by setting 
\begin{equation}
\delta_1 = O\left( \frac{\delta}{s^{3/2} \log^3(n)} \right),
\end{equation}
and 
\begin{equation}
\delta_2 = O\left( \frac{\delta}{s \log^{5/2}(n) \sqrt{\max(n,s)}} \right). 
\end{equation}
This satisfies the requirement (\ref{eqn-delta-normCF}).

In addition, one can easily check that the estimate $\mathbf{g}$ for $\diag(C)$ satisfies the requirement (\ref{eqn-delta-normdiagC2}).

Then the sample complexity is as follows (see the discussion following (\ref{eqn-norm-uj}) and (\ref{eqn-norm-vj})): the method performs single-qubit spectroscopy on 
\begin{equation}
O(n / \delta_1^2) = O(n s^3 \log^6(n) / \delta^2)
\end{equation}
samples, and multi-qubit spectroscopy on 
\begin{equation}
O(m / \delta_2^2) \leq O(s^3 \max(n,s) \log^6(n) / \delta^2)
\end{equation}
samples. Hence the total sample complexity is at most 
\begin{equation}
O(s^3 \max(n,s) \log^6(n) / \delta^2).
\end{equation}
This is less than the sample complexity of the naive method, provided the off-diagonal part of the correlation matrix is sufficiently sparse, i.e., when $s \leq O(n^{2/3} / \log^2 n)$.


	\section{Choosing the Evolution Time $t$}
	\label{sec-evolution-time}
	
	We now discuss a technical detail involving the physical implementation of our measurements of the correlation matrix $C$. As described in Section \ref{sec-estimating-decay-rates}, this requires estimating certain decay rates $\Gamma_\mathbf{ab}$. To do this, we prepare quantum states $\ket{\psi_\mathbf{ab}}$, allow them to evolve for some time $t$, and then measure them in an appropriate basis. This works well when $t$ is chosen appropriately, so that $\Gamma_\mathbf{ab} t \sim 1$.
	
	In this section, we sketch one way of choosing the evolution time $t$ such that $\frac{1}{2} \leq \Gamma_\mathbf{ab} t \leq 2$. The basic idea is to start with some initial guess for $t$ (call it $\tau_0$), then perform ``binary search,'' i.e., run a sequence of experiments, where one observes the dephasing of the state $\ket{\psi_\mathbf{ab}}$ for some time $t$, and after each experiment, one adjusts the time $t$ adaptively, multiplying and dividing by factors of 2, in order to get the ``right amount'' of dephasing. We claim that this requires $\sim \abs{\log(\Gamma_\mathbf{ab} \tau_0)}$ experiments. 
	
	More precisely, we consider the following procedure:
	\begin{enumerate}
		\item Fix some $\tau_0 > 0$; this is our initial guess for the evolution time $t$. 
		\item For $r = 1,2,\ldots,N_\text{trials}$, do the following: (we set $N_\text{trials}$ according to equation (\ref{eqn-Ntrials}) below)
		\begin{enumerate}
			\item Set $s_0 = 0$ and $t_0 = 2^{s_0}\tau_0$. (This is our initial guess for $t$.)
			\item For $j = 0,1,2,\ldots,N_\text{steps}-1$, do the following: (we set $N_\text{steps}$ according to equation (\ref{eqn-Nsteps}) below)
			\begin{enumerate}
				\item Prepare the state $\ket{\psi_\mathbf{ab}} = \frac{1}{\sqrt{2}} (\ket{\mathbf{a}} + \ket{\mathbf{b}})$, allow the state to dephase for time $t_j$, then measure in the basis $\frac{1}{\sqrt{2}} (\ket{\mathbf{a}} \pm \ket{\mathbf{b}})$
				\item If the measurement returns $\frac{1}{\sqrt{2}} (\ket{\mathbf{a}} + \ket{\mathbf{b}})$, then set 
				\begin{equation}\label{eqn-sprob}
					s_{j+1} = \begin{cases}
						s_j + 1 & \text{with probability } \tfrac{e-1}{e+1}, \\
						s_j & \text{otherwise.}
					\end{cases}
				\end{equation}
				If the measurement returns $\frac{1}{\sqrt{2}} (\ket{\mathbf{a}} - \ket{\mathbf{b}})$, then set $s_{j+1} = s_j - 1$.
				\item Set $t_{j+1} = 2^{s_{j+1}} \tau_0$. (This is our next guess for $t$.)
			\end{enumerate}
			\item Define $\xi_r$ to be the value of $s_j$ from the last iteration of the loop, that is, $\xi_r = s_{N_\text{steps}}$.
		\end{enumerate}
		\item Compute the average $\xi = \frac{1}{N_\text{trials}} \sum_{r=1}^{N_\text{trials}} \xi_r$. Return $\hat{t} = 2^\xi \tau_0$. (This is our estimate for $t$.)
	\end{enumerate}
	
\begin{figure}
		\centering
		\includegraphics[width=1\columnwidth]{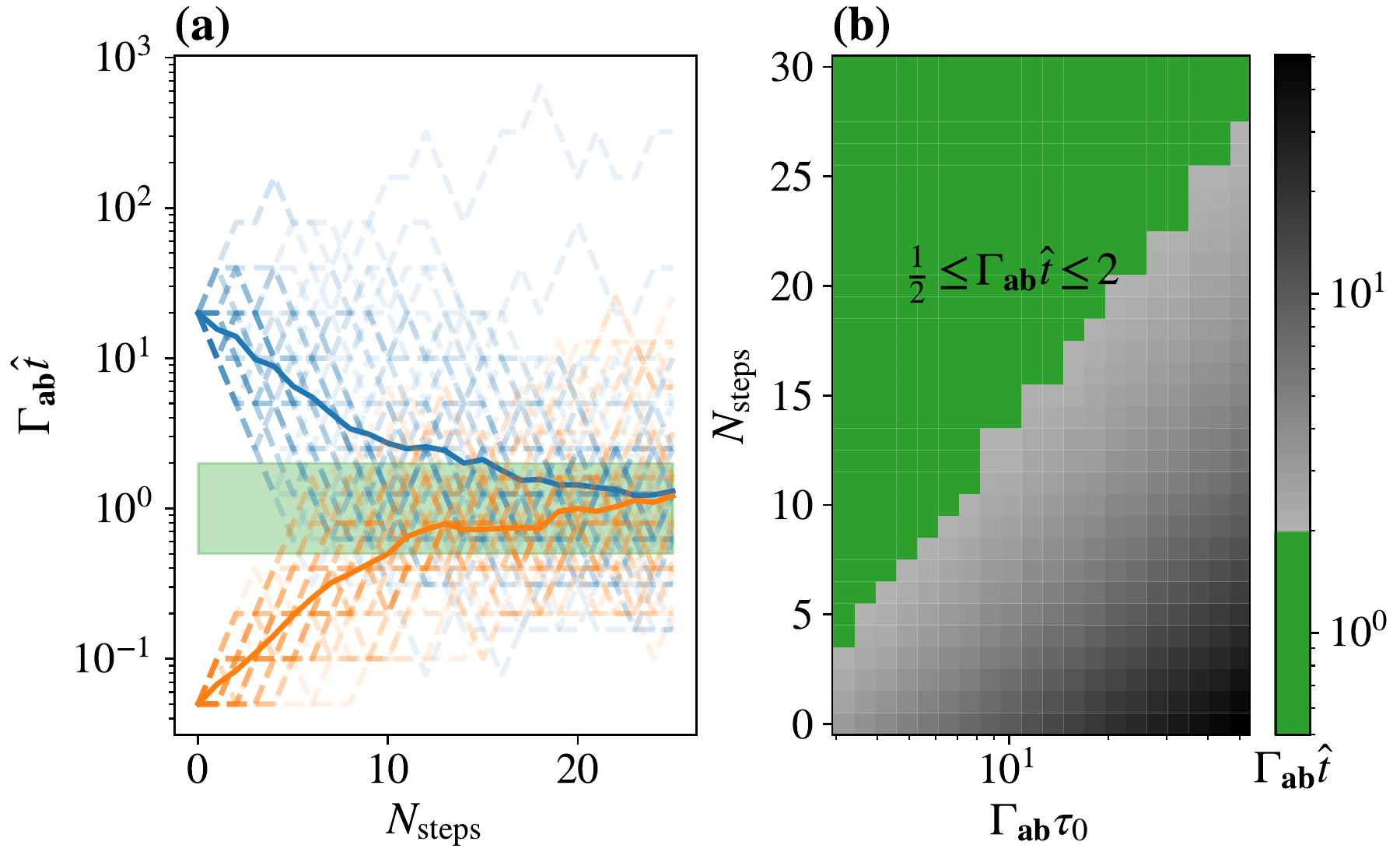}
		\caption{Choosing the evolution time $t$. (a) Trajectories of the random walk. We start with an initial  guess $\tau_0$, that can either be shorter (orange) or longer (blue) than the optimal time. The evolution time is then stochastically halved or doubled over $N_\text{steps}$ iterations, according to the algorithm outlined in Section~\ref{sec-evolution-time}. This procedure is repeated (dashed curves) and the outcome is averaged (solid curves) to obtain an estimate $\hat{t}$.  The region in which $\frac{1}{2} < \Gamma_\mathbf{ab} \hat{t} < 2$ is shaded in green. (b) The accuracy of the final estimate $\hat{t}$, as a function of the number of steps $N_\text{steps}$ and the starting point $\tau_0$. The green shading shows the region where $\hat{t}$ satisfies the bound $\frac{1}{2} < \Gamma_\mathbf{ab} \hat{t} < 2$. We see that, at the boundary of the green region, $N_\text{steps}$ scales logarithmically with $\abs{\Gamma_\mathbf{ab} \tau_0}$, as predicted by Eq.~\eqref{eqn-Nsteps}.}
		\label{fig:chooset}
	\end{figure}

	This procedure can be described in an intuitive way as follows. The inner loop of this procedure (the loop indexed by $j$) can be viewed as a kind of stochastic gradient descent, which behaves like a random walk on real numbers of the form $t = 2^s \tau_0$ ($s \in \mathbb{Z}$) (see the dashed curves in Fig.~\ref{fig:chooset}(a)). 
	
	We will show that this random walk has a single basin of attraction at a point $t^* = 2^{s^*} \tau_0$ that satisfies $\Gamma_\mathbf{ab} t^* \approx 1$, that is, $s^* \approx -\log_2(\Gamma_\mathbf{ab} \tau_0)$. We claim that the random walk converges to this point: with high probability, the sequence $s_0, s_1, s_2, \ldots$ will reach the point $s^*$ after $O(|s^*|) = O(|\log(\Gamma_\mathbf{ab} \tau_0)|)$ steps; after that point, the sequence will remain concentrated around $s^*$, with exponentially-decaying tail probabilities (see Fig.~\ref{fig:chooset}(b)). 
	This claim is made precise in Section \ref{sec-bounding-random-walk}, equations (\ref{eqn-Nsteps}) and (\ref{eqn-walk-tail-prob}).
	
	Finally, the outer loop of this procedure (the loop indexed by $r$) computes an estimate $\xi$ of $s^*$, by averaging over several independent trials (see the solid curves in Fig.~\ref{fig:chooset}(a)). This then yields an estimate $\hat{t}$ of $t^*$.
	The required number of trials, and the accuracy of the resulting estimate $\hat{t}$, are analyzed in Section \ref{sec-error-bound-t-hat}, equations (\ref{eqn-Ntrials}) and (\ref{eqn-t-hat}).
		
	\subsection{Convergence of the Random Walk}
	\label{sec-bounding-random-walk}

	We now give a rigorous analysis of our procedure for choosing $t$. We begin by describing the random walk in more detail. We will work with the variables $s_j$, which are related to the $t_j$ via the identity $t_j = 2^{s_j} \tau_0$.
	It is easy to see that $s_0 = 0$, $s_{j+1} \in \set{s_j, s_j-1, s_j+1}$, and 
	\begin{equation}
		\begin{split}
			\EE[s_{j+1} \;|\; s_j] &= s_j + \tfrac{1}{2}(1+e^{-\Gamma_\mathbf{ab} t_j}) \tfrac{e-1}{e+1} - \tfrac{1}{2}(1-e^{-\Gamma_\mathbf{ab} t_j}) \\
			&= s_j + \tfrac{1}{e+1} (e^{1-\Gamma_\mathbf{ab} t_j} - 1).
		\end{split}
	\end{equation}
	Hence the sequence $s_0, s_1, s_2, \ldots$ can be viewed as the trajectory of a random walk on a 1-dimensional chain, beginning at $s_0$, with transition probabilities that vary along the chain. The expected behavior of the random walk can be bounded as follows:
	\begin{equation}\label{eqn-mu-bound}
		\EE[s_{j+1} \;|\; s_j] \begin{cases}
			\geq s_j + \mu &\text{when } \Gamma_\mathbf{ab} t_j \leq \tfrac{1}{\sqrt{2}}, \\
			\leq s_j - \mu &\text{when } \Gamma_\mathbf{ab} t_j \geq \sqrt{2}, \\
			\approx s_j + \frac{e-1}{e+1} &\text{when } \Gamma_\mathbf{ab} t_j \ll 1, \\
			\approx s_j - \frac{1}{e+1} &\text{when } \Gamma_\mathbf{ab} t_j \gg 1, 
		\end{cases}
	\end{equation}
	where $\mu$ is a numerical constant, 
	\begin{equation}
	\mu = 0.09. 
	\end{equation}
	Hence the random walk will tend to converge towards some integer $s^*$ such that $t^* = 2^{s^*} \tau_0$ satisfies $\frac{1}{\sqrt{2}} \leq \Gamma_\mathbf{ab} t^* \leq \sqrt{2}$.

	Note that the expected position of the random walk moves towards $s^*$ at a rate that is lower-bounded by $\mu$. So, in order to go from $s_0 = 0$ to $s^* \approx -\log_2(\Gamma_\mathbf{ab} \tau_0)$, we expect that the random walk will take roughly $\tfrac{1}{\mu} |s^*| = \tfrac{1}{\mu} |\log_2(\Gamma_\mathbf{ab} \tau_0)|$ steps. It is easy to see that the stationary distribution of the random walk is centered around $s^*$, with exponentially decaying tails; hence, once the walk reaches $s^*$, it will remain concentrated around that point.

	We now explain how to set the parameter $N_\text{steps}$ so that, with high probability, after $N_\text{steps}$ steps, the random walk will converge. Say we are given an upper bound $h$ on the magnitude of $s^*$, i.e., we are promised that $|s^*| \leq h$, or equivalently, we are promised that $2^{-h} \leq \Gamma_\mathbf{ab} \tau_0 \leq 2^h$. Then we will run the random walk for a number of steps 
	\begin{equation}
		\label{eqn-Nsteps}
		N_\text{steps} = \tfrac{h}{\mu} + \eta, 
	\end{equation}
	where $\eta \geq 0$. Here, $\frac{h}{\mu}$ is (an upper bound on) the expected number of steps needed to reach $s^*$. We take an additional $\eta$ steps to ensure that the walk does indeed reach $s^*$ with high probability (we will show that the probability of failure decreases exponentially with $\eta$).
	
	We claim that, after $N_\text{steps}$ steps, the final position of the walk is close to $s^*$, with exponentially decaying tail probabilities: for any $\ell \geq 1$, 
	\begin{equation}\label{eqn-walk-tail-prob}
	\begin{split}
	\Pr[&|s_{N_\text{steps}} - s^*| \geq \ell \;|\; s_0 = 0] \\
	&\leq \tfrac{16}{\mu^2} \exp(-\tfrac{\mu(\mu+1)}{8}\ell + \tfrac{\mu^2}{4}) \\
	&+ 2\exp(-\tfrac{\mu}{16} \min\set{\tfrac{\mu\eta}{h},1} (\ell+\mu\eta)). 
	\end{split}
	\end{equation}

	In particular, when $\eta \geq \frac{h}{\mu}$, this bound can be slightly simplified:

	\begin{equation}
		\begin{split}
		\Pr[&|s_{N_\text{steps}} - s^*| \geq \ell \;|\; s_0 = 0] \\
		&\leq \tfrac{16}{\mu^2} \exp(-\tfrac{\mu(\mu+1)}{8}\ell + \tfrac{\mu^2}{4}) 
		+ 2\exp(-\tfrac{\mu}{16} (\ell+\mu\eta)).
		\end{split}
	\end{equation}

	The bound \eqref{eqn-walk-tail-prob} is proved in Appendix~\ref{sec-proof-tbound}, using martingale techniques. 
	
	
	\subsection{Error Bound for $\hat{t}$}
	\label{sec-error-bound-t-hat}
	
	We now explain how to set the parameter $N_\text{trials}$, and derive an error bound on our estimator $\hat{t}$. The bound (\ref{eqn-walk-tail-prob}) implies that the random variables $\xi_r = s_{N_\text{steps}}$ are sub-exponential (see \cite{vershynin2010introduction}, section 5.2.4), and their sub-exponential norms are bounded by some constant $\norm{\xi_r}_{\psi_1} \leq K$. Hence their average $\xi$ satisfies a Bernstein-type concentration inequality (see \cite{vershynin2010introduction}, corollary 5.17): for every $\delta \geq 0$, 
	\begin{equation}
		\Pr[|\xi - s^*| \geq \delta] \leq 2\exp(-c \min\set{\tfrac{\delta^2}{K^2}, \tfrac{\delta}{K}} N_\text{trials}),
	\end{equation}
	where $c > 0$ is a universal constant. 
	
	Now, for any $\epsilon > 0$, we set 
	\begin{equation}
		\label{eqn-Ntrials}
		N_\text{trials} = \tfrac{1}{c} \max\set{\tfrac{K}{\delta}, \tfrac{K^2}{\delta^2}} \log(\tfrac{2}{\epsilon}). 
	\end{equation}
	Then we have the following error bound on our estimator $\hat{t} = 2^\xi \tau_0$: with probability $\geq 1-\epsilon$, we have $|\xi - s^*| < \delta$, which implies that 
	\begin{equation}
		\label{eqn-t-hat}
		2^{-0.5-\delta} < \Gamma_\mathbf{ab} \hat{t} < 2^{0.5+\delta}.
	\end{equation}
	Assuming $\delta < 1/2$, this implies that $\tfrac{1}{2} < \Gamma_\mathbf{ab} \hat{t} < 2$, as desired.


	\section{Effect of SPAM Errors\label{sec-spam-errors}}
	When the measurement protocols described in this paper are implemented in an experiment, errors may occur during state preparation and measurement (SPAM errors). We investigate the effect of these errors on estimating the decay rates $\Gamma_\mathbf{ab}$. Let $\rho_0$ and $E_0$ denote the noiseless initial state and observable of interest, respectively. We have 
	\begin{equation}
		\rho_0 = E_0 = \tfrac{1}{2}(\ket{\mathbf{a}}\bra{\mathbf{a}}+\ket{\mathbf{b}}\bra{\mathbf{b}}+\ket{\mathbf{b}}\bra{\mathbf{a}}+\ket{\mathbf{a}}\bra{\mathbf{b}}).
	\end{equation}
	We consider error channels $\mathcal{E}_s$ and $\mathcal{E}_m$ that act on state preparation and measurement operations as
	\begin{align}
		\label{eqn-pertrho}
		\tilde{\rho} &= \mathcal{E}_s(\rho_0)= \rho_{0} +\delta\rho \\
		\label{eqn-pertmeas}
		\tilde{E} &= \mathcal{E}_m(E_0)= E_{0} + \delta E,
	\end{align}
	where $\norm{\delta\rho}_\text{tr}\leq\epsilon_s$ and $\norm{\delta E}\leq\epsilon_m$, and $\epsilon_s$ and $\epsilon_m$ are small parameters.   
	The outcome of the protocol is now given by 
	\begin{equation}
		\tilde{P}_{\mathbf{ab}}(t) = {\rm{Tr}}[\tilde{E} \mathcal{E}_t (\tilde{\rho})].
	\end{equation}
	where $\mathcal{E}_t = \exp(\mathcal{L}t)$ is the evolution under the correlated dephasing noise \eqref{eqn-master}. 
	
	We show that our protocol is robust against these kinds of errors, and for short times $t$ the decay of $\tilde{P}_{\mathbf{ab}}(t)$ is still dominated by $\Gamma_\mathbf{ab}$. Using Eqs.~\eqref{eqn-pertrho} and \eqref{eqn-pertmeas} we find that 
	\begin{align}
		\Tr[\tilde{E}\mathcal{E}_t (\tilde{\rho})] &= \Tr[{E}_0\mathcal{E}_t ({\rho}_0)] \\
		\nonumber &+\Tr[{E}_0\mathcal{E}_t (\delta\rho)]+ \Tr[\delta E \mathcal{E}_t ({\rho}_0)]\\
		\nonumber &+ \Tr[\delta E\mathcal{E}_t (\delta\rho)]
	\end{align}
	The first term is the outcome without errors, and we have 
	\begin{equation}
		\Tr[{E}_0\mathcal{E}_t ({\rho}_0)]=\frac{1+e^{-t\Gamma_\mathbf{ab}}}{2}.
	\end{equation}
	We can find the effect of errors on the second and third terms, by considering the effect of $\mathcal{E}_t$ on $\rho_1$ and $E_1$. Specifically, we find 
	\begin{align}
		\Tr[{E}_0\mathcal{E}_t (\delta\rho)]&= \eta_{s} + \zeta_s e^{-\Gamma_\mathbf{ab}t}, \label{eqn-spam-1}\\
		\Tr[\delta E\mathcal{E}_t ({\rho}_0)]&= \eta_{m} + \zeta_m e^{-\Gamma_\mathbf{ab}}\label{eqn-spam-2},
	\end{align}
	where  $\eta_{s,m}$ and $\zeta_{s,m}$ are constants that are determined by $\delta\rho$ and $\delta E$, for $s$ and $m$, respectively. Therefore, these terms decay with the same rate as the first case and do not affect the exponential decay. However, the last term can, in principle, contain different decay rates and can cause deviation from a single exponential decay. We can bound the rate at which $R(t)=\Tr[\delta E \mathcal{E}_t(\delta \rho)]$ grows:
	\begin{align}\label{eqn-spam-growth}
		\abs{\dot{R}(t)} &= \abs{\tfrac{\partial}{\partial t} \Tr[\delta E \mathcal{E}_t(\delta \rho)]} \\
		&= \abs{\Tr[\delta E \mathcal{L}(\delta \rho)]} \\
		&\leq 2 \epsilon_m \epsilon_s (n+s),  \label{eqn-rdot}
	\end{align}
	see Appendix~\ref{sec-proof-singval-bound} for the proof.	Therefore, we find 
	\begin{equation}
		\label{eqn-gateerrorbound}
		\tilde{P}_{\mathbf{ab}}(t) =\frac{1}{2}(1 + \eta_s +\eta_m + (1+\zeta_s +\zeta_m) e^{-\Gamma_\mathbf{ab} t}) + R(t) ,
	\end{equation} 
	The deviations from a single exponential decay are attributed to $R(t)$. Using Eqs.~\eqref{eqn-rdot} and \eqref{eqn-gateerrorbound} we can see that the decay rate of $\tilde{P}_{\mathbf{ab}}(t)$ is dominated by $\Gamma_\mathbf{ab}$ for evolution times $t \lesssim 1/ (2\epsilon_s \epsilon_m (n+s))^{-1}$. 
	\begin{figure}
		\centering
		\includegraphics[width=1\columnwidth]{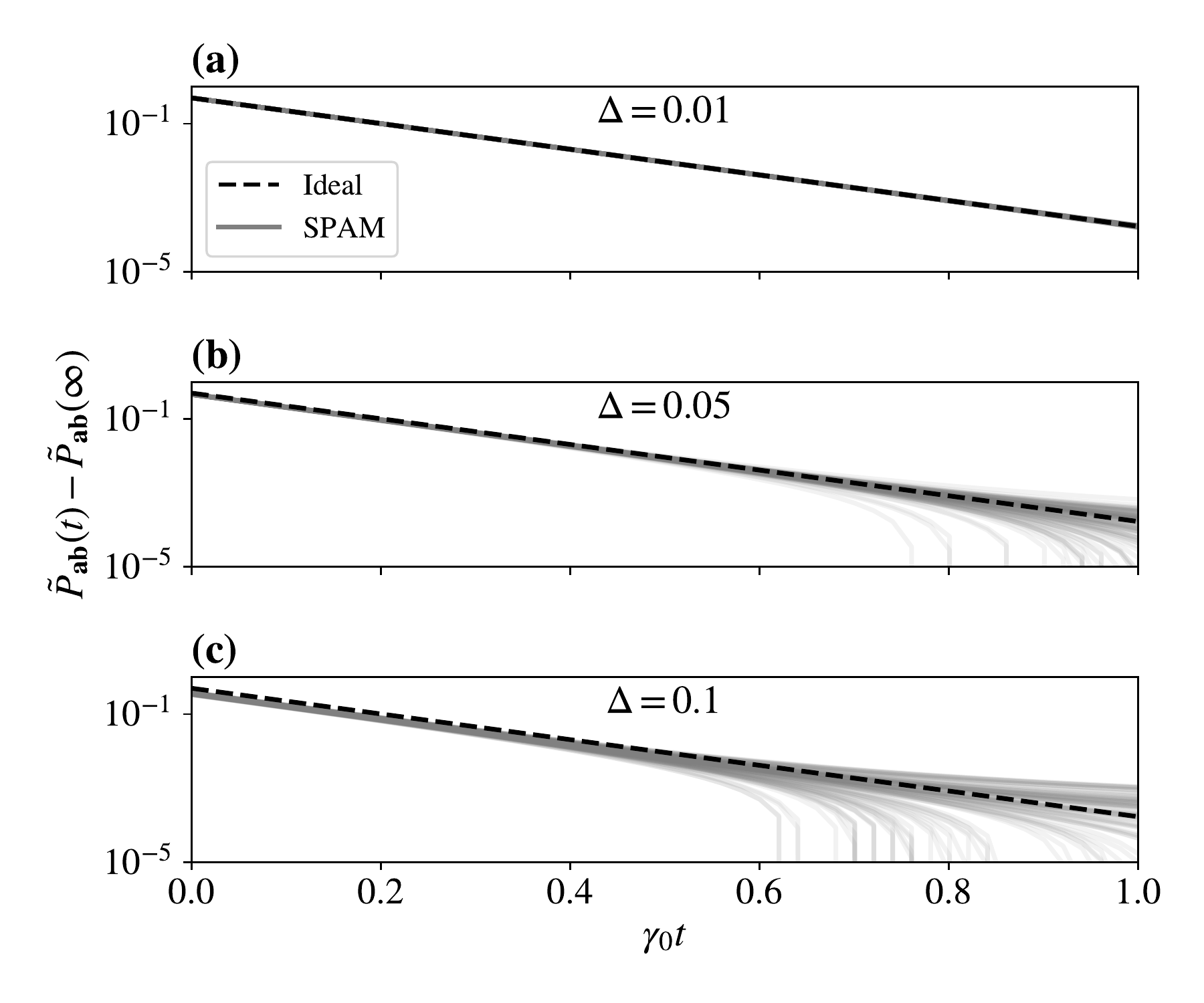}
		\caption{Exponential decay of $\tilde{P}_{\mathbf{ab}}(t)$ when there are SPAM errors. We simulate the decay of a 3-qubit GHZ state. We choose a correlation matrix $C$ with uniform single-qubit dephasing rate ($c_{ii}=\gamma_0$) and nearest-neighbor correlations $c_{i,i\pm1}=\gamma_0/4$. The panels show the decay of $\tilde{P}_{\mathbf{ab}}(t)$ under different noise strengths: (a) $\Delta=0.01$, (b) $\Delta=0.05$, (c) $\Delta = 0.1$. (See Appendix~\ref{app:numdetails} for the precise definition of $\Delta$.) The dashed lines show the decay with no SPAM errors, and solid lines show the decay with SPAM errors from randomly-sampled error channels. The solid lines resemble the dashed lines for short evolution times $t$. }
		\label{fig:gateerrors}
	\end{figure}
	
	We also use numerical simulations to investigate the effect of SPAM errors on estimates of the decay rate $\Gamma_\mathbf{ab}$. We simulate SPAM errors by applying random error channels $\mathcal{E}_s$ and $\mathcal{E}_m$, whose strengths are controlled by a parameter $\Delta$ (see Appendix~\ref{app:numdetails} for details). We then compare the decay of $\tilde{P}_{\mathbf{ab}}(t)$ with and without SPAM errors, for different values of $\Delta$.  We observe that, for short times $t$, the decay with SPAM errors matches the decay without SPAM errors, i.e., the decay rate is dominated by $\Gamma_\mathbf{ab}$, see Fig.~\ref{fig:gateerrors}. This is consistent with our theoretical analysis.

	\section{Generalizations \label{sec-complex-c}}
	
	We now sketch how our method for learning sparse correlated dephasing noise can be extended to the most general case of the master equation (\ref{eqn-master}), where the matrix $C$ is complex, and there is an additional Hamiltonian term $H_s$.
	
	\subsection{Complex Decay Rates}
	
	The complete dynamics imposed by the environment on the system can have a coherent evolution in addition to the decay. The evolution of the system $\frac{d\rho}{dt} = \mathcal{L}(\rho)$  is in general given by the Lindblad generator
	\begin{equation}\label{eqn-fulllindblad}
		\mathcal{L}(\rho)= -i[\rho,H_s] + \sum_{l,m} {c}_{lm}\left(Z_l \rho Z_m -\frac{1}{2} \{Z_l Z_m,\rho\}\right),
	\end{equation}
	where 
	\begin{equation}
	H_s=\sum_{l,m} {r}_{lm} Z_l Z_m 
	\end{equation}
	is sometimes called the (generalized) Lamb shift Hamiltonian, and $C = (c_{lm})$ is now a complex matrix~\cite{breuer2002theory} (see Appendix \ref{sec:physderiv}).  We can decompose $C$ as 
	\begin{equation}
	C = V+iT, 
	\end{equation}
	where the real and imaginary parts are separated into $V=(v_{lm})$, a real symmetric matrix, and $T=(t_{lm})$, a real skew-symmetric matrix, respectively. Moreover, the Lamb shift Hamiltonian can be encoded in the symmetric matrix $R = (r_{lm})$. 
	
	We now show how the operator $\mathcal{L}$ acts on the off-diagonal matrix elements $\ket{\mathbf{a}}\bra{\mathbf{b}}$. This involves ``decay rates'' that depend on $R$, $V$ and $T$ in a simple way, although these ``decay rates'' now complex rather than real. Specifically,  
	\begin{equation}
	\mathcal{L}(\ket{\mathbf{a}}\bra{\mathbf{b}}) = (-\Gamma_{\mathbf{ab}} + i \Omega_{\mathbf{ab}})\ket{\mathbf{a}}\bra{\mathbf{b}}, 
	\end{equation}
	where $\Gamma_{\mathbf{ab}}$ and $\Omega_{\mathbf{ab}}$ are real numbers that capture the decay and oscillations of the matrix element, respectively. We follow the convention defined in Section~\ref{sec:prelims} for states $\ket{\mathbf{a}}$ and $\ket{\mathbf{b}}$. We remind the reader that $Z_i \ket{\mathbf{a}} = \alpha_i \ket{\mathbf{a}}$, and similarly for $\ket{\mathbf{b}}$ and $\beta_i$. Therefore, separating the real and imaginary part of $\mathcal{L}(\ket{\mathbf{a}}\bra{\mathbf{b}})$ we find that 
	\begin{align}\label{eqn-generalized-Gamma-ab}
		\Gamma_{\mathbf{ab}} &= -\sum_{l,m} v_{lm}( \alpha_l \beta_m-\tfrac{1}{2}\alpha_l \alpha_m- \tfrac{1}{2}\beta_l \beta_m ) \nonumber\\
		&= \tfrac{1}{2} (\boldsymbol{\alpha}-\boldsymbol{\beta})^T V (\boldsymbol{\alpha}-\boldsymbol{\beta}), 
	\end{align}
	where we used the fact that $V$ is symmetric. Similarly, we find
	\begin{align}\label{eqn-generalized-Omega-ab}
	\Omega_{\mathbf{ab}} = &-\sum_{l,m} r_{lm} \left(\alpha_l \alpha_m -\beta_l \beta_m\right) \nonumber\\
	&+ \sum_{l,m} v_{lm}\left( \alpha_l \beta_m-\tfrac{1}{2}\alpha_l \alpha_m- \tfrac{1}{2}\beta_l \beta_m \right) \nonumber\\
	&= - (\boldsymbol{\alpha}^T R \boldsymbol{\alpha}-\boldsymbol{\beta}^T R \boldsymbol{\beta})+\tfrac{1}{2} (\boldsymbol{\alpha}^T T \boldsymbol{\beta}-\boldsymbol{\beta}^T T \boldsymbol{\alpha}),
	\end{align}
	where we used the fact that $R$ and $T$, are symmetric and skew-symmetric matrices, respectively. (See Appendix \ref{sec-derivation-dynamics} for details.)
	
	Therefore, the coherences, that is, the $\ket{\mathbf{a}}\bra{\mathbf{b}}$ elements of the density matrix $\rho(t)$, exhibit both oscillations (at a frequency $\Omega_\mathbf{ab}$) and exponential decay (at a rate $\Gamma_\mathbf{ab}$) as a function of $t$. Note that it is possible to measure both $\Gamma_{\mathbf{ab}}$ and $\Omega_{\mathbf{ab}}$, by estimating the matrix elements of $\rho(t)$ at different times $t$, using the same types of Ramsey experiments outlined in Section~\ref{sec-ramsey}. (In particular, one can extract these parameters using standard techniques for analyzing spectral lines in atomic physics. Here, the squared magnitude of the Fourier transform of the measurement time series is a Lorentzian function, the center of the Lorentzian peak gives the oscillation frequency, and the width of the peak gives the decay rate~\cite{Thorne1988}.) 

	Given estimates of $\Gamma_{\mathbf{ab}}$ and $\Omega_{\mathbf{ab}}$, we can extend our compressed sensing method to recover both the correlation matrix $C$ and the Hamiltonian $H_s$. As before, we use $m \sim s\log n$ or $m \sim s\log^4 n$ measurement settings; for each measurement setting, we choose $\mathbf{a}$ and $\mathbf{b}$ uniformly at random in $\set{0,1}^n$. From estimates of $\Gamma_\mathbf{ab}$, we can recover $V$, the real part of $C$, exactly as before (this follows from equation (\ref{eqn-generalized-Gamma-ab})). In a similar manner, given estimates of $\Omega_\mathbf{ab}$, we want to recover $T$, the imaginary part of $C$, as well as $R$, the matrix that encodes the Hamiltonian $H_s$. This is possible, because equation (\ref{eqn-generalized-Omega-ab}) represents a measurement of $R$ and $T$ that has the needed isotropy and incoherence properties, as we will show below.

	\subsection{Isotropy and Incoherence of the Measurements $\Omega_\mathbf{ab}$\label{sec:measpropcomplex}}
	We consider $\Omega_\mathbf{ab}$, viewed as a measurement of $R$ and $T$ in equation (\ref{eqn-generalized-Omega-ab}), with $\mathbf{a}$ and $\mathbf{b}$ chosen uniformly at random in $\set{0,1}^n$. We show that this random measurement has the same isotropy and incoherence properties as before, and hence our compressed sensing method will still succeed using these measurements. The incoherence property is easy to see, but some work is required to show the isotropy property.
	
	The measurement $\Omega_\mathbf{ab}$ acts on $T$ and $R$ as 
	\begin{equation}\label{eqn:imagmeas}
		\Omega_\mathbf{ab} =  \begin{bmatrix}
			\boldsymbol{\alpha}^T & \boldsymbol{\beta}^T
		\end{bmatrix} 
		\begin{bmatrix}
			-R & \frac{1}{2}T \\
			-\frac{1}{2}T & R
		\end{bmatrix} 
		\begin{bmatrix}
			\boldsymbol{\alpha} \\ \boldsymbol{\beta}
		\end{bmatrix} .
	\end{equation}
	Similar to the analysis in Section~\ref{sec-isotropy-incoherence}, it is advantageous to consider the effect of measurements if we enforce the symmetries of $R$ and $T$. Note that $T$ and $R$ are real skew-symmetric and symmetric matrices, respectively. Moreveover, they are both traceless. Therefore, we have 
	\begin{equation}
		\Omega_\mathbf{ab} =  \sum_{i<j} (\alpha_i \beta_j - \beta_i \alpha_j) T_{ij} + 2 \sum_{i<j} (\alpha_i \alpha_j - \beta_i \beta_j) R_{ij} 
	\end{equation}
	
	As before, let $\uvec$ be the linear operator that returns the upper-triangular part of a matrix (not including the diagonal elements), that is,  
	\begin{equation}
		\uvec:\: R \mapsto (R_{ij})_{i<j}.
	\end{equation}
	Then $\Omega_\mathbf{ab}$ can be expressed as 
	\begin{equation}
	\Omega_\mathbf{ab} = 
	\mathbf{q} \begin{bmatrix}
		{\rm{uvec}}(T) \\ 
		2\, {\rm{uvec}}(R)
	\end{bmatrix}, 
	\end{equation}
	similar to Eq.~(\ref{eqn-phi-and-Q}), where $\mathbf{q}$ is a row vector of the form 
	\begin{equation}\label{eqn-def-qcomplex}
		\mathbf{q} =  
		\begin{bmatrix}
			{\rm{uvec}}(\boldsymbol{\alpha} \boldsymbol{\beta}^T-\boldsymbol{\beta} \boldsymbol{\alpha}^T), & 
			{\rm{uvec}}(\boldsymbol{\alpha} \boldsymbol{\alpha}^T-\boldsymbol{\beta}\boldsymbol{\beta}^T )
		\end{bmatrix} .
	\end{equation}
	
	Note that as described in Section~\ref{sec:randommeas}, $\boldsymbol{\alpha}$ and $\boldsymbol{\beta}$ are chosen uniformly and independently at random from $\{1,-1\}^n$. It is then straightforward to see that  $\mathbf{q}$ in this case satisfies the incoherence  property \eqref{eqn-incoherence} with $\mu = 1$. Furthermore, one can check that $\mathbf{q}$ is centered and isotropic (up to a normalization factor of $\sqrt{2}$), since: 
	\begin{align}\label{eqn-covariance-complex}
		\begin{cases}
			\EE[\alpha_i \alpha_j] = \EE[\alpha_i \beta_j] = \EE[\beta_i \beta_j] = 0 \\
			\EE[(\alpha_i \beta_j - \beta_i \alpha_j)(\alpha_k \beta_l - \beta_k \alpha_l)] = 2\delta_{ik}\delta_{jl} \\
			\EE[(\alpha_i \alpha_j-\beta_i \beta_j)(\alpha_k \alpha_l-\beta_k \beta_l)] = 2\delta_{ik}\delta_{jl}\\
			\EE[(\alpha_i \beta_j - \beta_i \alpha_j)(\alpha_k \alpha_l - \beta_k \beta_l)] = 0
		\end{cases},
	\end{align}
	where it is assumed that $i<j$ and $k<l$ in all cases. The second and the third lines in the above equation capture the correlations in the measurements of $T$ and $R$, respectively, and the last line captures the cross-correlations between the two measurements (see Appendix~\ref{sec-isotropy-complex} for more details). 
	
	
	\section{Outlook}
	In this paper we have demonstrated a new paradigm for characterizing spatially-correlated noise in many-body quantum systems, where one assumes that the correlations are sparse, and one uses compressed sensing to characterize them efficiently. Our method applies to correlated dephasing noise, and is capable of detecting long-range correlations, with provable recovery guarantees, even in the presence of SPAM errors. 
	
	There are several possible generalizations of our method. First, we expect that our method can be applied to a larger class of Markovian noise processes, whose Lindblad operators are Hermitian and commuting. It may also be possible to combine our methods with multi-qubit noise spectroscopy \cite{paz2017multiqubit}, in order to estimate spatial correlations in non-Markovian noise processes. Finally, it is an interesting question whether our methods can be combined with the more combinatorial techniques of \cite{harper2020fast} for estimating sparse Pauli channels.
	
	Our work touches on several other broad questions. First, we expect our protocol to be useful for characterizing dephasing in the presence of weak relaxation processes, i.e., when the relaxation time is much longer than the dephasing time. However, the harder case of concurrent relaxation and dephasing remains open, and is a subject of future research. 
	
	Also, while we use entangled states to probe the correlations in the noise process, it is intriguing to ask whether it is possible to achieve a similar scaling using non-entangled states. This motivates the more general question of whether entanglement can help noise spectroscopy in the same way that it is helpful in metrology \cite{giovannetti2006quantum}.

\section*{Acknowledgements}
We thank Paola Cappellaro, Alexey Gorshkov, Amir Kalev, David Layden, Ingo Roth, and Lorenza Viola for helpful discussions. We thank Victor Albert, Richard Kueng and Zachary Levine for helpful comments on an earlier draft of this paper. 

This research was supported in part by the AFOSR MURI project ``Scalable Certification of Quantum Computing Devices and Networks,'' ARO-MURI, and Northrop Grumman. This work was completed while AS was at the University of Maryland, before moving to the University of Chicago.

\appendix

\section{Sparse Correlated Dephasing\label{sec:physderiv} }
\subsection{Microscopic Derivation }
We consider a model of spatially correlated noise, which is described by the following master equation:
\begin{equation}\label{eqn-master2}
	\frac{d\rho}{dt} = \mathcal{L}(\rho) =  i[\rho, H_s] + 
	\sum_{j,k=1}^n c_{jk} \left( s_k\rho s_j^\dagger - 
	\tfrac{1}{2} \lbrace s_j^\dagger s_k, \rho \rbrace \right).
\end{equation}
Here the system consists of $n$ qubits, $H_s$ is the system Hamiltonian, and $s_j$ and $s_k$ are operators that act on the $j$'th and $k$'th qubits, respectively. We consider the case of dephasing noise, where $H_s = 0$, and $s_j = Z_j$ (the Pauli $\sigma_z$ operator acting on the $j$'th qubit). The noise is then fully described by the matrix of coupling coefficients $C = (c_{jk}) \in \CC^{n\times n}$. In order for the time evolution of the system to be a completely positive map, the matrix $C$ must be positive semidefinite \cite{jeske2013derivation}.

We will be interested in the case where the matrix $C$ is \textit{sparse}, in the sense that most of the off-diagonal elements are 0. This can arise when the qubits are coupled to a bath, in such a way that most of the time, different qubits are coupled to \textit{different} degrees of freedom in the bath, but occasionally it happens that two or more qubits couple to the \textit{same} degree of freedom in the bath. 

To motivate the above intuition, we follow Ref.~\cite{reina2002decoherence} and show how the master equation (\ref{eqn-master2}) can be derived from a commonly-occurring microscopic model. Here, multiple qubits are coupled to a bosonic bath through the interaction Hamiltonian 
\begin{equation}
	H_I(t) = \sum_{l,{\mathbf{k}}} Z_l \left(g_{\mathbf{k}}^l b^\dagger_{\mathbf{k}}(t) + g^{*l}_{\mathbf{k}} b_{\mathbf{k}}(t)\right),
\end{equation}
where $b_\mathbf{k}(t)$ are bath operators. If the bath is quantum mechanical, then $b_{\mathbf{k}}(t) = e^{-i\omega_{\mathbf{k}} t} b_{\mathbf{k}}$, where $b_{\mathbf{k}}$ is a bosonic annihilation operator corresponding to the bath mode ${\mathbf{k}}$, with frequency $\omega_{\mathbf{k}}$. If the bath is classical, then $b_{\mathbf{k}}(t)$ is a stochastic variable corresponding to the fluctuations of the external magnetic field~\cite{paz2017multiqubit}. 

The time evolution operator is then given by~\cite{reina2002decoherence}
\begin{equation}
	U_I(t) = \exp[i \Phi_{\omega_{\mathbf{k}}}(t)]\exp[\sum_{l,{\mathbf{k}}} Z_l \{\xi^l_{\mathbf{k}}(t) b^\dagger_{\mathbf{k}}-\xi^{*l}_{\mathbf{k}}(t) b_{\mathbf{k}}\}]
\end{equation}
where 
\begin{align}
	\Phi_{\omega_{\mathbf{k}}}(t) &= \sum_{l,{\mathbf{k}}} |Z_l g_{\mathbf{k}}^l|^2 \frac{\omega_{\mathbf{k}} t-\sin(\omega_{\mathbf{k}} t)}{\omega_{\mathbf{k}}^2},\\
	\xi_{\mathbf{k}}^l &= g_{\mathbf{k}}^l \frac{1-e^{i\omega_{\mathbf{k}} t}}{\omega_{\mathbf{k}}}.
\end{align}
The phase term $\Phi_{\omega_{\mathbf{k}}}(t)$ only appears when the bath is quantum mechanical, and is the result of the non-commutative nature of the bath operators. This term is also known as the Lamb shift. We assume that if qubit $l$ at position $\mathbf{x}_l$ is coupled to the bath mode $k$, then the coupling is of the form $g_{\mathbf{k}}^l = g_{\mathbf{k}} e^{i {\mathbf{k}}.{\mathbf{x}_l}}$, and if the qubit is not coupled to that mode, then $g_{\mathbf{k}}^l = 0$. 

Defining $\ket{\mathbf{a}} = \ket{a_1, a_2, \ldots, a_n}$,  $\ket{\mathbf{b}} = \ket{b_1, b_2, \ldots, b_n}$, with $a_m,b_m\in\{0,1\}$, it follows that 
\begin{align}\label{eqn-dephevol}
	\bra{\mathbf{a}}\rho(t)\ket{\mathbf{b}} &= \exp[-\tfrac{1}{2}\sum_{m,l} \tilde{v}_{m,l}(t) (\alpha_m - \beta_m)(\alpha_l - \beta_l )]\\
	&\times \exp[i\sum_{m,l} \tilde{t}_{m,l}(t)(\alpha_m \beta_l)] \nonumber\\
	&\times \exp[-i\sum_{m,l} \tilde{r}_{m,l}(t) (\alpha_m \alpha_l-\beta_m \beta_l )]\nonumber\\
	&\times \bra{\mathbf{a}}\rho(0)\ket{\mathbf{b}},
\end{align}
where we defined $\alpha_m = (-1)^{a_m}$, $\beta_m=(-1)^{b_m}$, and 
\begin{align}
	\tilde{v}_{l,m}(t) &=2 \sum_{k} |g_{\mathbf{k}}|^2 \frac{1-\cos(\omega_{\mathbf{k}} t)}{\omega_{\mathbf{k}}^2} \coth(\tfrac{\beta\omega_{\mathbf{k}}}{2}) \cos(\mathbf{k}.\mathbf{x}_{l,m}) \\
	\tilde{t}_{l,m}(t) &= -2\sum_{k} |g_{\mathbf{k}}|^2 \frac{1-\cos(\omega_{\mathbf{k}} t)}{\omega_{\mathbf{k}}^2}\sin(\mathbf{k}.\mathbf{x}_{l,m})\\
	\tilde{r}_{l,m}(t) &=-\sum_{k} |g_{\mathbf{k}}|^2 \frac{\omega_{\mathbf{k}} t-\sin(\omega_{\mathbf{k}} t)}{\omega_{\mathbf{k}}^2} \cos(\mathbf{k}.\mathbf{x}_{l,m}),
\end{align}
where $\mathbf{x}_{l,m}=\mathbf{x}_l - \mathbf{x}_m$. Note that $\tilde{t}_{l,m}(v)=\tilde{v}_{m,l}(t)$, $\tilde{t}_{l,m}(t)=-\tilde{t}_{m,l}(t)$, and $\tilde{r}_{l,m}(t)=\tilde{r}_{m,l}(t)$. Therefore, it is illustrative to express Eq.~\eqref{eqn-dephevol} as~\cite{addis2013two} 
\begin{equation}\label{eqn-tdeplind}
	\begin{split}
		\frac{d\rho}{dt} &=  i[\rho,\sum_{l,m}{r}_{l,m}(t)Z_l Z_m] \\&+ \sum_{l,m} {c}_{l,m}(t)\left(Z_l \rho Z_m -\tfrac{1}{2} \{Z_l Z_m,\rho\}\right),
	\end{split}
\end{equation}
where we defined $\tilde{c}_{l,m}(t)= \tfrac{1}{2}\tilde{v}_{l,m}(t)-i \tilde{t}_{l,m}(t)$ and ${o}_{l,m}(t)= \tfrac{\partial}{\partial t}\tilde{o}_{l,m}(t)$ for $o\in\{v,t,r,c\}$. We can see that Eq.~\eqref{eqn-tdeplind}, is in the Lindblad form of Eq.~\eqref{eqn-master}, but with time dependent coefficients. 

It should be noted that $v_{l,m}(t)$, $t_{l,m}(t)$, and $r_{l,m}(t)$ vanish if qubits $m$ and $n$ do not share a bath mode, that is, if there does not exist any $\mathbf{k}$ such that $g_{\mathbf{k}}^l$ and $g_{\mathbf{k}}^m$ are both non-zero. Therefore, in cases where only a few qubit pairs have shared bath modes, these coefficients are \textit{sparse}. 

A special case that is extensively studied in the literature~\cite{reina2002decoherence,palma1996quantum,duan1998reducing} is collective decoherence, where the distance between the qubits is much smaller than the wavelength of bath modes, $\exp(i \mathbf{k}.\mathbf{x}_{l,m})\approx1$, and consequently $v_{l,m}\approx0$. In this limit, qubits that are coupled to the same modes are maximally correlated. 

Moreover, in the high-temperature limit, where the temperature of the environment is much larger than the energy of the highest-frequency bath modes, the decaying part of the evolution due to $\tilde{t}_{l,m}$ dominate the coherent evolution that comes from the Lamb shift term $\tilde{r}_{l,m}$ \cite{reina2002decoherence}. In that limit, the decoherence dynamics is solely determined  by the \textit{real} coefficients $c_{l,m}$. This is similar to the case of a classical bath (with scalar stochastic $b_{\mathbf{k}}$'s), where again the dynamics solely depend on \textit{real} coefficients $c_{l,m}$ without the Lamb shift Hamiltonian.

Finally, we note that within Born-Markov approximation~\cite{breuer2002theory}, the coefficients $t_{l,m}$, $v_{l,m}$, and $r_{l,m}$ are time-independent and we recover Eq.~\ref{eqn-master} with the Lamb shift Hamiltonian $H_s = \sum_{l,m} r_{l,m} Z_l Z_m$\cite{breuer2002theory}. The validity of this approximation, depends on the details of the bath and the time scales under which the system is being studied. 

In this work we assume that the Markovian approximation is valid and the master equation has a time-independent generator $\mathcal{L}$. In this regime, Eq.~\eqref{eqn-tdeplind} coincides with Eq.~\eqref{eqn-fulllindblad}. 

\subsection{Derivation of the Dynamics}
\label{sec-derivation-dynamics}
In this section we present the full details of the derivation of the dynamics under the master equation \eqref{eqn-fulllindblad}. Note that $\mathcal{L}$ does not map $\ket{\mathbf{a}}\bra{\mathbf{b}}$ to another $\ket{\mathbf{a}'}\bra{\mathbf{b}'}$. Additionally, $\mathcal{L}$ is linear.  Therefore we can treat the evolution independently for each element $\ket{\mathbf{a}}\bra{\mathbf{b}}$ of the density matrix $\rho$, and obtain $\rho(t) = \sum_{\mathbf{a},\mathbf{b}} \bra{\mathbf{a}}\rho(0)\ket{\mathbf{b}} \exp[(i\Omega_{\mathbf{ab}}t-\Gamma_{\mathbf{ab}})t]\ket{\mathbf{a}}\bra{\mathbf{b}}$. 

\begin{widetext}
	\begin{align}
		\mathcal{L}(\ket{\mathbf{a}}\bra{\mathbf{b}}) &=  -i[\ket{\mathbf{a}}\bra{\mathbf{b}},\sum_{l,m} {r}_{lm}Z_l Z_m] + \sum_{l,m} {c}_{lm}\left(Z_l \ket{\mathbf{a}}\bra{\mathbf{b}} Z_m -\tfrac{1}{2} \{Z_l Z_m,\ket{\mathbf{a}}\bra{\mathbf{b}}\}\right)\\
		&= \left[-i \sum_{l,m} r_{lm} \left(\alpha_l \alpha_m -\beta_l \beta_m\right) + \sum_{l,m} {c}_{lm}\left( \alpha_l \beta_m-\tfrac{1}{2}\alpha_l \alpha_m- \tfrac{1}{2}\beta_l \beta_m \right)\right]\ket{\mathbf{a}}\bra{\mathbf{b}}.
	\end{align}
\end{widetext}
We now use the symmetries of $C$ and express the second sum as
\begin{widetext}
	\begin{align}
		&\sum_{l,m} {c}_{lm}\left( \alpha_l \beta_m-\tfrac{1}{2}\alpha_l \alpha_m- \tfrac{1}{2}\beta_l \beta_m \right)= \sum_{l,m} (v_{lm}+i t_{lm})\left( \alpha_l \beta_m-\tfrac{1}{2}\alpha_l \alpha_m- \tfrac{1}{2}\beta_l \beta_m \right)\\
		&= \sum_{l,m} \tfrac{1}{2}(v_{lm}+i t_{lm})\left( \alpha_l \beta_m-\tfrac{1}{2}\alpha_l \alpha_m- \tfrac{1}{2}\beta_l \beta_m \right)+ \sum_{l,m} \tfrac{1}{2}(v_{lm}-i t_{lm})\left( \alpha_m \beta_l-\tfrac{1}{2}\alpha_m \alpha_l- \tfrac{1}{2}\beta_m \beta_l \right)\\
		&=\tfrac{v_{lm}}{2} (\alpha_l \beta_m + \alpha_m\beta_l-\alpha_l \alpha_m- \beta_l \beta_m )+\tfrac{i t_{lm}}{2} (\alpha_l \beta_m - \alpha_m\beta_l)\\
		&=-\tfrac{1}{2} v_{lm} (\alpha_l-\beta_l)(\alpha_m-\beta_m)+\tfrac{i}{2} t_{lm}(\alpha_l \beta_m - \alpha_m\beta_l)
	\end{align}
\end{widetext}
We express the above results compactly as $\mathcal{L}(\ket{\mathbf{a}}\bra{\mathbf{b}}) = (-\Gamma_{\mathbf{ab}} + i \Omega_{\mathbf{ab}})\ket{\mathbf{a}}\bra{\mathbf{b}}$, where
\begin{align}
	\Gamma_{\mathbf{ab}} &= \frac{1}{2} (\boldsymbol{\alpha}-\boldsymbol{\beta})^T V (\boldsymbol{\alpha}-\boldsymbol{\beta}), \\
	\Omega_{\mathbf{ab}} &= \frac{1}{2} (\boldsymbol{\alpha}^T T \boldsymbol{\beta}-\boldsymbol{\beta}^T T \boldsymbol{\alpha})- (\boldsymbol{\alpha}^T R \boldsymbol{\alpha}-\boldsymbol{\beta}^T R \boldsymbol{\beta}).
\end{align}


\section{Details of Numerical Simulations}\label{app:numdetails}
\subsection{Compressed Sensing of Correlated Dephasing Noise}
Here we provide some additional details about the numerical simulations in Section \ref{sec-num-examples}.

Assume that we have $n$ qubits, whose individual dephasing rates is fully characterized. However, correlations in noise cannot be observed in single qubit measurements. We assume that there are $s'$ pairs of qubits that are correlated. To generate a correlation matrix $C$ that is positive semidefinite with a controllable number of  non-zero off-diagonal elements $s=2s'$, we choose 
\begin{equation}
	\label{eqn-cmatrix}
	\begin{cases}
		c_{ii} = 2 & i=1,\dots,n\\
		c_{ij} = c_{ji} = \frac{1}{2} &1\leq i\leq s' \text{ and }  j=i+ 1.
	\end{cases}
\end{equation}
We then remove the spatial structure in the matrix by randomly permuting the rows and columns of $C$, that is  we map $c_{ij}$ to $c_{\pi(i),\pi(j)}$ with $\pi \in S_n$. This procedure ensures that the eigenvalues of $C$ are non-negative. 

We then generate noiseless measurements by choosing $m$ samples of $\mathbf{a}$ and $\mathbf{b}$. Finally, we assume that the diagonal elements of $C$ are known, and we solve the convex optimization problem \eqref{eqn-min-W}-\eqref{eqn-Phi-W} using CVXPY, a convex optimization package for Python \cite{cvxpy,cvxpy_rewriting}. The scaling of the reconstruction error $\norm{W^{(opt)} - C}_\infty$ as a function of $m$ for different values of $n$ and $s$ is illustrated in Fig.~\ref{fig:scaling} (a-b).

We also investigate the effect of noisy measurements on the recovery error. We add a random vector $\mathbf{e}$, whose entries are independent Gaussian random variables with mean 0 and standard deviation $\sigma_\epsilon$, to the measurement vector $\mathbf{h}$. Therefore we replace the constraint \eqref{eqn-Phi-W} with \eqref{eqn-Phi-W-2}. Additionally, we assume that the diagonals of $C$ are known and hence $\epsilon_1=0$ in \eqref{eqn-diag-W-2}. We choose $\epsilon_2=\sqrt{m}\sigma_\epsilon$ and  solve the convex program. The scaling of the reconstruction error $\norm{W^{(opt)} - C}_\infty$ as a function of $\sigma_\epsilon$ is shown in Fig.~\ref{fig:scaling} (c).

\subsection{State Preparation and Measurement Errors}	
We provide some additional details about the simulations in Section \ref{sec-spam-errors}, showing the effects of SPAM errors. We generate error channels $\mathcal{E}_s$ and $ \mathcal{E}_m$ that act on $n$ qubits, by adding $n$ auxiliary qubits in the state $\ket{0}^{\otimes n}$, applying unitary operations $U_s$ and $U_m$ on all $2n$ qubits, and tracing out the $n$ auxiliary qubits. We control the strength of these error channels by a parameter $\Delta$, and we set $U_s = \exp(i \Delta H)$ where $H_s$ is a Hermitian operator on $2n$ qubits. To generate different error channels at random, we choose $H_s=\tfrac{1}{2}(M_s+M_s^\dagger)$, where elements of $M_s$ are complex numbers whose real and imaginary parts are chosen uniformly at random from $[0,1]$. We construct $U_m$ in a similar way.


\section{Setting the Regularization Parameter $\lambda$ \label{sec-proof-lasso-lambda}}

Here we show that, if $\lambda$ is set according to equation (\ref{eqn-lambda}), then it satisfies equations (\ref{eqn-lasso-noise-1}) and (\ref{eqn-lasso-noise-2}). 

In order to prove these bounds, we will want to use Hoeffding-type large deviation bounds on $\mathbf{u}$ and $\mathbf{v}$. To this end, we first need to upper-bound the quantities $\max_{1\leq j\leq n} \norm{u_j}_{\psi_2}$ and $\max_{1\leq j\leq m} \norm{v_j}_{\psi_2}$. We can do this as follows. First write 
\begin{equation}\label{eqn-max-norm-uj}
	\max_{1\leq j\leq n} \norm{u_j}_{\psi_2} \leq \delta_1 \norm{\diag(C)}_\infty,
\end{equation}
and 
\begin{equation}\label{eqn-norm-diagc-infty}
	\norm{\diag(C)}_\infty \leq \norm{\mathbf{g}}_\infty + \norm{\mathbf{u}}_\infty.
\end{equation}
We can upper-bound $\norm{\mathbf{u}}_\infty$ by using the union bound over all $j \in \set{1,\ldots,n}$, and bounding each coordinate $u_j$ using the definition of the sub-Gaussian norm (see Definition 5.7 in \cite{vershynin2010introduction}), as follows. Let $c_0 > 0$ be the universal constant in Definition 5.7 in \cite{vershynin2010introduction}. For any $t \geq 0$, we have 
\begin{equation}
	\begin{split}
		&\Pr[\norm{\mathbf{u}}_\infty > (t+1) \sqrt{\ln(n)/c_0} \; \delta_1 \norm{\diag(C)}_\infty] \\
		&\leq \sum_{j=1}^n \Pr[\abs{u_j} > (t+1) \sqrt{\ln(n)/c_0} \; \delta_1 \norm{\diag(C)}_\infty] \\
		&\leq \sum_{j=1}^n \Pr[\abs{u_j} > (t+1) \sqrt{\ln(n)/c_0} \; \delta_1 c_{jj}] \\
		&\leq n\cdot e\cdot \exp(-(t+1)^2 \ln(n)) \\
		&= e\cdot n^{-t(t+2)}.
	\end{split}
\end{equation}
Now let $t = 1$. Then we have that 
\begin{equation}
\label{eqn-u-ell-infty}
	\norm{\mathbf{u}}_\infty \leq \epsilon''_1 \norm{\diag(C)}_\infty, \quad \epsilon''_1 := 2\sqrt{\ln(n)/c_0} \; \delta_1,
\end{equation}
with failure probability at most $e/n^3$. Combining this with (\ref{eqn-norm-diagc-infty}) and (\ref{eqn-max-norm-uj}), we get:
\begin{equation}
	\norm{\diag(C)}_\infty \leq \frac{1}{1-\epsilon''_1} \norm{\mathbf{g}}_\infty,
\end{equation}
provided that $\epsilon''_1 < 1$, and
\begin{equation}
	\max_{1\leq j\leq n} \norm{u_j}_{\psi_2} \leq \frac{\delta_1}{1-\epsilon''_1} \norm{\mathbf{g}}_\infty =: \epsilon'''_1,
\end{equation}
which is our desired result. We can use a similar argument to bound the $v_j$, which yields:
\begin{equation}
\label{eqn-v-ell-infty}
	\norm{\mathbf{v}}_\infty \leq \epsilon''_2 \norm{\Phi(C)}_\infty, \quad \epsilon''_2 := 2\sqrt{\ln(m)/c_0} \; \delta_2,
\end{equation}
with failure probability at most $e/m^3$,
\begin{equation}
	\norm{\Phi(C)}_\infty \leq \frac{1}{1-\epsilon''_2} \norm{\mathbf{h}}_\infty,
\end{equation}
provided that $\epsilon''_2 < 1$, and
\begin{equation}
	\max_{1\leq j\leq m} \norm{v_j}_{\psi_2} \leq \frac{\delta_2}{1-\epsilon''_2} \norm{\mathbf{h}}_\infty =: \epsilon'''_2.
\end{equation}

We are now ready to prove (\ref{eqn-lasso-noise-1}). We will bound $\norm{\uvec(\Phi_D^\dagger \mathbf{z})}_\infty$, by splitting it into two terms, using (\ref{eqn-zuv-2}), as follows: 

\begin{equation}
	\label{eqn-catfish}
	\begin{split}
		\norm{\uvec(\Phi_D^\dagger \mathbf{z})}_\infty &\leq \norm{\uvec(\Phi_D^\dagger \mathbf{v})}_\infty 
		\\& +\norm{\uvec(\Phi_D^\dagger \Phi(\diag(\mathbf{u})))}_\infty.
	\end{split}
\end{equation}

We can bound the first term in (\ref{eqn-catfish}) as follows:
\begin{equation}
	\begin{split}
		\norm{\uvec(\Phi_D^\dagger \mathbf{v})}_\infty
		&= \max_{1\leq j<j'\leq n} \abs{\mathbf{e}_j^T (\Phi_D^\dagger \mathbf{v}) \mathbf{e}_{j'}} \\
		&= \max_{j<j'} \abs{\mathbf{v}^T \Phi_D(\mathbf{e}_j \mathbf{e}_{j'}^T)} \\
		&= \max_{j<j'} \abs{\mathbf{v}^T \mathbf{\xi}^{(j,j')}},
	\end{split}
\end{equation}
where we defined
\begin{equation}
	\label{eqn-xi}
	\mathbf{\xi}^{(j,j')} = \Phi_D(\mathbf{e}_j \mathbf{e}_{j'}^T) \in \RR^m, \quad 1\leq j<j'\leq n,
\end{equation}
where $\mathbf{e}_j$ denotes the $j$'th standard basis vector in $\RR^n$. Note that the $k$'th entry of $\mathbf{\xi}^{(j,j')}$ is given by 
\begin{equation}
	\mathbf{\xi}^{(j,j')}_k = 4(\mathbf{r}^{(k)})^T \mathbf{e}_j \mathbf{e}_{j'}^T \mathbf{r}^{(k)} = 4 r^{(k)}_j r^{(k)}_{j'},
\end{equation}
hence $\mathbf{\xi}^{(j,j')}$ is bounded by 
\begin{equation}
	\norm{\mathbf{\xi}^{(j,j')}}_2 \leq 4\sqrt{m}.
\end{equation}
We can then upper-bound $\norm{\uvec(\Phi_D^\dagger \mathbf{v})}_\infty$, by using a union bound over all $1\leq j<j'\leq n$, and a Hoeffding-type large deviation bound (Proposition 5.10 in \cite{vershynin2010introduction}). Let $c'>0$ be the universal constant in Proposition 5.10 in \cite{vershynin2010introduction}. Then for any $t\geq 0$, we have:
\begin{equation}
	\begin{split}
		\Pr[&\norm{\uvec(\Phi_D^\dagger \mathbf{v})}_\infty > 4\sqrt{m} (t+\sqrt{2}) \epsilon'''_2 \sqrt{\ln(n)/c'}] \\
		&\leq \sum_{j<j'} \Pr[\abs{\mathbf{v}^T \mathbf{\xi}^{(j,j')}} > 4\sqrt{m} (t+\sqrt{2}) \epsilon'''_2 \sqrt{\ln(n)/c'}] \\
		&\leq \tfrac{1}{2} n(n-1) \cdot e\cdot \exp(-(t+\sqrt{2})^2 \ln n) \\
		&\leq \tfrac{1}{2} e\cdot n^{-t(t+2\sqrt{2})}.
	\end{split}
\end{equation}
Now let $t = 1$. Then we have 
\begin{equation}
	\label{eqn-catfish-1}
	\norm{\uvec(\Phi_D^\dagger \mathbf{v})}_\infty \leq 4\sqrt{m} (1+\sqrt{2}) \epsilon'''_2 \sqrt{\ln(n)/c'}, 
\end{equation}
with failure probability at most $\tfrac{1}{2} e\cdot n^{-(1+2\sqrt{2})}$.

We now bound the second term in (\ref{eqn-catfish}), using a similar approach:
\begin{equation}
	\begin{split}
		\norm{\uvec&(\Phi_D^\dagger \Phi(\diag(\mathbf{u})))}_\infty \\
		&= \max_{1\leq j<j'\leq n} \abs{\mathbf{e}_j^T (\Phi_D^\dagger \Phi(\diag(\mathbf{u}))) \mathbf{e}_{j'}} \\
		&= \max_{j<j'} \abs{\mathbf{u}^T \Phi_G^\dagger \Phi_D(\mathbf{e}_j \mathbf{e}_{j'}^T)} \\
		&= \max_{j<j'} \abs{\mathbf{u}^T \Phi_G^\dagger \mathbf{\xi}^{(j,j')}},
	\end{split}
\end{equation}
where $\Phi_G$ is the submatrix of $\Phi$ that contains those columns of $\Phi$ that are indexed by the subset $G = \set{(j,j) \;|\; j=1,\ldots,n}$. We can bound $\Phi_G^\dagger \mathbf{\xi}^{(j,j')}$ in the following way. For $\ell = 1,\ldots,m$, let $\mathbf{\nu}_\ell \in \RR^n$ be the $\ell$'th row of $\Phi_G$. Note that $\mathbf{\nu}_\ell$ can be written as 
\begin{equation}
	\mathbf{\nu}_\ell = 4( (r^{(\ell)}_1)^2, (r^{(\ell)}_2)^2, \ldots, (r^{(\ell)}_n)^2 ),
\end{equation}
which implies that $\norm{\mathbf{\nu}_\ell}_2 \leq 4\sqrt{n}$. We can then write $\Phi_G^\dagger \mathbf{\xi}^{(j,j')} = \sum_{\ell=1}^m \mathbf{\nu}_\ell (\mathbf{\xi}^{(j,j')})_\ell$, which can be bounded by:
\begin{equation}
	\begin{split}
		\norm{\Phi_G^\dagger \mathbf{\xi}^{(j,j')}}_2 &\leq \sum_{\ell=1}^m \norm{\mathbf{\nu}_\ell}_2 \abs{(\mathbf{\xi}^{(j,j')})_\ell} \\
		&\leq 4\sqrt{n} \norm{\mathbf{\xi}^{(j,j')}}_1 \\
		&\leq 4\sqrt{nm} \norm{\mathbf{\xi}^{(j,j')}}_2 \\
		&\leq 16m\sqrt{n}.
	\end{split}
\end{equation}
We can then upper-bound $\norm{\uvec(\Phi_D^\dagger \Phi(\diag(\mathbf{u})))}_\infty$ using a similar approach as before, i.e., using a union bound over all $1\leq j<j'\leq n$, and a Hoeffding-type large deviation bound. This yields:
\begin{multline}
	\label{eqn-catfish-2}
	\norm{\uvec(\Phi_D^\dagger \Phi(\diag(\mathbf{u})))}_\infty \\
	\leq 16m\sqrt{n} (1+\sqrt{2}) \epsilon'''_1 \sqrt{\frac{\ln(n)}{c'}}, 
\end{multline}
with failure probability at most $\tfrac{1}{2} e\cdot n^{-(1+2\sqrt{2})}$. Finally, we combine (\ref{eqn-catfish}), (\ref{eqn-catfish-1}) and (\ref{eqn-catfish-2}) to prove (\ref{eqn-lasso-noise-1}), as desired.

Next, we will prove (\ref{eqn-lasso-noise-2}), using the same approach we used to prove (\ref{eqn-lasso-noise-1}). We will bound $\norm{\uvec(\Phi_{D,T^c}^\dagger (I-P) \mathbf{z})}_\infty$, by splitting it into two terms, using (\ref{eqn-zuv-2}), as follows: 
\begin{equation}
	\begin{split}
		\label{eqn-dogfish}
		\norm{\uvec&(\Phi_{D,T^c}^\dagger (I-P) \mathbf{z})}_\infty \\
		&\leq  \norm{\uvec(\Phi_{D,T^c}^\dagger (I-P) \mathbf{v})}_\infty \\
		&+ \norm{\uvec(\Phi_{D,T^c}^\dagger (I-P) \Phi(\diag(\mathbf{u})))}_\infty.
	\end{split}
\end{equation}
We can bound the first term in (\ref{eqn-dogfish}) as follows:
\begin{equation}
\begin{split}
	\norm{\uvec&(\Phi_{D,T^c}^\dagger (I-P) \mathbf{v})}_\infty \\
	&= \max_{(j,j')\in T^c} \abs{\mathbf{e}_j^T (\Phi_{D,T^c}^\dagger (I-P) \mathbf{v}) \mathbf{e}_{j'}} \\
	&\leq \max_{j<j'} \abs{\mathbf{v}^T (I-P) \Phi_D(\mathbf{e}_j \mathbf{e}_{j'}^T)} \\
	&= \max_{j<j'} \abs{\mathbf{v}^T (I-P) \mathbf{\xi}^{(j,j')}},
\end{split}
\end{equation}
where we defined $\mathbf{\xi}^{(j,j')}$ previously in equation (\ref{eqn-xi}). Note that 
\begin{equation}
	\norm{(I-P) \mathbf{\xi}^{(j,j')}}_2 \leq \norm{\mathbf{\xi}^{(j,j')}}_2 \leq 4\sqrt{m},
\end{equation}
since $P$ is a projector. We can then repeat the same argument that led to equation (\ref{eqn-catfish-1}). This yields:
\begin{equation}
	\label{eqn-dogfish-1}
	\norm{\uvec(\Phi_{D,T^c}^\dagger (I-P) \mathbf{v})}_\infty \leq 4\sqrt{m} (1+\sqrt{2}) \epsilon'''_2 \sqrt{\frac{\ln(n)}{c'}}, 
\end{equation}
with failure probability at most $\tfrac{1}{2} e\cdot n^{-(1+2\sqrt{2})}$.

We now bound the second term in (\ref{eqn-dogfish}), using a similar approach:
\begin{equation}
	\begin{split}
		\norm{\uvec&(\Phi_{D,T^c}^\dagger (I-P) \Phi(\diag(\mathbf{u})))}_\infty\\
		&= \max_{(j,j')\in T^c} \abs{\mathbf{e}_j^T (\Phi_{D,T^c}^\dagger (I-P) \Phi(\diag(\mathbf{u}))) \mathbf{e}_{j'}} \\
		&\leq \max_{j<j'} \abs{\mathbf{u}^T \Phi_G^\dagger (I-P) \Phi_D(\mathbf{e}_j \mathbf{e}_{j'}^T)} \\
		&= \max_{j<j'} \abs{\mathbf{u}^T \Phi_G^\dagger (I-P) \mathbf{\xi}^{(j,j')}}.
	\end{split}
\end{equation}
We can bound $\Phi_G^\dagger (I-P) \mathbf{\xi}^{(j,j')}$ in the same way as before:
\begin{equation}
	\begin{split}
		\norm{\Phi_G^\dagger (I-P) \mathbf{\xi}^{(j,j')}}_2 &\leq \sum_{\ell=1}^m \norm{\mathbf{\nu}_\ell}_2 \abs{((I-P) \mathbf{\xi}^{(j,j')})_\ell} \\
		&\leq 4\sqrt{n} \norm{(I-P) \mathbf{\xi}^{(j,j')}}_1 \\
		&\leq 4\sqrt{nm} \norm{(I-P) \mathbf{\xi}^{(j,j')}}_2 \\
		&\leq 16m\sqrt{n}.
	\end{split}
\end{equation}
We can then repeat the same argument that led to equation (\ref{eqn-catfish-2}). This yields:
\begin{equation}
	\label{eqn-dogfish-2}
	\begin{split}
		\norm{\uvec&(\Phi_{D,T^c}^\dagger (I-P) \Phi(\diag(\mathbf{u})))}_\infty \\ &\leq 16m\sqrt{n} (1+\sqrt{2}) \epsilon'''_1 \sqrt{\ln(n)/c'}, 
	\end{split}
\end{equation}
with failure probability at most $\tfrac{1}{2} e\cdot n^{-(1+2\sqrt{2})}$. Finally, we combine (\ref{eqn-dogfish}), (\ref{eqn-dogfish-1}) and (\ref{eqn-dogfish-2}) to prove (\ref{eqn-lasso-noise-2}), as desired.

\section{Convergence of the Random Walk\label{sec-proof-tbound}}

Here we prove equation (\ref{eqn-walk-tail-prob}), which shows the convergence of the random walk that is used to choose the evolution time $t$.

We will bound the upper tail probability, $\Pr[s_{N_\text{steps}} \geq s^* + \ell \;|\; s_0 = 0]$. (The lower tail, $\Pr[s_{N_\text{steps}} \leq s^* - \ell \;|\; s_0 = 0]$, can be handled in a similar way.) 

The basic idea is to argue that, if $s_{N_\text{steps}} \geq s^* + \ell$, then the random walk must spend a significant amount of time in the region above $s^*$, and this is unlikely to happen, because the walk tends to drift down towards $s^*$. More precisely, let us define the events 
\begin{equation}\label{eqn-E-1}
	E_{-1} = \biggl(\bigwedge_{k=0}^{N_\text{steps}-1} (s_k \geq s^*+1)\biggr) \wedge (s_{N_\text{steps}} \geq s^*+\ell), 
\end{equation}
and for $0 \leq j \leq N_\text{steps}-1$, 
\begin{equation}\label{eqn-Ej}
	\begin{split}
		E_j &= (s_j = s^*) \wedge \biggl(\bigwedge_{k=j+1}^{N_\text{steps}-1} (s_k \geq s^*+1)\biggr)\\
		& \wedge (s_{N_\text{steps}} \geq s^*+\ell). 
	\end{split}
\end{equation}
That is, the event $E_{-1}$ occurs if the walk is strictly above $s^*$ at all times $\geq 0$, and the walk ends at or above $s^* + \ell$ at time $N_\text{steps}$. The event $E_j$ occurs if the walk touches $s^*$ at time $j$, the walk remains strictly above $s^*$ after time $j$, and the walk ends at or above $s^* + \ell$ at time $N_\text{steps}$. (Note that the event $E_j$ can only occur if $j \leq N_\text{steps}-\ell$, since it takes at least $\ell$ steps for the walk to move from $s^*$ to $s^*+\ell$.) We can bound the upper tail probability in terms of the events $E_{-1}$ and $E_j$, as follows:

\begin{equation}\label{eqn-split-into-E}
	\begin{split}
		&\Pr[s_{N_\text{steps}} \geq s^* + \ell \;|\; s_0 = 0] \\ &\leq \Pr[E_{-1} \;|\; s_0 = 0] + \sum_{j=0}^{N_\text{steps}-\ell} \Pr[E_j \;|\; s_0 = 0].
	\end{split}
\end{equation}
(Note that some of these events may occur with probability 0. For instance, the event $E_0$ (conditioned on $s_0 = 0$) can only occur if $s^* = 0$. Similarly, the event $E_{-1}$ (conditioned on $s_0 = 0$) can only occur if $s^* \leq -1$.)

We now use martingale techniques to bound the probabilities of the events $E_{-1}$ and $E_j$. For $j \geq 1$, let us define random variables 
\begin{equation}
	z_j = s_j - \EE[s_j \;|\; s_{j-1},\ldots,s_0].
\end{equation}
Note that the $z_j$ form a martingale difference sequence with respect to the $s_j$, that is, $\EE[z_j \;|\; s_{j-1},\ldots,s_0] = 0$, and the $z_j$ are bounded, $|z_j| \leq 2$. In addition, for $j' \geq 1$, let us define random variables 
\begin{equation}
	\gamma_{j'} = \sum_{j=1}^{j'} z_j. 
\end{equation}
(We also define $\gamma_0 = 0$.) Note that the $\gamma_j$ form a martingale with respect to the $s_j$, that is, $\EE[\gamma_j \;|\; s_{j-1},\ldots,s_0] = \gamma_{j-1}$, and the $\gamma_j$ have bounded differences, $|\gamma_j - \gamma_{j-1}| \leq 2$. 

Using these definitions, we can write the following expression for $s_{j''} - s_{j'}$ (for any $j'' \geq j'$):
\begin{equation}\label{eqn-sj-gammaj}
	\begin{split}
		s_{j''} &- s_{j'} = \sum_{k=j'+1}^{j''} z_k + \EE[s_k \;|\; s_{k-1},\ldots,s_0] - s_{k-1} \\
		&= \gamma_{j''} - \gamma_{j'} + \sum_{k=j'+1}^{j''} \EE[s_k \;|\; s_{k-1},\ldots,s_0] - s_{k-1}.
	\end{split}
\end{equation}
Furthermore, via Azuma's inequality, we have the following large deviation bound: for any $\lambda \geq 0$, 
\begin{equation}\label{eqn-azuma}
	\Pr[\gamma_{j''} - \gamma_{j'} \geq \lambda] \leq \exp(-\tfrac{\lambda^2}{8(j''-j')}).
\end{equation}

We will now bound the probability $\Pr[E_{-1} \;|\; s_0 = 0]$. As noted earlier, this event implies that $s^* \leq -1$. We know that $s^* \geq -h$, hence $s_0 \leq s^* + h$. Using equations (\ref{eqn-mu-bound}), (\ref{eqn-E-1}) and (\ref{eqn-sj-gammaj}), this implies:
\begin{equation}
	\begin{split}
		\ell - h &\leq s_{N_\text{steps}} - s_0 \\
		&\leq \gamma_{N_\text{steps}} - \gamma_0 - \mu N_\text{steps}.
	\end{split}
\end{equation}
Hence, using (\ref{eqn-azuma}), and the fact that $N_\text{steps} = \frac{h}{\mu} + \eta$, we have: 

\begin{equation}\label{eqn-bound-E-1}
	\begin{split}
		&\Pr[E_{-1} \;|\; s_0 = 0] \\
		&\leq \Pr[\gamma_{N_\text{steps}} - \gamma_0 \geq \ell - h + \mu N_\text{steps} \;|\; s_0 = 0] \\
		&\leq \exp\biggl(-\frac{(\ell - h + \mu N_\text{steps})^2}{8N_\text{steps}}\biggr) \\
		&= \exp\biggl(-\frac{(\ell + \mu\eta)^2}{8(\frac{h}{\mu}+\eta)}\biggr) \\
		&\leq \exp\biggl(-\frac{\mu\eta}{16\max\set{\frac{h}{\mu},\eta}} (\ell + \mu\eta)\biggr) \\
		&= \exp(-\tfrac{\mu}{16} \min\set{\tfrac{\mu\eta}{h},1} (\ell + \mu\eta)).
	\end{split}
\end{equation}

In a similar way, we can bound the probability $\Pr[E_j \;|\; s_0 = 0]$. Suppose that the event $E_j$ happens. Using equations (\ref{eqn-mu-bound}), (\ref{eqn-Ej}) and (\ref{eqn-sj-gammaj}), this implies:
\begin{equation}
	\begin{split}
		\ell &\leq s_{N_\text{steps}} - s_j \\
		&\leq \gamma_{N_\text{steps}} - \gamma_j - \mu (N_\text{steps}-j-1).
	\end{split}
\end{equation}
Hence, using (\ref{eqn-azuma}), and the fact that $\ell \geq 1 \geq \mu$, we have: 
\begin{equation}
	\begin{split}
		&\Pr[E_j \;|\; s_0 = 0] \\
		&\leq \Pr[\gamma_{N_\text{steps}} - \gamma_j \geq \ell + \mu (N_\text{steps}-j-1) \;|\; s_0 = 0] \\
		&\leq \exp\biggl(-\frac{(\ell + \mu (N_\text{steps}-j-1))^2}{8(N_\text{steps}-j)}\biggr) \\
		&\leq \exp(-\tfrac{\mu}{8} (\ell + \mu (N_\text{steps}-j-1))).
	\end{split}
\end{equation}

We can now bound the sum of these probabilities as follows:
\begin{widetext}
\begin{align}\label{eqn-bound-sum-Ej}
	\sum_{j=0}^{N_\text{steps}-\ell} \Pr[E_j \;|\; s_0 = 0] &\leq \sum_{j=0}^{N_\text{steps}-\ell} \exp(-\tfrac{\mu}{8} (\ell + \mu (N_\text{steps}-j-1))) \\
	&=\exp({-\tfrac{\mu}{8} \ell - \tfrac{\mu^2}{8} (N_\text{steps}-1)}) \sum_{j=0}^{N_\text{steps}-\ell} \exp({\tfrac{\mu^2}{8} j}) \\
	&= \exp({-\tfrac{\mu}{8} \ell - \tfrac{\mu^2}{8} (N_\text{steps}-1)}) \frac{\exp({\tfrac{\mu^2}{8} (N_\text{steps}-\ell+1)}) - 1}{\exp({\tfrac{\mu^2}{8}} )- 1} \\
	&\leq \exp({-\tfrac{\mu}{8} \ell - \tfrac{\mu^2}{8} (N_\text{steps}-1)})  \tfrac{8}{\mu^2} \exp({\tfrac{\mu^2}{8} (N_\text{steps}-\ell+1)}) \\
	&\leq \tfrac{8}{\mu^2} \exp(-\tfrac{\mu(\mu+1)}{8} \ell + \tfrac{\mu^2}{4}).
\end{align}
\end{widetext}
Finally, combining equations (\ref{eqn-split-into-E}), (\ref{eqn-bound-E-1}) and (\ref{eqn-bound-sum-Ej}), we get the desired bound on the upper tail $\Pr[s_{N_\text{steps}} \geq s^* + \ell \;|\; s_0 = 0]$. The bound on the lower tail $\Pr[s_{N_\text{steps}} \leq s^* - \ell \;|\; s_0 = 0]$ is proved in a similar way. This completes the proof of (\ref{eqn-walk-tail-prob}).

\section{Bounding SPAM errors \label{sec-proof-singval-bound}}
We first give a derivation of equations \eqref{eqn-spam-1} and  \eqref{eqn-spam-2}. Note that the evolution of $E_0$ and $\rho_0$ is known. Additionally, we have  $\Tr[{E}_0\mathcal{E}_t (\delta\rho)] = \Tr[\mathcal{E}_t^\dagger({E}_0) \delta\rho]$, where $\mathcal{E}^\dagger_t = \exp(\mathcal{L}^\dagger t)$ is the adjoint dephasing map, which coincides with $\mathcal{E}_t$ in our case as the Lindblad operators are all Hermitian.  Therefore, 
\begin{widetext}
	\begin{align}
		\Tr[{E}_0\mathcal{E}_t (\delta\rho)]&=\Tr[\mathcal{E}^\dagger_t(\ket{\mathbf{a}}\bra{\mathbf{a}}) \delta\rho] +\Tr[\mathcal{E}^\dagger_t(\ket{\mathbf{b}}\bra{\mathbf{b}})\delta\rho] +\Tr[\mathcal{E}^\dagger_t (\ket{\mathbf{a}}\bra{\mathbf{b}})\delta\rho] +\Tr[\mathcal{E}_t^\dagger (\ket{\mathbf{b}}\bra{\mathbf{a}})\delta\rho] \\
		&= \Tr[\ket{\mathbf{a}}\bra{\mathbf{a}} \delta\rho] +\Tr[\ket{\mathbf{b}}\bra{\mathbf{b}}\delta\rho] +e^{-\Gamma_\mathbf{r}t}\Tr[\ket{\mathbf{a}}\bra{\mathbf{b}}\delta\rho] +e^{-\Gamma_\mathbf{r}t}\Tr[\ket{\mathbf{b}}\bra{\mathbf{a}}\delta\rho] \\
		&= \eta_{s} + \zeta_s e^{-\Gamma_\mathbf{ab}t}
	\end{align}
	\begin{align}
		\Tr[\delta E\mathcal{E}_t ({\rho}_0)]&=\Tr[\delta E\mathcal{E}_t (\ket{\mathbf{a}}\bra{\mathbf{a}})] +\Tr[\delta E\mathcal{E}_t (\ket{\mathbf{b}}\bra{\mathbf{b}})] +\Tr[\delta E\mathcal{E}_t (\ket{\mathbf{a}}\bra{\mathbf{b}})] +\Tr[\delta E\mathcal{E}_t (\ket{\mathbf{b}}\bra{\mathbf{a}})] \\
		&= \Tr[\delta E\ket{\mathbf{a}}\bra{\mathbf{a}}] +\Tr[\delta E\ket{\mathbf{b}}\bra{\mathbf{b}}] +e^{-\Gamma_\mathbf{r}t}\Tr[\delta E\ket{\mathbf{a}}\bra{\mathbf{b}}] +e^{-\Gamma_\mathbf{r}t}\Tr[\delta E\ket{\mathbf{b}}\bra{\mathbf{a}}] \\
		&= \eta_{m} + \zeta_m e^{-\Gamma_\mathbf{ab}}
	\end{align}
\end{widetext}
We see that both terms decay with the rate $\Gamma_\mathbf{ab}$. Therefore, as noted in Section~\ref{sec-spam-errors}, the outcome of the protocol in the presence of error is given by $\tilde{P}_{\mathbf{ab}}(t) =\frac{1}{2}(1 + \eta_s +\eta_m + (1+\zeta_s +\zeta_m) e^{-\Gamma_\mathbf{ab} t}) + R(t)$, where deviations form a single exponential decay $e^{-\Gamma_\mathbf{ab}t}$ are captured in $R(t) =  \Tr[\delta E \mathcal{E}_t(\delta \rho)]$. To bound the growth of this term we have
\begin{align}
	\label{eqn-lindbladnorm1}
		\abs{\dot{R}(t)} &= \abs{\tfrac{\partial}{\partial t} \Tr[\delta E \mathcal{E}_t(\delta \rho)]} \\
		&= \abs{\Tr[\delta E \mathcal{L}(\delta \rho)]}  \label{eqn-lindbladnorm2} \\ 
		&\leq \norm{\delta E} \norm{\mathcal{L}(\delta \rho)}_\text{tr} \label{eqn-lindbladnorm3}\\
		&\leq \norm{\delta E} \norm{\sum_{ij}L_i \delta \rho L_j^\dagger-\tfrac{1}{2}\{ L_j^\dagger L_i,\delta \rho\}}_\text{tr} \\
		&\leq \norm{\delta E} \sum_{ij}\norm{L_i \delta \rho L_j^\dagger-\tfrac{1}{2}\{ L_j^\dagger L_i,\delta \rho\}}_\text{tr} \label{eqn-lindbladnorm5}\\
		&\leq \norm{\delta E} \norm{\delta \rho}_\text{tr} \sum_{ij}2\norm{L_i}\norm{L_j^\dagger} \label{eqn-lindbladnorm6} \\
		&\leq 2 \epsilon_m \epsilon_s (n+s) 
\end{align}
where we used the H\"older's inequality~\cite{baumgartner2011inequality} in going from \eqref{eqn-lindbladnorm2} to \eqref{eqn-lindbladnorm3}. Additionally, in deriving \eqref{eqn-lindbladnorm6} from \eqref{eqn-lindbladnorm5} we used the fact that $\norm{AB}_\text{tr} \leq \norm{A} \norm{B}_\text{tr}$, which we will prove in the following. Finally, the last line is obtained by noting that $L_i = Z_i$ in our problem and $\norm{Z_i}=1$. 

We can obtain $\norm{AB}_\text{tr} \leq \norm{A} \norm{B}_\text{tr}$, by first showing that for any matrix $A$ and $B$, $\sigma(AB)_i \leq \sigma_{\max}(A)\sigma_{i} (B)$, where $\sigma_i$ are the ordered singular values \cite{zhang2011matrix}. We then use that to obtain
\begin{align}
	\label{eqn-norm1-infinity}
		\norm{AB}_\text{tr} &= \sum_i \sigma_i(AB) \\
		&\leq \sigma_{\max} (B)\sum_i \sigma_i(A) \\
		&=\norm{B}\norm{A}_\text{tr}.
\end{align}
To show that $\sigma(AB)_i \leq \sigma_{\max}(A)\sigma_{i} (B)$, first note that for Hermitian matrices $A$ and $B$ we have:
\begin{equation}\label{eqn-sumeigvals}
	A \geq B \implies \lambda_i(A)\geq \lambda_i(B) 
\end{equation}
where $\lambda_i$'s are ordered eigenvalues. We then show that $\lambda_i(AB)\leq \lambda_{\max}(A)\lambda_{i}(B)$ if $A\geq 0$ and $B\geq0$. Note that $\lambda_i(AB)=\lambda_i(B^{\frac{1}{2}}AB^{\frac{1}{2}})$, which can be seen by
\begin{equation}
	AB \ket{i} = \lambda_i \ket{i} \implies B^{\frac{1}{2}}A B^{\frac{1}{2}}(B^{\frac{1}{2}}\ket{i}) = \lambda_i (B^{\frac{1}{2}}\ket{i})
\end{equation}  
Next, note that $\lambda_{\max}(A)I-A\geq 0$. Therefore, $B^{\frac{1}{2}}(\lambda_{\max}(A)I-A)B^{\frac{1}{2}}\geq 0$. This is because $\bra{x} B^{\frac{1}{2}}(\lambda_{\max}(A)I-A)B^{\frac{1}{2}}\ket{x} = \bra{y} \lambda_{\max}(A)I-A\ket{y}\geq 0$ for all $\ket{y} = B^{\frac{1}{2}}\ket{x}$. We then obtain
\begin{equation}
	\begin{split}
		B^{\frac{1}{2}} A B^{\frac{1}{2}} &\leq B^{\frac{1}{2}}AB^{\frac{1}{2}} + B^{\frac{1}{2}}(\lambda_{\max}(A)I-A)B^{\frac{1}{2}}\\
		&= \lambda_{\max}(A)  B
	\end{split}
\end{equation}
Therefore, $B^{\frac{1}{2}} A B^{\frac{1}{2}}\leq \lambda_{\max}(A) B$ together with \eqref{eqn-sumeigvals} implies that 
\begin{equation}\label{eqn-prodeigvals}
	\lambda_{i}(AB) = \lambda_{i}(B^{\frac{1}{2}}A B^{\frac{1}{2}})\leq \lambda_{\max}(A)\lambda_i(B). 
\end{equation}
Now for arbitrary square matrices $A$ and $B$ we have 
\begin{equation}
	\sigma_i(AB) = \sqrt{\lambda_i (B^\dagger A^\dagger A B)} = \sqrt{\lambda_i (A^\dagger A B B^\dagger )}.
\end{equation}
Therefore, we can use \eqref{eqn-prodeigvals} for positive-semidefinite matrices $A^\dagger A$ and $B B^\dagger$ to obtain 
\begin{equation}
	\lambda_i (A^\dagger A B B^\dagger ) \leq \lambda_{\max}(A^\dagger A) \lambda_{i}(B B^\dagger).
\end{equation}
Since both sides of the inequality are positive, we can take their square root to obtain
\begin{equation}
	\sigma_i(AB) \leq  \sigma_{\max}(A) \sigma_{i}(B).
\end{equation}


\section{Isotropy of the measurements $\Omega_\mathbf{ab}$\label{sec-isotropy-complex}}
In this section we show that the random vectors $\mathbf{q}$ in equation~\eqref{eqn-def-qcomplex} are centered and isotropic (up to a normalization factor of $\sqrt{2}$), as claimed in equation~\eqref{eqn-covariance-complex}. Noting that $\alpha_i$ and $\beta_i$ are chosen uniformly and independently at random from $\{-1,1\}$ we find that
\begin{align}
	\EE[\alpha_i \alpha_j] = \EE[\alpha_i \beta_j] = \EE[\beta_i \beta_j] = 0,
\end{align}
which shows that $\EE[\mathbf{q}]=0$. 

To show that $\EE(\mathbf{q}\mathbf{q}^T) = I$, we have to consider three types of correlations: ($i$) correlations between the elements of ${\rm{uvec}}(\boldsymbol{\alpha} \boldsymbol{\beta}^T-\boldsymbol{\beta} \boldsymbol{\alpha}^T)$, ($ii$) correlations between the elements of ${\rm{uvec}}(\boldsymbol{\alpha} \boldsymbol{\alpha}^T-\boldsymbol{\beta}\boldsymbol{\beta}^T )$, and ($iii$) cross-correlations between these two types of elements.  

We start with correlations of type ($i$). We consider
\begin{multline}
	\EE[(\alpha_i \beta_j - \beta_i \alpha_j)(\alpha_k \beta_l - \beta_k \alpha_l)] \\
	= \EE[\alpha_i \beta_j\alpha_k \beta_l - \alpha_i \beta_j\beta_k \alpha_l - \beta_i \alpha_j\alpha_k \beta_l + \beta_i \alpha_j\beta_k \alpha_l].
\end{multline}
Similar to the analysis in Section~\ref{sec-isotropy-incoherence}, we consider three different cases for how the sets of indices can intersect. Note that in the following we assume $i<j$ and $k< l$, as we are only considering the off-diagonal upper-triangular parts of $V$. 
If $\{i,j\}\cap\{k,l\}=\emptyset$ we have 
\begin{equation}
	\EE[\alpha_i \beta_j\alpha_k \beta_l - \alpha_i \beta_j\beta_k \alpha_l - \beta_i \alpha_j\alpha_k \beta_l + \beta_i \alpha_j\beta_k \alpha_l] = 0, 
\end{equation}
because of the symmetry of the terms.
If $|\{i,j\}\cap\{k,l\}|=1 $, without the loss of generality we assume that $i=k$ and find
\begin{equation}
	\begin{cases}
		\EE[\alpha_i \beta_j\alpha_k \beta_l] &=\EE[\alpha_i^2 \beta_j \beta_l] = \EE[ \beta_j \beta_l] =  0\\
		\EE[\alpha_i \beta_j\beta_k \alpha_l ] &=\EE[\alpha_i \beta_i \beta_j \beta_l] = 0\\
		\EE[\beta_i \alpha_j\alpha_k \beta_l] &= \EE[\beta_i\alpha_i  \alpha_j\beta_l] =0\\
		\EE[\beta_i \alpha_j\beta_k \alpha_l] &= \EE[\beta_i^2 \alpha_j \alpha_l] =  \EE[\alpha_j \alpha_l] = 0
	\end{cases}. 
\end{equation}
Therefore, we find
\begin{equation}
	\EE[\alpha_i \beta_j\alpha_k \beta_l - \alpha_i \beta_j\beta_k \alpha_l - \beta_i \alpha_j\alpha_k \beta_l + \beta_i \alpha_j\beta_k \alpha_l] = 0. 
\end{equation}
Note that if we had assumed $j=l$ we would have obtained the same results.
Finally, when $|\{i,j\}\cap\{k,l\}|=2 $, we have that $i=k$ and $j=l$ and find
\begin{align}
	&\EE[\alpha_i \beta_j\alpha_k \beta_l - \alpha_i \beta_j\beta_k \alpha_l - \beta_i \alpha_j\alpha_k \beta_l + \beta_i \alpha_j\beta_k \alpha_l] \\
	&= \EE[\alpha_i^2 \beta_j^2 - \alpha_i \beta_i \alpha_j\beta_j - \alpha_i \beta_i \alpha_j\beta_j+ \beta_i^2 \alpha_j^2]\\
	&= 2 
\end{align}

Next we consider correlations of type ($ii$), that is, 
\begin{multline}
	\EE[(\alpha_i \alpha_j-\beta_i \beta_j)(\alpha_k \alpha_l-\beta_k \beta_l)] \\
	= \EE[\alpha_i\alpha_j\alpha_k\alpha_l -\alpha_i \alpha_j\beta_k \beta_l-\beta_i \beta_j\alpha_k \alpha_l+\beta_i \beta_j\beta_k \beta_l],
\end{multline}
where $i< j$ and $k< l$. Similarly, we consider different cases: 
If $\{i,j\}\cap\{k,l\}=\emptyset$ we have 
\begin{equation}
	\EE[\alpha_i\alpha_j\alpha_k\alpha_l -\alpha_i \alpha_j\beta_k \beta_l-\beta_i \beta_j\alpha_k \alpha_l+\beta_i \beta_j\beta_k \beta_l] = 0, 
\end{equation}
because of the symmetries.
If $|\{i,j\}\cap\{k,l\}|=1 $, without the loss of generality we assume that $i=k$ and find
\begin{equation}
	\begin{cases}
		\EE[\alpha_i\alpha_j\alpha_k\alpha_l] &=\EE[\alpha_i^2 \alpha_j \alpha_l]  = \EE[ \alpha_j \alpha_l] = 0\\
		\EE[\alpha_i \alpha_j\beta_k \beta_l] &=\EE[\alpha_i \beta_i \alpha_j \beta_l] = 0\\
		\EE[\beta_i \beta_j\alpha_k \alpha_l] &= \EE[\beta_i\alpha_i  \beta_j\alpha_l] = 0\\
		\EE[\beta_i \beta_j\beta_k \beta_l] &=  \EE[\beta_i^2 \beta_j \beta_l] =  \EE[\beta_j \beta_l] =0
	\end{cases}. 
\end{equation}
Therefore, we find
\begin{equation}
	\EE[\alpha_i\alpha_j\alpha_k\alpha_l -\alpha_i \alpha_j\beta_k \beta_l-\beta_i \beta_j\alpha_k \alpha_l+\beta_i \beta_j\beta_k \beta_l]= 0.
\end{equation}
Finally, when $|\{i,j\}\cap\{k,l\}|=2$, we have $i=k$ and $j=l$ and find
\begin{align}
	&\EE[\alpha_i\alpha_j\alpha_k\alpha_l -\alpha_i \alpha_j\beta_k \beta_l-\beta_i \beta_j\alpha_k \alpha_l+\beta_i \beta_j\beta_k \beta_l] \\&= \EE[\alpha_i^2 \alpha_j^2 - \alpha_i \beta_i \alpha_j\beta_j - \alpha_i \beta_i \alpha_j\beta_j+ \beta_i^2 \beta_j^2]\\
	&= 2.
\end{align}

Finally, we consider cross-correlations of type ($iii$), which are given by 
\begin{multline}
	\EE[(\alpha_i \beta_j - \beta_i \alpha_j)(\alpha_k \alpha_l - \beta_k \beta_l)] \\
	= \EE[\alpha_i \beta_j\alpha_k \alpha_l - \alpha_i \beta_j\beta_k \beta_l - \beta_i \alpha_j\alpha_k \alpha_l + \beta_i \alpha_j\beta_k \beta_l],
\end{multline} 
with $i<j$ and $k<l$. 
If $\{i,j\}\cap\{k,l\}=\emptyset$ we have 
\begin{equation}
	\EE[\alpha_i \beta_j\alpha_k \alpha_l - \alpha_i \beta_j\beta_k \beta_l - \beta_i \alpha_j\alpha_k \alpha_l + \beta_i \alpha_j\beta_k \beta_l] = 0, 
\end{equation}
because of the symmetries.
If $|\{i,j\}\cap\{k,l\}|=1 $, without the loss of generality we assume that $i=k$ and find
\begin{equation}
	\begin{cases}
		\EE[\alpha_i \beta_j\alpha_k \alpha_l ] &=\EE[\alpha_i^2 \beta_j \alpha_l]  = \EE[ \beta_j \alpha_l] = 0\\
		\EE[\alpha_i \beta_j\beta_k \beta_l] &=\EE[\alpha_i \beta_j \beta_i \beta_l] = 0\\
		\EE[\beta_i \alpha_j\alpha_k \alpha_l] &= \EE[\beta_i\alpha_j  \alpha_i\alpha_l] = 0\\
		\EE[\beta_i \alpha_j\beta_k \beta_l] & = \EE[\beta_i^2 \alpha_j \beta_l] =  \EE[\alpha_j \beta_l] = 0
	\end{cases}. 
\end{equation}
Therefore, we find
\begin{equation}
	\EE[\alpha_i \beta_j\alpha_k \alpha_l - \alpha_i \beta_j\beta_k \beta_l - \beta_i \alpha_j\alpha_k \alpha_l + \beta_i \alpha_j\beta_k \beta_l]= 0.
\end{equation}
Finally, when $|\{i,j\}\cap\{k,l\}|=2$, we assume $i=k$ and $j=l$ and find
\begin{align}
	&\EE[\alpha_i \beta_j\alpha_k \alpha_l - \alpha_i \beta_j\beta_k \beta_l - \beta_i \alpha_j\alpha_k \alpha_l + \beta_i \alpha_j\beta_k \beta_l] \\
	&= \EE[\alpha_i^2 \beta_j \alpha_j - \alpha_i \beta_i \beta_j^2 - \beta_i \alpha_i \alpha_j^2 + \beta_i^2 \alpha_j\beta_j]\\
	&= 0
\end{align}

Therefore, given all cases considered here we arrive at equation~\eqref{eqn-covariance-complex}.


\bibliographystyle{apsrev}
\bibliography{correlatednoise-prx}

\end{document}